\documentclass[twocolumn]{aa}

\usepackage[pdftex,dvipsnames]{xcolor}  % Coloured text etc.
\usepackage{graphicx}
\usepackage{txfonts}
\usepackage{lipsum}                     % Dummytext
\usepackage{xargs}                      % Use more than one optional parameter in a new commands
\usepackage{soul}

\usepackage{hyperref}
\usepackage[nameinlink]{cleveref}
\usepackage[normalem]{ulem}
\usepackage{float}
\usepackage{placeins}
\usepackage{bm}
\usepackage{academicons}

\usepackage{xcolor}

\usepackage{hyperref}
\hypersetup{
    colorlinks=true,
    linkcolor=blue,
    filecolor=blue,      
    urlcolor=blue,
    citecolor=blue,
    }

\usepackage{natbib}
\bibpunct{(}{)}{;}{a}{}{,} % to follow the A&A style

\newcommand{\Computer}{\footnote{All calculations involving $R^{\scriptscriptstyle \mathrm{III}}$ (in its angle-dependent formulation) were performed on the Piz Daint supercomputer of the Swiss National Supercomputing Center (CSCS) on Cray XC40 nodes (\href{https://www.cscs.ch/computers/piz-daint}{https://www.cscs.ch/computers/piz-daint}).
An XC40 compute node is equipped with two Intel® Xeon® E5-2695 v4 microprocessors at 2.10GHz (2x18 cores, 64/128 GB RAM).
}}

\begin{document}

\title{
Assessment of the CRD approximation for the\\ observer's frame $R^{\scriptscriptstyle \rm III}$ redistribution matrix
} 

\titlerunning{Assessment of CRD approximation for $R^{\scriptscriptstyle \rm III}$}

\authorrunning{Riva et al.}

\author{Simone Riva \inst{1,2} \orcid{0000-0003-1032-0609}
        \and
        Nuno Guerreiro\inst{2,1}
        \and
        Gioele Janett\inst{2,1} \orcid{0000-0003-3247-6612}
        \and
        Diego Rossinelli\inst{3} \orcid{0000-0003-1600-4068}
        \and
        Pietro Benedusi\inst{4,1} \orcid{0000-0001-7799-5999}
        \and
        \\ Rolf Krause\inst{1,5}
        \and
        Luca Belluzzi\inst{2,6,1}  \orcid{0000-0002-8775-0132}
}

\institute{
        Euler Institute, Universit\`a della Svizzera italiana (USI), East Campus, Sector D, Via la Santa 1 6962 Lugano, Switzerland
        \and
        Istituto ricerche solari Aldo e Cele Daccò (IRSOL), Faculty of Informatics, Università della Svizzera italiana (USI), Via Patocchi 57, 6605 Locarno-Monti, Switzerland
        \and 
        Institute for Computational and Mathematical Engineering, Stanford University, Center for Turbulence Research, Room 204, 481 Panama Mall, Stanford, CA 94305-3024
        \and
        Simula Research Laboratory, Kristian Augusts gate 23, 0164 Oslo, Norway
        \and
        FernUni, Schinerstrasse 18, 3900 Brig
        \and
        Leibniz-Institut f\"ur Sonnenphysik (KIS), Schöneckstr. 6, D-79104, Freiburg, Germany\\
        \email{simone.riva@usi.ch}
}

\abstract
{
Approximated forms of the $R^{\scriptscriptstyle \mathrm{II}}$ and $R^{\scriptscriptstyle \mathrm{III}}$ redistribution matrices are frequently applied to simplify the numerical solution of the radiative transfer problem for polarized radiation, taking partial frequency redistribution (PRD) effects into account.
A widely used approximation for $R^{\scriptscriptstyle \mathrm{III}}$ is to consider its expression under the assumption of complete frequency redistribution (CRD) in the observer's frame ($R^{\scriptscriptstyle \mathrm{III-CRD}}$).
The adequacy of this approximation for modeling the intensity profiles has been firmly established. By contrast, its suitability for modeling scattering polarization signals has only been analyzed in a few studies, %which only 
considering simplified settings.
}
{
In this work, we aim at quantitatively assessing the impact and the range of validity of the $R^{\scriptscriptstyle \mathrm{III-CRD}}$ approximation in the modeling of scattering polarization.
} 
{
We first present an analytic comparison between $R^{\scriptscriptstyle \mathrm{III}}$ and $R^{\scriptscriptstyle \mathrm{III-CRD}}$.
We then compare the results of radiative transfer calculations, out of local thermodynamic equilibrium, performed with $R^{\scriptscriptstyle \mathrm{III}}$ and $R^{\scriptscriptstyle \mathrm{III-CRD}}$ in realistic one-dimensional atmospheric models.
We focus on the chromospheric Ca~{\sc i} line at 4227\,{\AA} and on the photospheric Sr~{\sc i} line at 4607\,{\AA}.
}
{
The $R^{\scriptscriptstyle \mathrm{III-CRD}}$ approximation provides accurate results for the Ca~{\sc i} 4227\,{\AA} line. 
Only when velocities are included, some appreciable discrepancies can be found, especially for lines of sight close to the disk center.
The approximation performs well also for the Sr~{\sc i} 4607\,{\AA} line, especially in the absence of magnetic fields or when a micro-turbulent field is included. 
However, some appreciable errors appear when deterministic magnetic fields or bulk velocities are considered.
}
{
Our results show that the $R^{\scriptscriptstyle \mathrm{III-CRD}}$ approximation is suited for the PRD modeling of the scattering polarization signals of strong chromospheric lines, both in the core and wings.
With a few minor exceptions, this approximation is also suitable for photospheric lines, %where 
although PRD effects generally play a minor role in their modeling.
}


\maketitle

%%%%%%%%%%%%%%%%%%%%%%%%%%%%%%%%%%%%%%%%%%%%%%%%%%%%%%%%%%%%%%%%%%%%%%%%%%%%%%%%%%%%%%%%%%%%%%%%%%%%%%%%%%%%%%%%%%%%
\section{Introduction}

Significant scattering polarization signals are observed in several strong resonance lines of the solar spectrum, such as H~{\sc i} Ly-$\alpha$ \citep{kano2017}, Mg~{\sc ii} k \citep{rachmeler2022}, Ca~{\sc ii} K, Ca~{\sc i} 4227\,{\AA}, and Na~{\sc i} D$_2$ \citep[e.g.,][]{stenflo1980,stenflo1997sss,gandorfer2000,gandorfer2002}.
These signals, which are characterized by broad profiles with large amplitudes in the line wings, encode a variety of information on the thermodynamic and magnetic properties of the upper layers of the solar atmosphere \citep[e.g.,][]{trujillo2014,trujillo2017}. 
A correct modeling of these profiles requires solving the radiative transfer (RT) problem for polarized radiation in non-local thermodynamic equilibrium (non-LTE) conditions, taking partial frequency redistribution (PRD) effects (i.e., correlations between the frequencies of incoming and outgoing photons in scattering processes) into account 
\citep[e.g.,][]{faurobert1992,stenflo1994,Holzreuter+al2005,Belluzzi2012}.

A powerful formalism to describe PRD phenomena is that of the redistribution function \citep[e.g.,][]{hummer1962non,mihalas1978stellar}, which is generalized to the redistribution matrix in the polarized case \citep[e.g.,][]{domke88redMat,stenflo1994,bommier1997masterI,bommier1997masterII}.
For resonance lines, the redistribution matrix is given by the sum of two terms, commonly labeled as $R^{\scriptscriptstyle \rm II}$ and $R^{\scriptscriptstyle \rm III}$ according to the notation introduced by \citet{hummer1962non}.
The $R^{\scriptscriptstyle \rm II}$ matrix describes scattering processes that are coherent in frequency in the atomic reference frame, while $R^{\scriptscriptstyle \rm III}$ describes scattering processes that are totally incoherent in the same frame \citep[e.g.,][]{bommier1997masterI,bommier1997masterII}.\footnote{In this work, the terms \emph{coherent} and \emph{totally incoherent} are used in the sense that the frequencies of the incident and scattered radiation are fully correlated and completely uncorrelated, respectively.}
The linear combination of $R^{\scriptscriptstyle \rm II}$ and $R^{\scriptscriptstyle \rm III}$ allows taking frequency redistribution effects due to elastic collisions with neutral perturbers into account \citep[e.g.,][]{bommier1997masterII}.
In the observer's frame, the Doppler shifts due to the thermal motions of the atoms are responsible for further frequency redistribution effects. 
The Doppler effect actually induces a complex coupling between the frequencies and propagation directions of the incident and scattered radiation, which makes the evaluation of both $R^{\scriptscriptstyle \rm II}$ and $R^{\scriptscriptstyle \rm III}$, as well as the solution of the whole non-LTE RT problem, notoriously challenging from the computational standpoint.
For this reason, approximate expressions of the redistribution matrices in the observer's frame, in which such coupling is loosened, have been proposed and extensively used.
In this context, a common choice is to use the angle-averaged (AA) expression of $R^{\scriptscriptstyle \rm II}$, hereafter $R^{\scriptscriptstyle \rm II-AA}$ \citep[e.g.,][]{mihalas1978stellar,rees1982,bommier1997masterII,leenaarts2012}, and the expression of $R^{\scriptscriptstyle \rm III}$ obtained under the assumption that the scattering processes described by this matrix are totally incoherent \emph{also} in the observer's frame, hereafter $R^{\scriptscriptstyle \rm III-CRD}$ \citep[e.g.,][]{mihalas1978stellar,bommier1997masterII,ballester2017transfer}.
The latter is also referred to as the assumption of complete frequency redistribution (CRD) in the observer's frame.

The impact and the range of validity of the $R^{\scriptscriptstyle \rm II-AA}$ approximation in the modeling of scattering polarization has been discussed by several authors \citep[e.g.,][]{faurobert1987,faurobert1988,nagendra2002Hanle,nagendra2011Spectal,sampoorna2011spectral,anusha2012,sampoorna2015,sampoorna2017,nagendra2020,delpinoaleman2020,janett2021a}. 
These studies showed that the use of $R^{\scriptscriptstyle \rm II-AA}$ can introduce significant (and hardly predictable) inaccuracies in the modeling of the line-core signals, while it seems to be suited for modeling the wing lobes and their magnetic sensitivity through the magneto-optical (MO) effects. 
By contrast, little effort has been directed to determine the suitability of the $R^{\scriptscriptstyle \rm III-CRD}$ assumption when modeling scattering polarization.
Although this approximation has no true physical justification, it proved to be suitable for modeling the intensity profiles of spectral lines \citep[e.g., Chapter 13 of][and references therein]{mihalas1978stellar}.
However, \citet{bommier1997masterII} pointed out that it may lead to appreciable errors when polarization phenomena are taken into account.
In that work, the author considered a $90^\circ$ scattering process of an unpolarized beam of radiation in the presence of a weak magnetic field, and compared the polarization of the scattered radiation calculated considering the exact expression of $R^{\scriptscriptstyle \rm III}$ and the $R^{\scriptscriptstyle \rm III-CRD}$ approximation, finding appreciable differences between the two cases.\footnote{It must be observed that the author applied the CRD approximation after introducing in $R^{\scriptscriptstyle \rm III}$ other simplifications suited for the weak-field regime \citep[see Sect.~4 of][]{bommier1997masterII}.}
The exact expression of $R^{\scriptscriptstyle \rm III}$ has been used in the non-LTE RT calculations with isothermal one-dimensional (1D) atmospheric models by \cite{sampoorna2011spectral} and \cite{supriya2012effect} for the non-magnetic case, \cite{nagendra2002Hanle} and \cite{supriya2013efficient} in the weak field Hanle regime and, more recently, \citet{sampoorna2017} in the more general Hanle-Zeeman regime.
%
% \cred{More recently, \citet{sampoorna2017} analyzed the suitability of the $R^{\scriptscriptstyle \rm III-CRD}$ approximation through non-LTE RT calculations in isothermal one-dimensional (1D) models of the solar atmosphere, considering an ideal spectral line.}
%
In this last paper, the authors analyzed the suitability of the $R^{\scriptscriptstyle \rm III-CRD}$ approximation for modeling scattering polarization, considering an ideal spectral line. They first concluded that the
 use of $R^{\scriptscriptstyle \rm III-CRD}$ introduces some inaccuracies, especially in the line wings, when considering optically-thin self-emitting slabs in the presence of weak magnetic fields (i.e., fields for which the ratio $\Gamma_B$ between the magnetic splitting of the Zeeman sublevels and the natural width of the upper level of the considered transition is in the order of unity).
Moreover, these discrepancies are much smaller when stronger magnetic fields ($\Gamma_B \approx 100$) are considered \citep[see Fig.~1 and Sect. 4 in][]{sampoorna2017}.
However, when considering an atmospheric model with a greater optical depth, they found that the use of $R^{\scriptscriptstyle \rm III-CRD}$ does not seem to produce any noticeable effect on the emergent Stokes profiles already for weak fields \citep[see Fig.~2  in][]{sampoorna2017}.
To the best of the authors' knowledge, the aforementioned works are the only ones reporting PRD calculations of scattering polarization performed using the exact expression of $R^{\scriptscriptstyle \rm III}$.

Taking advantage of a new solution strategy for the non-LTE RT problem for polarized radiation, tailored for including PRD effects \citep[see][]{benedusi2022}, we provide a quantitative analysis of the suitability of the $R^{\scriptscriptstyle \rm III-CRD}$ approximation, considering more general and realistic settings than in previous studies.
In particular, we show the results of non-LTE RT calculations of the scattering polarization signals of two different spectral lines (i.e., Ca~{\sc i} 4227\,{\AA} and Sr~{\sc i} 4607\,{\AA}) in 1D models of the solar atmosphere, both semi-empirical and extracted from 3D magneto-hydrodynamic (MHD) simulations, and accounting for the impact of realistic magnetic and bulk velocity fields.

The paper is structured as follows. 
Section~\ref{sec:transfer} presents the basic equations for the RT problem for polarized radiation, as well as the redistribution matrix formalism for describing PRD effects.
Section~\ref{sec:discretization} briefly presents the discretization and algebraic formulation of the problem, together with the 
considered numerical solution strategy.
Section~\ref{sec:RIIIvsRIIICRD} exposes analytical and numerical 
comparisons between $R^{\scriptscriptstyle \mathrm{III}}$ and $R^{\scriptscriptstyle \mathrm{III-CRD}}$.
Section~\ref{sec:prior_analysis} provides some general considerations on the role of $R^{\scriptscriptstyle \rm III}$ in the RT modeling of scattering polarization, and on the expected impact of $R^{\scriptscriptstyle \rm III-CRD}$.
Sections~\ref{sec:FALC_atmos} and~\ref{sec:bifrost_atmos} report comparisons of non-LTE RT calculations of scattering polarization signals in realistic 1D settings, performed with $R^{\scriptscriptstyle \mathrm{III}}$ and $R^{\scriptscriptstyle \mathrm{III-CRD}}$. 
Finally, Section~\ref{sec:conclusions} provides remarks and conclusions.

%%%%%%%%%%%%%%%%%%%%%%%%%%%%%%%%%%%%%%%%%%%%%%%%%
\section{Transfer problem for polarized radiation}
\label{sec:transfer}
%%%%%%%%%%%%%%%%%%%%%%%%%%%%%%%%%%%%%%%%%%%%%%%%%
The intensity and polarization of a radiation beam are fully described by the four Stokes parameters $I$, $Q$, $U$, and $V$.
The Stokes parameter $I$ is the intensity, $Q$ and $U$ quantify the linear polarization, while $V$ quantifies the circular polarization \citep[e.g.,][hereafter LL04]{degl2006polarization}. 
The Stokes parameters and other physical quantities involved in the RT problem are generally functions of the frequency and propagation direction of the radiation beam under consideration, as well as of the spatial point and time. 
Hereafter, we will assume stationary conditions so that all quantities are time-independent.

\newcommand{\azimuthHO}{\footnote{
The azimuth is defined in the half-open interval $\chi = [0,2\pi)$ because it has a periodicity of $2\pi$ and the two limits refer to the same coordinate.
%Since the azimuth $\chi$ has a periodicity of $2\pi$, it is defined in the half-open interval $[0,2\pi)$ because the two limits refer to the same coordinate.
}}

A beam of radiation propagating in a medium (e.g., the plasma of a stellar atmosphere) is modified as a result of the interaction with the particles therein %(e.g., atoms, ions, molecules, etc.), 
and the possible presence of magnetic, electric, and bulk velocity fields. 
This modification is fully described by the RT equation for polarized radiation, consisting of coupled systems of first-order, inhomogeneous, ordinary differential equations.
Defining the Stokes parameters as the four components of the Stokes vector $\vec{I}=(I,Q,U,V)^T=(I_1,I_2,I_3,I_4)^T\in \mathbb{R}^4$, the RT equation for a beam of radiation of frequency $\nu$, propagating along the direction specified by the unit vector $\vec{\Omega}=(\theta,\chi)\in[0, \pi]\times[0, 2\pi)$ can be written as\azimuthHO{}
\begin{equation}
 	\vec{\nabla}_{{\vec{\Omega}}} \vec{I}(\vec{r},\vec{\Omega},\nu) = 
	- K(\vec{r},\vec{\Omega},\nu) \vec{I}(\vec{r},\vec{\Omega},\nu) + 
	\bm{\varepsilon}(\vec{r},\vec{\Omega},\nu) \, ,
	\label{eq:rte}
\end{equation}
where $\vec{\nabla}_{\vec{\Omega}}$ denotes the directional derivative along $\vec{\Omega}$, and $\vec{r}\in D$ is the spatial point in the domain $D\subset \mathbb{R}^3$. 
The propagation matrix $K \in \mathbb{R}^{4 \times 4}$ is given by
\begin{equation}
	K = \begin{pmatrix}
%		\eta_I & \eta_Q & \eta_U & \eta_V \\
%		\eta_Q & \eta_I & \rho_V & -\rho_U \\
%		\eta_U & -\rho_V & \eta_I & \rho_Q \\
%		\eta_V & \rho_U & -\rho_Q & \eta_I \\
%	\end{pmatrix} = 
%	\begin{pmatrix}
		\eta_1 & \eta_2 & \eta_3 & \eta_4 \\
		\eta_2 & \eta_1 & \rho_4 & -\rho_3 \\
		\eta_3 & -\rho_4 & \eta_1 & \rho_2 \\
		\eta_4 & \rho_3 & -\rho_2 & \eta_1 \\
	\end{pmatrix} \, . 
\label{eq:prop_matrix}
\end{equation}
The elements of $K$ describe absorption ($\eta_1$), dichroism ($\eta_2$, $\eta_3$, and $\eta_4$), and anomalous dispersion ($\rho_2$, $\rho_3$, and $\rho_4$) phenomena.
The emission vector $\bm{\varepsilon}=(\varepsilon_1,\varepsilon_2,\varepsilon_3,\varepsilon_4)^T\in \mathbb{R}^4$
describes the radiation emitted by the plasma in the four Stokes parameters.

In this work, we consider the contribution to $K$ and $\vec{\varepsilon}$ brought by both line and continuum processes.
These contributions, labelled with the superscripts $\ell$ and $c$, respectively, simply add to each other:
\begin{align*}
    K(\vec{r},\vec{\Omega},\nu) &= K^{\ell}(\vec{r},\vec{\Omega},\nu) + K^c(\vec{r},\vec{\Omega},\nu) \, , \\
    \vec{\varepsilon}(\vec{r},\vec{\Omega},\nu) &= \vec{\varepsilon}^{\ell}(\vec{r},\vec{\Omega},\nu) + \vec{\varepsilon}^{c}(\vec{r},\vec{\Omega},\nu) \, .
\end{align*}
%
%The former will be labeled with superscript $\ell$ and the latter with the superscript $c$.}
%
In the frequency interval of a given spectral line, the line contribution to the elements of $K$ and $\vec{\varepsilon}$ %generally 
depends on the state of the atom (or molecule) giving rise to that line.
In general, this state has to be determined by solving a set of rate equations (statistical equilibrium equations), which describe the interaction of the atom with the 
radiation field (radiative processes), other particles present in the plasma (collisional processes), and external magnetic and electric fields.
When the statistical equilibrium equations have an analytic solution, the line contribution to the emission vector can be directly related to the radiation field that illuminates the atom (incident radiation) through the redistribution matrix formalism. 
More precisely, the line contribution to the emission vector can be written as the sum of two terms, namely,
\begin{equation}
	\label{eq:emissionVec}
%	\begin{aligned}
	\vec{\varepsilon}^{\ell}(\vec{r},\vec{\Omega},\nu) = %\, &
	\vec{\varepsilon}^{\ell, \mathrm{sc}}(\vec{r},\vec{\Omega},\nu) +
	\vec{\varepsilon}^{\ell, \mathrm{th}}(\vec{r},\vec{\Omega},\nu) %\\ & + 
	% \vec{\varepsilon}^{c, \mathrm{sc}}(\vec{r},\vec{\Omega},\nu) +
	% \vec{\varepsilon}^{c, \mathrm{th}}(\vec{r},\vec{\Omega},\nu)
	\, ,
%\end{aligned}
\end{equation}
% %
where $\vec{\varepsilon}^{\ell, \mathrm{sc}}$ describes the contribution from atoms that are radiatively excited (scattering term), and $\vec{\varepsilon}^{\ell, \mathrm{th}}$ describes the contribution from atoms that are collisionally excited (thermal term).
% %
%
Following the convention that primed and unprimed quantities refer to the incident and scattered radiation, respectively, the line scattering term can be written through the scattering integral
\begin{equation}
%	\varepsilon_i^{\mathrm{sc}}(\vec{r},&\vec{\Omega},\nu) = k_L(\vec{r}) 
%	\nonumber \\
%	& \times 
%	\int_{\mathbb{R}_{+}} \mathrm{d} \nu' \, \oint 
%	\frac{\mathrm{d} \vec{\Omega}'}{4 \pi}
%	\sum_{j=1}^4 R_{ij}(\vec{r},\vec{\Omega}',\vec{\Omega},\nu',\nu ) \, 
%	I_{j}(\vec{r},\vec{\Omega}',\nu') \, ,
	\vec{\varepsilon}^{\ell, \mathrm{sc}}(\vec{r},\vec{\Omega},\nu) = k_L(\vec{r}) 
%	\nonumber \\
%	& \times 
	\int_{\mathbb{R}_{+}} \!\! \mathrm{d} \nu' \oint 
	\frac{\mathrm{d} \vec{\Omega}'}{4 \pi} R(\vec{r},\vec{\Omega},\vec{\Omega}',\nu,\nu' ) \, \vec{I}(\vec{r},\vec{\Omega}',\nu') \, ,
	\label{eq:scat_int}
\end{equation}
where the factor $k_L$ is the frequency-integrated absorption coefficient, and $R \in \mathbb{R}^{4 \times 4}$ the redistribution matrix.
The redistribution matrix element $R_{ij}$ relates the emissivity in the $i$-th Stokes parameter in direction $\vec{\Omega}$ and frequency $\nu$ to the $j$-th Stokes parameter of the incident radiation with direction $\vec{\Omega}'$ and frequency $\nu'$.

In this work, we consider an atomic system composed of two-levels (two-level atom) with an unpolarized and infinitely-sharp lower level (see Appendix~\ref{sec:atomic_model} for more details), and we apply the corresponding redistribution matrix as derived in \cite{bommier1997masterII}:
\begin{equation*}
	R(\vec{r},\vec{\Omega},\vec{\Omega}',\nu,\nu') = 
	R^{\scriptscriptstyle \mathrm{II}}(\vec{r},\vec{\Omega},
	\vec{\Omega}',\nu,\nu')+ 
	R^{\scriptscriptstyle \mathrm{III}}(\vec{r},\vec{\Omega},
	\vec{\Omega}',\nu,\nu') \, .
\end{equation*}
The explicit expressions of $R^{\scriptscriptstyle \mathrm{II}}$ and $R^{\scriptscriptstyle \mathrm{III}}$ in both the atomic and observer's reference frames %, taking the quantization axis along the local vertical 
are given in Appendix~\ref{sec:R_analytic}.
%The expressions of 
The line contribution to the elements of the propagation matrix $K$, and %of
the line thermal %contribution to the 
emissivity $\vec{\varepsilon}^{\ell, \mathrm{th}}$
in the observer's frame can be found in %{\textst{Eq.}~\eqref{eq:prop_matrix}}\textst{, and}
Appendix~\ref{sec:thermal}.
%The continuum contributions to $K$ and $\vec{\varepsilon}$}
%In this work, we also consider the contribution brought by continuum processes to the elements of the propagation matrix and the emission vector.
%These continuum contributions 
%are discussed in Appendix~\ref{sec:continuum}.
The continuum contribution to the emissivity can also be written as the sum of a thermal and a scattering term (see Eq.~\eqref{eq:epsilon_c}), where the latter can be expressed as a scattering integral fully analogous to Eq.~\eqref{eq:scat_int}.
A detailed discussion of the continuum terms is provided in Appendix~\ref{sec:continuum}.
For simplicity, hereafter we will use the notation $\vec{\varepsilon}^{\textrm{th}}$ and $\vec{\varepsilon}^{\textrm{sc}}$ to refer to the thermal and scattering contributions to the emissivity, including both line and continuum.

The RT problem consists in finding a self-consistent solution for the RT equation (\ref{eq:rte}) and the equation for the scattering contribution to the emissivity (\ref{eq:scat_int}). %which can be easily generalized in order to include continuum processes (see Appendix~\ref{sec:continuum} for more details).
This problem is in general non-linear because of the factor $k_L$ appearing in the line contribution to both the elements of the propagation matrix and the emission coefficients.
This factor is proportional to the population of the lower level, which in turn depends non-linearly on the incident radiation field through the statistical equilibrium equations.

We linearize the problem with respect to $\vec{I}$, by fixing a-priori the population of the lower level, and thus the factor $k_L$.
In such scenario, whose suitability is discussed in \cite{janett2021b} and \cite{benedusi2021}, the propagation matrix $K$ and the thermal %contributions to the emissivity
%\cblue{term}
contribution to the emissivity $\vec{\varepsilon}^{\mathrm{th}}$ is independent of $\vec{I}$, while the %\textst{line} 
scattering 
term $\vec{\varepsilon}^{\mathrm{sc}}$
% and $\vec{\varepsilon}^{c, \mathrm{sc}}$
depends on it linearly through the scattering integral. %~\eqref{eq:scat_int}.
The population of the lower level can be taken either from the atmospheric model (if provided) or from independent calculations. 
The latter can be carried out with available non-LTE RT codes that possibly neglect polarization (which is expected to have a minor impact on the population of ground or metastable levels), but allow considering multi-level atomic models. %
In this way, accurate estimates of the lower level population can be used, and reliable results can be obtained in spite of the simplicity of the considered two-level atomic model \citep[e.g.,][]{janett2021a,alsina2021}.

%%%%%%%%%%%%%%%%%%%%%%%%%%%%%%%%%%%%%%%%%%%%%%%%%%%%%%%%%%%%
\section{Numerical solution strategy}
\label{sec:discretization}

Following the works by \citet{janett2021b} and \citet{benedusi2021,benedusi2022}, we first present an algebraic formulation of the considered linearized RT problem for polarized radiation.
Starting from this formulation, we then apply a parallel solution strategy, based on Krylov iterative methods with physics-based preconditioning.
This strategy allows us to %routinely 
solve the problem in semi-empirical 1D models of the solar atmosphere, considering the exact expression of both $R^{\scriptscriptstyle \rm II}$ and $R^{\scriptscriptstyle \rm III}$ redistribution matrices.
The same approach, coupled with a new domain decomposition technique, has been recently generalized to the 3D case \citep{benedusi3Drt2023}.

\begin{figure}
    \centering
    \includegraphics[width=0.25\textwidth]{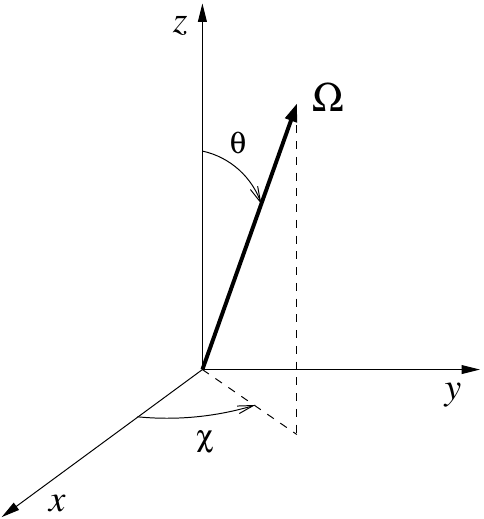}
    \caption{Right-handed Cartesian reference system considered in the problem. The $z$-axis is directed along the local vertical. 
    Any vector is specified through its polar angles $\theta$ (inclination) and $\chi$ (azimuth).}
    \label{fig:ref_sys}
\end{figure}

The problem presented in Sect.~\ref{sec:transfer} is discretized by introducing suitable grids for the continuous variables $\vec{r}$, $\vec{\Omega}$, and $\nu$.
The angular grid $\{\vec{\Omega}_i\}_{i=1}^{N_\Omega}$ is determined by the quadrature rule chosen to solve the angular integral in Eq.~(\ref{eq:scat_int}). 
We consider a right-handed Cartesian reference system with $z$-axis directed along the vertical. 
A given direction $\vec{\Omega}$ is %defined accordingly.
specified by the inclination $\theta$ and the azimuth $\chi$, defined as shown in Fig.~\ref{fig:ref_sys}.
We discretize the inclination through two Gauss-Legendre grids with  $N_\theta/2$ points each, one for $\mu=\cos(\theta) \in (-1,0)$, and one for $\mu \in (0, 1)$.
For the azimuth, we use a grid of $N_\chi$ equidistant points. 
The angular quadrature is then the spherical Cartesian product \cite[e.g.,][]{davis2007methods} of the Gauss-Legendre rule for the inclination and the trapezoidal rule for the azimuth.
This approach is commonly used in RT applications and allows for the implementation of fast algorithms.
The frequency grid $ \{ \nu_j \}_{j=1}^{N_\nu}$ is chosen to adequately sample the spectral line under investigation. 

In a 1D (plane-parallel) setting, the spatial coordinate $\vec{r}=(x,y,z)$ can be replaced by the vertical coordinate $z \in [z_{\mathrm{min}},z_{\mathrm{max}}]$, and the 
RT equation \eqref{eq:rte} can be rewritten as
\begin{equation}
%	\cos{(\theta)} \frac{\mathrm{d}}{\mathrm{d}z} I_i(z,\theta,\chi,\nu) 
%	= -\sum_{j=1}^4K_{ij}&(z,\theta,\chi,\nu) I_j(z,\theta,\chi,\nu)\\ 
%	&+ \varepsilon_i(z,\theta,\chi,\nu) \, .
 	\cos{(\theta)} \frac{\mathrm{d}}{\mathrm{d}z} \vec{I}(z,\theta,\chi,\nu) 
	= -K(z,\theta,\chi,\nu) \vec{I}(z,\theta,\chi,\nu) + \bm{\varepsilon}(z,\theta,\chi,\nu) \, .
	\label{eq:rte_1D}
\end{equation}
The spatial grid $\{z_k\}_{k=1}^{N_z}$ is provided by the considered atmospheric model.
%Moreover, 
We assume that the radiation entering the atmosphere from the lower boundary is isotropic, unpolarized, and equal to the Planck function, and that no radiation is entering from the upper boundary.
Equation~\eqref{eq:rte_1D} is thus equipped with the following boundary conditions:
\begin{align*}
%	& I_i(z_{\rm min},\theta,\chi,\nu) = %I_1^{\mathrm{in}}(\nu) 
%	B(\nu) \, \delta_{i,1}
%	\quad \mathrm{for} \; \theta \in [0,\pi/2), \, \forall \chi, \, 
%	\forall \nu \, , \\
%	& I_i(z_{\rm max},\theta,\chi,\nu) = 0
%	\qquad \qquad \mathrm{for} \; \theta \in (\pi/2,\pi], \, \forall \chi, \, \forall \nu \, ,
    & \vec{I}\left(z_{\rm min},\theta,\chi,\nu \right) = %I_1^{\mathrm{in}}(\nu) 
	\left[ B_{P}\left(\nu, T(z_{\rm{min}}) \right), 0, 0, 0 \right]^T
    % \quad 
	& %\mathrm{for} 
    \theta \in [0,\pi/2), \, \forall \chi, \, \forall \nu \, , \\
	& \vec{I}\left(z_{\rm max},\theta,\chi,\nu \right) = \vec{0}
	% \qquad \qquad \qquad \;\; \; 
	% & \mathrm{for}
 & \theta \in (\pi/2,\pi], \, \forall \chi, \, \forall \nu \, ,
\end{align*}
%
%
%with $I_1^{\mathrm{in}}(\nu) = B(\nu)$, 
where $B_{P}$ is the Planck function and $T$ the effective temperature $T$ at $z_{\mathrm{min}}$.
Given the propagation matrices and the emission vectors at all height points $\{z_k\}_{k=1}^{N_z}$, for a given direction $(\theta,\chi)$ and frequency $\nu$, the RT equation \eqref{eq:rte} can be numerically solved along that direction and at that frequency by applying a suitable integrator. 
This process is generally known as \emph{formal solution}.
In this work, we use the $L$-stable DELO-linear method combined with a linear conversion to optical depth \citep[e.g.,][]{janettI,janettIV}.
An analysis of the stability properties of this method can be found in \citet{janettIII}.

Being $N=4 N_z N_\theta N_\chi N_\nu$ the total number of unknowns, we introduce the collocation vectors $\vec{\mathrm{I}}\in\mathbb R^N$ and $\vec{\epsilon}\in\mathbb R^N$, which contain the numerical approximations of the Stokes vector and the emission vector, respectively, at all nodes.
Given $\vec{\epsilon}$, the solution of all discretized RT equations \eqref{eq:rte_1D} can be written as
\begin{equation}
	\vec{\mathrm{I}} = \Lambda \vec{\epsilon} + \vec{t} \, ,
	\label{eq:tran_op}
\end{equation}
where $\Lambda: \mathbb{R}^N \rightarrow \mathbb{R}^N$ is the transfer operator, which encodes the formal solver and the propagation matrix, and $\vec{t}\in\mathbb R^N$ is a vector encoding the boundary conditions.

Similarly, given $\vec{\mathrm{I}}$, the discrete computation of the emission vector can be written as
\begin{equation}
	\vec{\epsilon} = \vec{\epsilon}^{\rm sc} + 
	\vec{\epsilon}^{\rm th} = \Sigma \vec{\mathrm{I}} + 
	\vec{\epsilon}^{\rm th} \, ,
	\label{eq:scat_op}
\end{equation}
%
% \cblue{where $ \vec{\epsilon}^{\rm sc} = \vec{\epsilon}^{\ell, \rm sc} + \vec{\epsilon}^{c, \rm sc}$ describes the discretized scattering term, while $\vec{\epsilon}^{\rm th} = \vec{\epsilon}^{\ell, \rm th} + \vec{\epsilon}^{c, \rm th}$  describes the discretized thermal term.}
%
where $\vec{\epsilon}^{\rm sc}$ and $\vec{\epsilon}^{\rm th}$ encode the scattering and thermal contributions (including both line and continuum processes), respectively, as described in Sect.~\ref{sec:transfer}.
The scattering operator $\Sigma: \mathbb{R}^N \rightarrow \mathbb{R}^N$ encodes the numerical evaluation of the scattering integral~\eqref{eq:scat_int} and thus depends on the chosen numerical quadratures.
In general, the operator $\Sigma$ is given by the sum of different components, namely,
\begin{equation*}
	\Sigma = \Sigma^{\scriptscriptstyle \rm II} + 
	\Sigma^{\scriptscriptstyle \rm III} + 
	\Sigma^{\rm c}\,,
\end{equation*}
where $\Sigma^{\scriptscriptstyle \rm II}$ and $\Sigma^{\scriptscriptstyle \rm III}$ encode the contributions from $R^{\scriptscriptstyle \rm II}$ and $R^{\scriptscriptstyle \rm III}$, respectively, and $\Sigma^{\rm c}$ the scattering contribution from the continuum. 
The vector $\vec{\epsilon}^{\rm th} \in \mathbb{R}^N$ encodes the thermal emissivity.

Under the assumption that $k_L$ is known a-priori (see end of Sect.~\ref{sec:transfer}), the operators $\Lambda$ and $\Sigma$ are linear with respect to $\vec{\mathrm{I}}$.
Combining \eqref{eq:tran_op} and \eqref{eq:scat_op}, the whole discrete RT problem can be formulated as a linear system of size $N$ with unknown $\vec{\mathrm{I}}$, namely
\begin{equation}
	(Id - \Lambda \Sigma) \vec{\mathrm{I}} = \Lambda \vec{\epsilon}^{\mathrm{th}} + \vec{t} \, ,
	\label{eq:lin_sys}
\end{equation}
where $Id\in\mathbb R^{N\times N}$ is the identity matrix.
The right-hand-side vector $\Lambda \vec{\epsilon}^{\mathrm{th}} + \vec{t}$ can be computed a priori by solving \eqref{eq:rte_1D} with thermal contributions only (i.e., by performing a single formal solution with $\vec{\epsilon}^{\mathrm{sc}}=\vec{0}$).
%An anylsis of the matrices $\Lambda$, $\Sigma$, and $Id-\Lambda\Sigma$ can 
%be found in \citet{benedusi2022}.
The action of the matrices $\Lambda$, $\Sigma$, and $Id-\Lambda\Sigma$ can be encoded in a matrix-free form
% as shown in algorithms 1, 2, and 3, respectively, in 
\citep[see][]{benedusi2022}.

We solve the linear system \eqref{eq:lin_sys} by applying a matrix-free, preconditioned GMRES method.
Preconditioning is performed by describing scattering processes in the limit of CRD.
% following the theoretical framework presented in LL04. 
This corresponds to substituting the operator $\Sigma^{\scriptscriptstyle \rm II} + \Sigma^{\scriptscriptstyle \rm III}$ with a new operator $\Sigma^{\scriptscriptstyle \rm CRD}$ that is much cheaper to evaluate.
The explicit expression of $\Sigma^{\scriptscriptstyle \rm CRD}$ for the considered atomic model can be found in Chap. 10 of LL04. %\citet{degl2006polarization}. 
The reader is referred to \citet{benedusi2022} for more details on this solution strategy.

We conclude this section by briefly describing the numerical strategy used to handle bulk velocities.
These velocities introduce Doppler shifts, which depend on the propagation direction of the considered radiation, in the expressions of the elements of $K$ and $\bm{\varepsilon}$ (see Appendix~\ref{sec:R_analytic}).
We compute the emission vector in a reference frame in which the bulk velocity is zero (comoving reference frame).
In this reference frame, there are no Doppler shifts, and this allows us to significantly reduce the number of evaluations of the redistribution matrix when performing the scattering integral \eqref{eq:scat_int}.
The drawback of this approach is that it requires transforming the incident radiation field from the observer's frame to the comoving frame and then transforming the emission vector back to the observer's frame.
These steps are performed by means of high-order interpolations (e.g., cubic splines) on the frequency axis.
The additional computational cost introduced by such interpolations is more than compensated by the reduced number of evaluations of the redistribution matrices.

%%%%%%%%%%%%%%%%%%%%%%%%%%%%%%%%%%%%%%%%%%%%%%%%%

\section{Comparison of $R^{\scriptscriptstyle \mathrm{III}}$ 
and $R^{\scriptscriptstyle \mathrm{III-CRD}}$}
\label{sec:RIIIvsRIIICRD}

In this section, we first present an analytical comparison between the general angle-dependent expression of the $R^{\scriptscriptstyle \mathrm{III}}$ matrix (hereafter denoted as $R^{\scriptscriptstyle \mathrm{III}}$) and its $R^{\scriptscriptstyle \mathrm{III-CRD}}$ approximation.
We then review the reasons why the $R^{\scriptscriptstyle \mathrm{III-CRD}}$ approximation allows for a significant simplification of the problem from a computational point of view, and also provide a presentation of the challenges faced when considering 
$R^{\scriptscriptstyle \mathrm{III}}$, focusing on algorithmic aspects.

%%%%%%%%%%%%%%%%%%%%%%%%
\subsection{Analytic considerations}

In the formalism of the irreducible spherical tensors for polarimetry (see Chapter~5 of LL04), the $R^{\scriptscriptstyle \mathrm{III}}$ redistribution matrix in the observer's reference frame \eqref{eq:RIII_obs_vertical} is given by the product between the scattering phase matrix $\mathcal{P}^{KK'}_{Q} \in \mathbb{C}^{4\times 4}$ \eqref{eq:PKKpQ}, which only depends on $\vec{r}$, $\vec{\Omega}$, and $\vec{\Omega}'$, and the redistribution function $\mathcal{R}^{{\scriptscriptstyle \mathrm{III}},KK'}_Q \in \mathbb{C}$ \eqref{eq:RIII_KKpQ_obs}, which also depends on $\nu$ and $\nu'$. 
This factorization also holds for the $R^{\scriptscriptstyle \mathrm{III-CRD}}$ matrix, leaving unchanged the $\mathcal{P}^{KK'}_{Q}$ matrix,
while replacing the redistribution function $\mathcal{R}^{{\scriptscriptstyle \mathrm{III}},KK'}_Q$ with the approximation $\mathcal{R}^{{\scriptscriptstyle \mathrm{III-CRD}},KK'}_Q$ given by~\eqref{eq:RIII_KKpQ_CRD}.
A fundamental difference between $\mathcal{R}^{{\scriptscriptstyle \mathrm{III}},KK'}_Q$ and $\mathcal{R}^{{\scriptscriptstyle \mathrm{III-CRD}},KK'}_Q$ is that the former depends on the scattering angle
$\Theta = \arccos{(\vec{\Omega} \cdot \vec{\Omega}')}$ (i.e., the angle between the directions $\vec{\Omega}$ and $\vec{\Omega}'$), while the latter does not.

\begin{figure*}
\includegraphics[width=\textwidth]{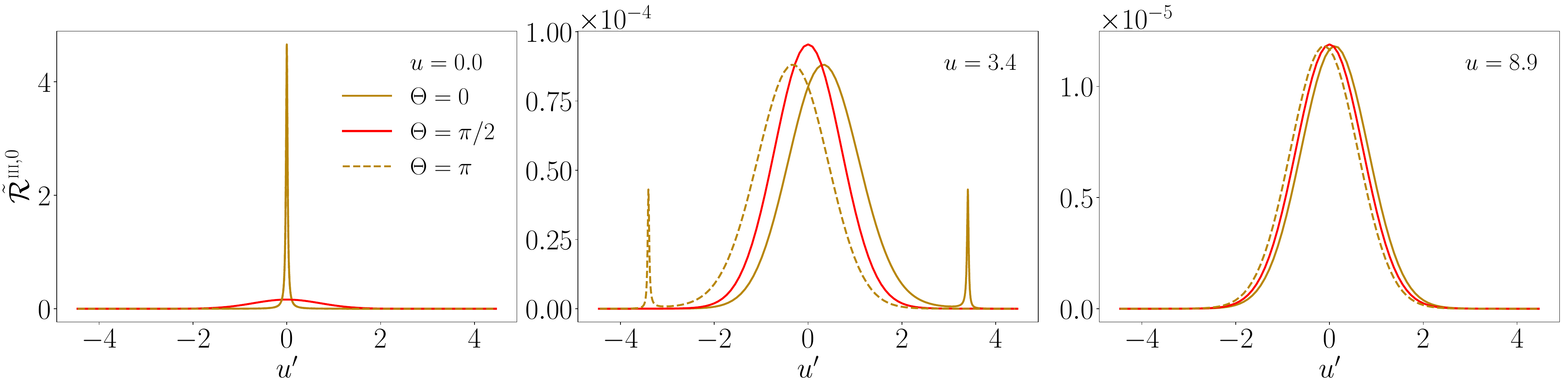}
\caption{
Comparison of $\tilde{\mathcal{R}}^{{\scriptscriptstyle \mathrm{III}},0}$ as a function of $u'$ for three different scattering angles $\Theta$ (see legends), for $u=0$ (\emph{left panel}), $u=3.4$ (\emph{middle panel}), and $u=8.9$ (\emph{right panel}).
The function is calculated considering a transition between levels with total angular momenta $J_\ell=0$ and $J_u=1$, and a damping parameter $a=0.01$. 
The factor $(\beta^0 - \alpha)$ is calculated setting $\alpha=0$ and using the value of $\beta^0$ corresponding to the Ca~{\sc i} 4227\,{\AA} line at a height of 300\,km in the FAL-C atmospheric model.}
\label{fig:RIII_KKpQZero}
\end{figure*}

To analyze the dependence on $\Theta$, we start considering the simpler expressions that $\mathcal{R}^{{\scriptscriptstyle \mathrm{III}},KK'}_Q$ and $\mathcal{R}^{{\scriptscriptstyle \mathrm{III-CRD}},KK'}_Q$ assume %when no bulk velocities and magnetic fields are present.
in the absence of magnetic fields ($B=0$).
In this case, the magnetic shifts $u_{M_u M_\ell}$ (see Eq.~\eqref{eq:reduced_shifts}) %\eqref{eq:reduced_freq} 
are zero, and the sums over the magnetic quantum numbers $M$ appearing in \eqref{eq:RIII_KKpQ_obs} and \eqref{eq:gen_prof_obs} can be performed analytically.
If we additionally assume that there are no bulk velocities ($v_b=0$) or, without loss of generality, we evaluate the redistribution matrix in the comoving frame (see Sect.~\ref{sec:discretization}), the Doppler shifts $u_b$ (see Eq.~\eqref{eq:reduced_shifts}) %\eqref{eq:reduced_freq}) 
also vanish, and the dependence on the propagation directions $\vec{\Omega}$ and $\vec{\Omega}'$ is only through the scattering angle $\Theta$.
Expressing the functions in terms of the reduced frequencies $u$ and $u'$ (see Eq.~\eqref{eq:reduced_freq}), in the limit of $B=0$ and $v_b=0$, one finds:
\begin{equation*}
        \mathcal{R}^{{\scriptscriptstyle {\rm III}},KK'}_Q
        (\vec{r},\vec{\Omega},\vec{\Omega}',u,u')
        \xrightarrow[B=0,\,v_b=0]{} \delta_{KK'} \,
        \tilde{\mathcal{R}}^{{\scriptscriptstyle {\rm III}},K}
        (\vec{r},\Theta,u,u') \, ,
\end{equation*}
with
\begin{equation}
        \tilde{\mathcal{R}}^{{\scriptscriptstyle {\rm III}},K}
        (\vec{r},\Theta,u,u') =
        \left( \beta^K(\vec{r}) - \alpha(\vec{r}) \right)
	W^K(J_\ell,J_u) \, \tilde{\mathcal{I}}(\vec{r},\Theta,u,u') \, .
 \label{eq:RIII_K}
\end{equation}
The quantities $\beta^K$ and $\alpha$ are given by \eqref{eq:beta_KQ} and \eqref{eq:alpha_Q}, respectively,
in the limit of no magnetic fields, while $W^K$ is defined as (see Eq.~(10.17) of LL04)
\begin{equation*}
    W^K \left( J_\ell , J_u \right ) =  3 \left( 2 J_u + 1 \right ) 
    \begin{Bmatrix}
    1 & 1 & K\\ 
    J_u & J_u  & J_\ell
    \end{Bmatrix}^2,
\end{equation*}
where $J_\ell$ and $J_u$ are the total angular momenta of the lower and upper level, respectively, and the operator in curly brackets is the Wigner's 6-$j$ symbol. 
The quantity $\tilde{\mathcal{I}}$ takes different analytical forms depending on the value of $\Theta$:
\begin{itemize}
    \item 
if $\Theta \in (0, \pi)$ (i.e., if $\vec{\Omega}' \ne \vec{\Omega},-\vec{\Omega}$) 
	\begin{align}
    \label{eq:int_final_zero} 
	\tilde{\mathcal{I}}(\vec{r},\Theta,u,u')
	& = \frac{1}{\pi^2 \, \sin(\Theta)} \,  
	\int \mathrm{d} y \, \mathrm{exp} \left( -y^2 \right) 
    \\
	& \quad \times 
	H \left( \frac{a(\vec{r})}{\sin(\Theta)} ,
	\frac{u + y \cos(\Theta)}{\sin(\Theta)} \right) 
	\phi\!\left( a(\vec{r}), u' + y \right), \nonumber
	\end{align}

 \item
if $\Theta = 0$ (i.e., forward scattering $\vec{\Omega}' = \vec{\Omega}$) 
\begin{align}
	\tilde{\mathcal{I}}(\vec{r},\Theta=0,u,u') 
	& = \frac{1}{\pi^{5/2}} \,  
	\int \mathrm{d} y \, \mathrm{exp} \left( -y^2 \right)
	\nonumber \\
	& \quad \times 
	\phi \left( a(\vec{r}), u + y\right) 
	\phi \left( a(\vec{r}), u' + y \right) \, ,
	\label{eq:int_final_T0_zero}
\end{align}

\item
if $\Theta = \pi$ (i.e., backward scattering $\vec{\Omega}' = -\vec{\Omega}$) 
	\begin{align}
	\tilde{\mathcal{I}}(\vec{r},\Theta=\pi,u,u')
	& = \frac{1}{\pi^{5/2}} \,  
	\int \mathrm{d} y \, \mathrm{exp} \left( -y^2 \right)
	\nonumber \\
	& \quad \times 
	\phi \left( a(\vec{r}), u - y \right) 
	\phi \left( a(\vec{r}), u' + y \right) \, ,
	\label{eq:int_final_Tpi_zero}
	\end{align}
\end{itemize}
In the previous equations, $H$ is the Voigt profile (i.e., the real part of the Faddeeva function, see Chapter~5 of LL04) and $\phi$ the Lorentzian profile 
%{from Eq.~\eqref{eq:phi_obs}}.
(i.e., the real part of the function $\varphi$ defined in Eq.~\eqref{eq:phi_obs}).
Similarly, it can be shown that
\begin{equation*}
    \mathcal{R}^{{\scriptscriptstyle {\rm III-CRD}},KK'}_Q(\vec{r},\vec{\Omega},\vec{\Omega}',u,u')
    \xrightarrow[B=0,\,v_b=0]{} \delta_{KK'} \, \tilde{\mathcal{R}}^{{\scriptscriptstyle {\rm III-CRD}},K}(\vec{r},u,u') \, ,
\end{equation*}
with
\begin{align*}
    \tilde{\mathcal{R}}^{{\scriptscriptstyle {\rm III-CRD}},K}(\vec{r},u,u') & =
    \left( \beta^K(\vec{r}) - \alpha(\vec{r}) \right) W^K(J_\ell,J_u) 
	\nonumber \\
	& \quad \times
	\frac{1}{\pi} \, H(a(\vec{r}),u) \, H(a(\vec{r}),u') \, .
\end{align*}
Recalling that the Voigt profile is, by definition, a convolution between a Gaussian and a Lorentzian distribution (e.g., Chapter 5 of LL04), from Eq. \eqref{eq:int_final_zero}, it can be easily seen that for $\Theta=\pi/2$
%
%{ %
\begin{equation}
\begin{aligned}
	\tilde{\mathcal{I}}(\vec{r},\Theta\!=\!\pi/2,u,u')
	&  = \frac{1}{\pi^2} H\left( a(\vec{r}) ,
	u \right)\!\!\int\!\! \mathrm{d} y \, \mathrm{exp} \left( -y^2 \right) 
	\phi\left( a(\vec{r}), u' + y \right) \nonumber \\ 
	& = \frac{1}{\pi} H \left( a(\vec{r}), u \right) 
	H \left( a(\vec{r}), u' \right) \, .
	\label{eq:RIII_equival}
\end{aligned}
\end{equation} %}
This shows that the approximate $\tilde{\mathcal{R}}^{{\scriptscriptstyle {\rm III-CRD}},K}$ function corresponds to the exact $\tilde{\mathcal{R}}^{{\scriptscriptstyle {\rm III}},K}$ for $\Theta=\pi/2$, namely,
\begin{equation}
	\tilde{\mathcal{R}}^{{\scriptscriptstyle {\rm III-CRD}},K}(\vec{r},u,u') =
    \tilde{\mathcal{R}}^{{\scriptscriptstyle {\rm III}},K}(\vec{r},\Theta=\pi/2,u,u') \, .
 \label{eq:RIIIADCRDEq}
\end{equation}
For scattering angles $\Theta \ne \pi/2$, the functions $\tilde{\mathcal{R}}^{{\scriptscriptstyle {\rm III}},K}$ and $\tilde{\mathcal{R}}^{{\scriptscriptstyle {\rm III-CRD}},K}$ are generally different, and this difference increases as $\Theta$ approaches the two limiting cases of forward and backward scattering (i.e., $\Theta=0$ and $\Theta=\pi$, respectively).
This can be clearly seen in Fig.~\ref{fig:RIII_KKpQZero}, where $\tilde{\mathcal{R}}^{{\scriptscriptstyle {\rm III}},0}$ is plotted as a function of $u'$ for different values of $u$ and $\Theta$. 
In particular, we compare the profiles for $\Theta=0$ and $\pi$, to that of $\Theta=\pi/2$, which, as shown above, corresponds to $\tilde{\mathcal{R}}^{{\scriptscriptstyle {\rm III-CRD}},0}$.
For any value of $u$, the function $\tilde{\mathcal{R}}^{{\scriptscriptstyle {\rm III-CRD}},0}$ shows a relatively broad profile (full width at half maximum of about 2), centered at $u'=0$.    
The amplitude of the peak is maximum at $u=0$ (left panel) and quickly decreases by various orders of magnitude already at $u \approx 3$ (middle panel). 
%For $\Theta=0$ and $\pi$, 
The behavior of the function $\tilde{\mathcal{R}}^{{\scriptscriptstyle {\rm III}},0}$ is much more complex. 
For $u=0$ (left panel) and $\Theta=0$ or $\pi$, %for both values of $\Theta$, 
the profile is centered at $u'=0$ and it is much sharper and with a larger amplitude than that of $\tilde{\mathcal{R}}^{{\scriptscriptstyle {\rm III-CRD}},0}$.
% \cred{
% 	For $u \approx 3$ (middle panel) and $\Theta=0\,(\pi)$, the function is characterized by a broad profile, similar in amplitude and width to that of $\tilde{\mathcal{R}}^{{\scriptscriptstyle {\rm III-CRD}},0}$, but with its maximum slightly shifted to positive (negative) values of $u'$. 
% 	Moreover, it shows a secondary peak \st{of similar amplitude}, but much sharper than the previous one, centered at $u'=u\,(-u)$.
% 	For $u \approx 9$ (right panel), the secondary peak becomes negligible, while the main one is very similar to that of $\tilde{\mathcal{R}}^{{\scriptscriptstyle {\rm III-CRD}},0}$.
% }
	For $u \approx 3$ (middle panel) and $\Theta=0\,(\mathrm{resp.\;} \Theta= \pi)$, the function is characterized by a broad profile, similar in amplitude and width to that of $\Theta=\pi / 2$ (which is equivalent to $\tilde{\mathcal{R}}^{{\scriptscriptstyle {\rm III-CRD}},0}$, see Eq.~\eqref{eq:RIIIADCRDEq}), but with its maximum slightly shifted to positive (resp. negative) values of $u'$. 
 Additionally, it shows a secondary sharp peak centered at $u'=u\,(\mathrm{resp.\;} u'=-u)$.
	For $u \approx 9$ (right panel), in the case of $\Theta=0\,(\mathrm{resp.\;} \Theta= \pi)$, the secondary peak becomes negligible, while the main one is very similar to that of $\tilde{\mathcal{R}}^{{\scriptscriptstyle {\rm III-CRD}},0}$.
    Noting that the dependence on $K$ is limited to the factors $\beta^K$ and $W^K$ (see Eq.~\eqref{eq:RIII_K}), it is possible to state that $\tilde{\mathcal{R}}^{{\scriptscriptstyle {\rm III-CRD}},K}$ and $\tilde{\mathcal{R}}^{{\scriptscriptstyle {\rm III}},K}$ are {\it de facto} equivalent in the wings, independently of the value of $\Theta$. By contrast, they differ in the core and near wings, where the magnitude of the discrepancies strongly depends on $\Theta$.

\begin{figure*}[ht!]
\centering
    \includegraphics[width=\textwidth]{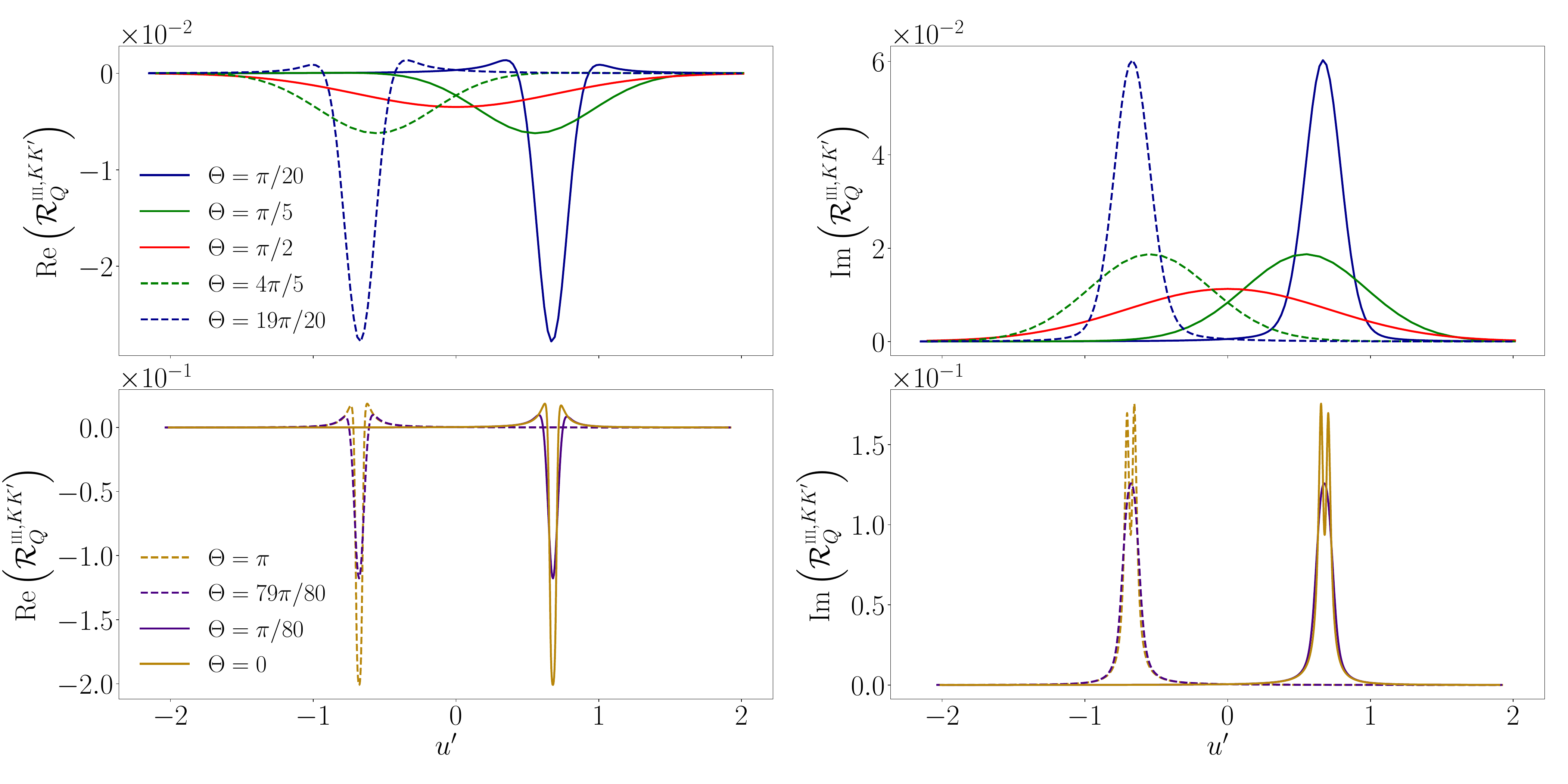}
    \caption{
		Real (\emph{left column}) and imaginary (\emph{right column}) parts of $\mathcal{R}^{{\scriptscriptstyle \mathrm{III}},KK'}_Q$ as a function of $u'$, for different scattering angles (see legend).
		We consider the component with $K=K'=2$ and $Q=-2$. The function is evaluated at $u=0.76$, including a magnetic field of 30\,G.
    	The other parameters are the same as in Fig. \ref{fig:RIII_KKpQZero}.
	}
    \label{fig:RIII_KKpQ_Tph}
\end{figure*}

In the presence of a magnetic field, the redistribution function $\mathcal{R}^{{\scriptscriptstyle \mathrm{III}},KK'}_Q$ \eqref{eq:RIII_KKpQ_obs} is given by a linear combination of the functions $\mathcal{I}_{(M_u M_\ell),(M'_u M'_\ell)}$ \eqref{eq:int_final} -- \eqref{eq:int_final_Tpi}.
These functions are fully analogous to $\tilde{\mathcal{I}}$ \eqref{eq:int_final_zero} -- \eqref{eq:int_final_Tpi_zero}, the only difference being that the second and third functions in the integrand (i.e., the profiles depending on the scattered and incident radiation, respectively) are shifted by $u_{M_u M_\ell}$ and $u_{M'_u M'_\ell}$, respectively.
It can be shown that also in the presence of magnetic fields (but still neglecting bulk velocities), the following relation holds:
\begin{equation*}
    \mathcal{R}^{{\scriptscriptstyle {\rm III-CRD}},KK'}_Q(\vec{r},u,u') =
    \mathcal{R}^{{\scriptscriptstyle {\rm III}},KK'}_Q(\vec{r},\Theta=\pi/2,u,u') \, .
\end{equation*}
%
%
% Figure~\ref{fig:RIII_KKpQ_Tph} shows the component $\mathcal{R}^{{\scriptscriptstyle \mathrm{III}},22}_{-2}$ 
%
When $\Theta \ne \pi/2$, $\mathcal{R}^{{\scriptscriptstyle \mathrm{III}},KK'}_Q$ (as function of $u'$) differs in general from $\mathcal{R}^{{\scriptscriptstyle \mathrm{III-CRD}},KK'}_Q$. 
As in the unmagnetized case, the difference is marginal for large values of $u$ (i.e., $u>6$) independently of the magnetic field strength, while it can be very significant in the core and near wings, especially when $\Theta$ is close to 0 or $\pi$. 
When approaching these limit cases, the curves for $\mathcal{R}^{{\scriptscriptstyle \mathrm{III}},KK'}_Q$ can show very sharp peaks and the contribution of the various Zeeman components, split in frequency by the magnetic field, can eventually be resolved. 
As an illustrative example, Fig.~\ref{fig:RIII_KKpQ_Tph} shows the component $\mathcal{R}^{{\scriptscriptstyle \mathrm{III}},22}_{-2}$ plotted as a function of $u'$, for $u = 0.76$, $B=30$\,G, and different values of $\Theta$. We note that this component has a non-zero imaginary part since $Q\ne 0$. 
%
% \textbf{It can be noted that also in the generic magnetized case in the wings (i.e. for large enoght values of $u$) the CRD approximation and the exact expressions of $\mathcal{R}^{{\scriptscriptstyle {\rm III}},KK'}_Q$ are practically equivalent, following the behavior shown in the right panel of Fig.~\ref{fig:RIII_KKpQZero}, where the effects of $\Theta$ are negligible.}}
%
% {\textbf{To illustrate a possible behavior of the function in the near wing (where $\Theta$ plays a key role),} we analyze the component with $K=K'=2$ and $Q=-2$, which has a non-vanishing imaginary part ($Q \ne 0$), and highlight some general properties of the functions in the magnetized case.}
%Figure~\ref{fig:RIII_KKpQ_Tph} shows \{a plot of} %the component 
%$\mathcal{R}^{{\scriptscriptstyle \mathrm{III}},22}_{-2}$
%, plotted 
%as a function of $u'$, for $u=0.76$ and different values of $\Theta$, \{considering a magnetic field of 30\,G.}
% \{This component is rather instructive because its imaginary part is not vanishing ($Q \ne 0$), and it highlights the general behavior of the functions in a magnetized scenario.}
%
As for the unmagnetized case, the curve for $\Theta=\pi/2$, which corresponds to the CRD approximation, shows a broad profile centered at $u'=0$.
As $\Theta$ departs from $\pi/2$, the corresponding profiles show increasingly large differences from the previous one. 
In particular, for $\Theta$ approaching 0 (resp. $\pi$), the position of the maximum moves from $u'=0$ towards $u'=u$ (resp. $u'=-u$), while the width of the profile becomes sharper and the amplitude larger, in both the real and imaginary parts.
As the profiles become sharper, the Zeeman components %, split in frequency by the magnetic field, 
become visible, producing small lobes of opposite signs with respect to the central peak in the real part (left panels) and small substructures in the central peak in the imaginary part (right panels).

%\{These analyses, limited to two specific components and settings,} %seem to 
In summary, the $R^{\scriptscriptstyle \mathrm{III-CRD}}$ approximation should be suitable in the far wings of the spectral lines (see right panel of Fig.~\ref{fig:RIII_KKpQZero}), while it can introduce inaccuracies in the core and near wings, 
%These possible inaccuracies would 
mainly due to scattering processes with $\Theta$ close to 0 or $\pi$, for which $R^{\scriptscriptstyle \mathrm{III-CRD}}$ and $R^{\scriptscriptstyle \mathrm{III}}$ differ significantly.

%%%%%%%%%%%%%%%%%%%%%%%%%%%%%%%%%%%%%%%%%%%%%%%%%%%%%%%%%%%%%%%%%%%%
\subsection{Computational considerations on $R^{\scriptscriptstyle \mathrm{III-CRD}}$}
The $R^{\scriptscriptstyle \mathrm{III-CRD}}$ approximation is based on the assumption that scattering processes are totally incoherent \emph{also} in the observer's frame, which in turn implies that in a scattering event, absorption and re-emission can be treated as completely independent processes. 
Consistently with this picture, Eq.~\eqref{eq:RIII_KKpQ_CRD} shows that in $R^{\scriptscriptstyle \mathrm{III-CRD}}$ the joint probability of absorbing radiation with frequency $\nu'$ and re-emitting radiation with frequency $\nu$ is simply given by the product of two generalized profiles $\Phi^{KK'}_Q$ \eqref{eq:gen_prof_obs}.
Following the approach discussed at the end of Sect.~\ref{sec:discretization}, we evaluate the emission vector in the comoving frame, where no bulk velocities are present. 
In this case, the generalized profiles do not depend on the propagation directions $\vec{\Omega}$ and $\vec{\Omega}'$, and from Eqs.~\eqref{eq:RIII_obs_vertical}, \eqref{eq:RIII_KKpQ_CRD}, and \eqref{eq:PKKpQ}, one finds that the emission coefficients corresponding to $R^{\scriptscriptstyle \mathrm{III-CRD}}$ are given by
%
%\cred{
%\begin{allowdisplaybreaks}
%\begin{align}
%    \varepsilon^{\mathrm{\scriptscriptstyle{III-CRD}}}_i(\vec{r},\vec{\Omega},&\nu) = 
%    k_L(\vec{r}) \nonumber \\ 
%    & \times \int_{\mathbb{R_+}} \!\!\! \mathrm{d} \nu'  
%    \oint \frac{\mathrm{d} \vec{\Omega}'}{4\pi} 
%    \sum_{j=1}^4
%    R^{\scriptscriptstyle \mathrm{III-CRD}}_{ij}(\vec{r},\vec{\Omega},\vec{\Omega}',\nu,\nu') \,
%	I_j(\vec{r},\vec{\Omega}',\nu') \nonumber \\ 
%    = & \, \frac{k_L(\vec{r})}{\Delta \nu_D^2(\vec{r})}
%    \sum_{K,K'=0}^{2} \sum_{Q=-K_{\min}}^{K_{\min}} 
%    \sum_{K'' = |Q|}^{2 J_u}
%    \left(\beta^{K''}_Q(\vec{r}) - \alpha_Q(\vec{r}) \right) \nonumber \\
%    & \times \sum_{Q' = -K}^{K} \sum_{Q'' = -K'}^{K'} \, 
%    \left( -1 \right)^{Q'}
%    \mathcal{T}^{K'}_{Q',i}(\vec{\Omega}) \,
%    \overline{\mathcal{D}}^{K'}_{QQ''}(\vec{r}) \,
%    {\mathcal{D}^{K}_{QQ'}}(\vec{r}) \nonumber \\
%    & \times 
%    \Phi^{K''K'}_Q\left(\vec{r},\nu \right)
%    \int_{\mathbb{R}_+} \mathrm{d} \nu' \, 
%    \Phi^{K''K}_Q\left(\vec{r},\nu'\right) \,
%    J^{K}_{-Q} \left( \vec{r},\nu' \right) \, ,
%    \label{eq:RIIIJkqRED}
%\end{align}
%\end{allowdisplaybreaks}
%}
%
	\begin{allowdisplaybreaks}
		\begin{align}
			\varepsilon^{\mathrm{\scriptscriptstyle{III-CRD}}}_i(\vec{r},& \, \vec{\Omega},\nu) = 
			k_L(\vec{r}) \nonumber \\ 
			& \times \int_{\mathbb{R_+}} \!\!\! \mathrm{d} \nu'  
			\oint \frac{\mathrm{d} \vec{\Omega}'}{4\pi} 
			\sum_{j=1}^4
			R^{\scriptscriptstyle \mathrm{III-CRD}}_{ij}(\vec{r},\vec{\Omega},\vec{\Omega}',\nu,\nu') \,
			I_j(\vec{r},\vec{\Omega}',\nu') \nonumber \\ 
			= & \, \frac{k_L(\vec{r})}{\Delta \nu_D^2(\vec{r})}
			\sum_{K,K'=0}^{2} \sum_{Q=-K_{\min}}^{K_{\min}} 
			\sum_{K'' = |Q|}^{2 J_u}
			\left(\beta^{K''}_Q(\vec{r}) - \alpha_Q(\vec{r}) \right) \nonumber \\
			& \times \sum_{Q' = -K}^{K} \sum_{Q'' = -K'}^{K'} \, 
			\left( -1 \right)^{Q'}
			\mathcal{T}^{K'}_{Q'',i}(\vec{\Omega}) \,
			\overline{\mathcal{D}}^{K'}_{QQ''}(\vec{r}) \,
			{\mathcal{D}^{K}_{QQ'}}(\vec{r}) \nonumber \\
			& \times 
			\Phi^{K''K'}_Q\left(\vec{r},\nu \right)
			\int_{\mathbb{R}_+} \mathrm{d} \nu' \, 
			\Phi^{K''K}_Q\left(\vec{r},\nu'\right) \,
			J^{K}_{-Q'} \left( \vec{r},\nu' \right) \, ,
			\label{eq:RIIIJkq}
		\end{align}
		\end{allowdisplaybreaks}	
where $i=1,...,4$, $K_{\mathrm{min}} = \min{(K,K')}$, and $\Delta \nu_D$, $\mathcal{T}^K_{Q,i}$, and $\mathcal{D}^K_{QQ'}$ are the Doppler width (in frequency units), the polarization tensor, and the rotation matrices, respectively (see Appendix~\ref{sec:R_analytic} for more details).
Finally, $J^{K}_{Q}$ is the radiation field tensor, defined as (see Eq.~(5.157) of LL04)
\begin{equation*}
% \label{eq:Jkq}
    J^{K}_{Q} \left( \vec{r}, \nu \right) = 
    \oint  \frac{\mathrm{d} \vec{\Omega}}{4 \pi} 
    \sum_{j=1}^4 \mathcal{T}^{K}_{Q,j} \left( \vec{r}, \vec{\Omega} \right)
    I_j \left(\vec{r},\vec{\Omega},\nu \right) \, .
\end{equation*}
Equation \eqref{eq:RIIIJkq} shows that the dependencies on the frequency and propagation direction of the incoming and outgoing radiation are completely decoupled.
This allows the implementation of simple and fast computational algorithms: once the values of $J^K_Q$ are obtained from the formal solution of the RT equation, it is possible to independently compute the values of $\Phi^{K'K}_Q$ for the frequencies of the incoming and outgoing radiation and combine them only during the final calculation of the emission coefficient. 
Using the big-$\text{O}$ notation, the time complexity \cite[e.g.,][]{sipser1996introduction} of the algorithm scales as O$\left(N_\nu^d \right)$, where $d \in (1,2)$ grows monotonically with $N_\nu$.
For the typical size of the frequency grids needed to synthesize one (or a few) spectral lines (i.e., $N_\nu \approx 100$), $d$ can be considered close to $1$. 
This kind of time complexity is justified because the calculation of the generalized profile is slow (complex algorithms are required),
while the subsequent combination of them is fast as it only implies products of complex numbers.

In numerical applications considering the $R^{\scriptscriptstyle \mathrm{III-CRD}}$ approximation, it is a common practice to perform the integration over the frequencies of the incoming radiation in \eqref{eq:RIIIJkq} using the frequency grid from the problem discretization (see \mbox{Sect.~\ref{sec:discretization}}) without further refinements, and applying the trapezoidal rule.
This methodology has the advantages of a decreased \textit{time-to-solution} (TTS) and implementation simplicity. 
On the other hand, it should be pointed out that the accuracy of the results can be improved by using denser and more specific grids for the frequencies of the incoming radiation without any major loss of overall performance.
This is because, even if a finer frequency grid is used in \eqref{eq:RIIIJkq}, in a PRD setting the TTS for evaluating the total emissivity remains mainly dominated by the contribution of the angle-dependent $R^{\scriptscriptstyle \mathrm{II}}$ redistribution matrix.

%%%%%%%%%%%%%%%%%%%%%%%%%%%%%%%%%%%%%%%%%%%%%%%%%%%%%%%%%%%%%%%%%%%%%%%%%%%%%%%%%%%%%%%%%%%%%%%%%%%%%%%%%%%%%%%%%%%%
\subsection{Computational considerations on
$R^{\scriptscriptstyle \mathrm{III}}$}

The evaluation of $\bm{\varepsilon}^{\mathrm{\scriptscriptstyle{III}}}$ considering the exact expression of $R^{\scriptscriptstyle \mathrm{III}}$ (see equations in Appendix \ref{sec:RIII_observer}) is computationally challenging for the following main reasons: \textbf{a)} it involves a 4-dimensional integration, leading to a very large number of evaluations of the integrands in Eqs. \eqref{eq:int_final} -- \eqref{eq:int_final_Tpi}. This issue is otherwise known as the curse of dimensionality; \textbf{b)} the integration variables ($\nu'$, $\vec{\Omega}'$, and $y$) cannot be algebraically decoupled; 
\textbf{c)} already for simple atomic transitions, the total number of combinations of magnetic quantum numbers coupled with the tensorial polarimetric indices is high, which adds another layer of complexity to the 4 dimensions of the overall problem;
\textbf{d)} the integrands include the {\it{Faddeeva}} function \cite[]{faddeeva1961tables} whose evaluation requires a large TTS \cite[e.g.,][]{oeftiger2016review};
\textbf{e)} the integrands show a very complex behavior with respect to the various integration variables and parameters, thus requiring the use of very fine, unstructured, and case-dependent grids.

Our overall approach for calculating $\bm{\varepsilon}^{\scriptscriptstyle \mathrm{III}}$ is to first evaluate the integral over $y$, followed by that over the frequencies $\nu'$, and finally, the one over the propagation direction $\bm{\Omega}'$.
The analytic expressions of $\mathcal{I}$ \eqref{eq:int_final} -- \eqref{eq:int_final_Tpi} show that in the absence of bulk velocities (or if the redistribution matrix is evaluated in the comoving frame) the coupling between the propagation directions of the incoming and outgoing radiation occurs through the scattering angle $\Theta$.
The integrand in $\mathcal{I}$ shows a complex behavior with respect to the integration variable $y$, especially when the scattering angle approaches 0 and $\pi$. 
In these cases, the integrand becomes close to a Lorentz distribution or its square. 
These functions are not easy to numerically integrate due to the presence of sharp peaks and extended wings. 
Indeed, the convergence rate for the numerical quadrature of the Lorentz distribution (and its square) is generally slow.
Furthermore, the computation of a single $\mathcal{I}$ (see \eqref{eq:int_final}) can require up to thousands of evaluations of the Faddeeva function (accounting for more than 70\% of the TTS).
In order to perform the quadratures over $y$, we applied an adaptive quadrature method based on the Gauss-Kronrod approach \cite[e.g.,][]{kronrod1965nodes, piessens2012quadpack}. 
The advantage of this method is that it is capable of automatically inferring the behavior of the integrand by achieving very high accuracy with a relatively low number of function evaluations.

In general, the time complexity for the computation of $\bm{\varepsilon}^{\scriptscriptstyle \mathrm{III}}$ is O$\left( N_\Omega^2 N_\nu^3 \right)$. 
The cubic contribution given by the number of frequency grid points is due to the fourth dimension induced by $y$.
It must be noted that the number of grid points needed to adequately perform the quadrature of the integral over $y$ is generally larger than $N_\nu$.

For the angular integral in \eqref{eq:scat_int}, it is convenient to apply a quadrature rule characterized by a regular angular grid,
because in this case, the effective number of different scattering angles is significantly lower than the total pairs of directions. 
The results reported in the next sections were obtained with the quadrature method described in Sect.~\ref{sec:discretization}, with 12 Gauss-Legendre inclinations and 8 equally spaced azimuths. 
In this case, the total number of scattering angles is limited to 200.
In our algorithm, the quantities $\mathcal{I}$ are pre-computed for the whole set of different scattering angles corresponding to the chosen angular grid, thus avoiding repeating the calculation of $\mathcal{I}$, which is rather expensive, for different pairs of directions having the same scattering angle.

The quantity $\mathcal{I}$ depends on two pairs of magnetic quantum numbers, $(M_u,M_\ell)$ and $(M_u',M_\ell')$, where the labels $u$ and $\ell$ indicate the upper and lower level, respectively.
Equation~\eqref{eq:RIII_KKpQ_obs} shows that the expression of $R^{\scriptscriptstyle \mathrm{III}}$ actually includes four quantities $\mathcal{I}$, which differ for the values of the magnetic quantum numbers and are coupled inside six nested loops over such quantum numbers.
It can be easily realized that in such loops several tuples of magnetic sublevels are repeated, which suggests the opportunity to increase the efficiency of the algorithm by pre-computing the quantities $\mathcal{I}$ for all possible combinations of magnetic quantum numbers, thus avoiding to re-calculate the same quantity $\mathcal{I}$ various times.
The pre-computation of $\mathcal{I}$ as described above leads to a drastic reduction of the total number of calculations needed to compute $\bm{\varepsilon}^{\scriptscriptstyle \mathrm{III}}$, which otherwise would be very significant due to the extra dimension in the integration.
The pre-computed values of $\mathcal{I}$ are stored out-of-core (e.g. on disk) because they require a large footprint, and are accessed during the calculation of $\bm{\varepsilon}^{\scriptscriptstyle \mathrm{III}}$ through a system of look-up tables.
It must be observed that the quantity $\mathcal{I}$ and, consequently, the grids used for performing the numerical integration, also depends on the spatial point through the damping parameter $a$, the magnetic field, and the Doppler width $\Delta \nu_D$.
Our strategy thus calls for relatively large data storage capabilities that, however, remain manageable in the case of 1D applications.

%%%%%%%%%%%%%%%%%%%%%%%%%%%%%%%%%%%%%%%%%%%%%%%%%%%%%%%%
%%%%%%%%%%%%%%%%%%%%%%%%%%%%%%%%%%%%%%%%%%%%%%%%%%%%%%%%
\section{Impact of $R^{\scriptscriptstyle \mathrm{III}}$ on spectral lines formation}
\label{sec:prior_analysis}

The $R^{\scriptscriptstyle \rm III}$ redistribution matrix describes scattering processes during which the atom undergoes elastic collisions with neutral perturbers (mainly hydrogen and helium atoms in the solar atmosphere) that completely relax any correlation between the frequencies of the incident and scattered radiation, thus making the scattering totally incoherent.
Informally speaking, due to such collisions, the atom does not keep any \emph{memory} of the frequency of the incident photon.
On the contrary, the $R^{\scriptscriptstyle \rm II}$ redistribution matrix describes scattering processes in which the atom is not subject to any elastic collisions so that the frequencies of the incident and scattered radiation remain fully correlated (coherent scattering).
A quantitative estimate of the relative weight of $R^{\scriptscriptstyle \rm III}$ with respect to $R^{\scriptscriptstyle \rm II}$ is provided by the branching ratio for $R^{\scriptscriptstyle \mathrm{II}}$ (see \eqref{eq:branching_ratio_RII_RIII}) in the absence of magnetic fields (a.k.a. coherence fraction):
\begin{equation*}
    \tilde{\alpha}(\vec{r}) = \frac{A_{u \ell} + C_{u \ell}(\vec{r})}{A_{u \ell} + C_{u \ell}(\vec{r}) + Q_{\mathrm{el}}(\vec{r})} \, ,
\end{equation*}
where $A_{u \ell}$ is the Einstein coefficient for spontaneous emission, $C_{u \ell}$ is the rate of inelastic de-exciting collisions, and $Q_{\mathrm{el}}$ is the rate of elastic collisions with neutral perturbers.
A value of $\tilde{\alpha}$ close to unity (corresponding to a very low rate of elastic collisions compared to the rates for spontaneous emission and collisional de-excitation) means that $R^{\scriptscriptstyle \mathrm{II}}$ dominates over $R^{\scriptscriptstyle \mathrm{III}}$, while a value of $\tilde{\alpha}$ close to zero (corresponding to a very high value of $Q_{\mathrm{el}}$ compared to $A_{u \ell}$ and $C_{u \ell}$) means instead that $R^{\scriptscriptstyle \mathrm{III}}$ dominates with respect to $R^{\scriptscriptstyle \mathrm{II}}$.

The impact of $R^{\scriptscriptstyle \rm III}$ is thus expected to be marginal in the core of strong spectral lines (i.e., lines showing broad intensity profiles with extended wings). 
Indeed, their line-core radiation generally originates from the upper layers of the solar atmosphere, where the number density of neutral perturbers (and thus the rate of elastic collisions) is relatively low.
The relative weight of $R^{\scriptscriptstyle \rm III}$ can, however, become significant in the wings of such lines, as they usually form much lower in the atmosphere. 
On the other hand, it must be observed that the profiles entering the definition of $R^{\scriptscriptstyle \mathrm{III}}$ are all centered around the line-center frequency, and therefore the net contribution of this redistribution matrix to the emissivity in the line wings is generally marginal with respect to that of $R^{\scriptscriptstyle \mathrm{II}}$.
This can be seen as an analytical proof of the well-known fact that scattering processes are mainly coherent in the line wings \citep[e.g.,][]{mihalas1978stellar}. 
Consistently with this picture, \citet{alsina2022} showed that in the wings of strong resonance lines, the contribution of $R^{\scriptscriptstyle \rm III}$ needs to be taken into account in order not to overestimate the weight of $R^{\scriptscriptstyle \rm II}$, but its net contribution to scattering polarization is fully negligible.
These considerations suggest that the exact analytical form of $R^{\scriptscriptstyle \rm III}$ should not be crucial for modeling the core and far wings of strong lines, and that the $R^{\scriptscriptstyle \rm III-CRD}$ approximation should therefore be suitable in these spectral ranges.
The most critical regime is that of the near wings, where strong resonance lines may show very significant scattering polarization signals.
There, $R^{\scriptscriptstyle \rm III}$ may bring a non-negligible contribution, and it is harder to estimate a-priori the suitability of the $R^{\scriptscriptstyle \rm III-CRD}$ approximation.

The relative weight of $R^{\scriptscriptstyle \rm III}$ is also non-negligible in the case of spectral lines forming in the deeper layers of the solar atmosphere (photosphere), where the number density of colliders is significant.
On the other hand, these lines are generally weak in the intensity spectrum, showing narrow absorption profiles with a Doppler core and no wings. 
Since Doppler redistribution is generally very efficient in the line-core, the limit of CRD (i.e., to assume that all scattering processes are totally incoherent) has always been considered a good approximation for modeling both the intensity \citep[e.g.,][]{mihalas1978stellar} and scattering polarization profiles of these lines (e.g., LL04).

In order to quantitatively verify these considerations, and assess the suitability of the $R^{\scriptscriptstyle \rm III-CRD}$ approximation for modeling scattering polarization, we will model the intensity and polarization of two different spectral lines, namely a strong spectral line with extended wings forming in the upper layers of the solar atmosphere, and a weaker line forming deeper in the atmosphere.
Excellent examples for these two typologies of spectral lines are, respectively, the Ca~{\sc i} line at 4227\,{\AA} and the Sr~{\sc i} line at 4607\,{\AA}.
Both lines result from a resonant transition between the ground level of the considered atomic species, which in both cases has total angular momentum $J_\ell = 0$, and an excited level with $J_u = 1$.
Both of them show conspicuous scattering polarization signals, which can be suitably modeled considering a two-level atom with an unpolarized and infinitely-sharp lower level.
\begin{figure}[ht!]
	\centering
	\includegraphics[page=1, width=\columnwidth]{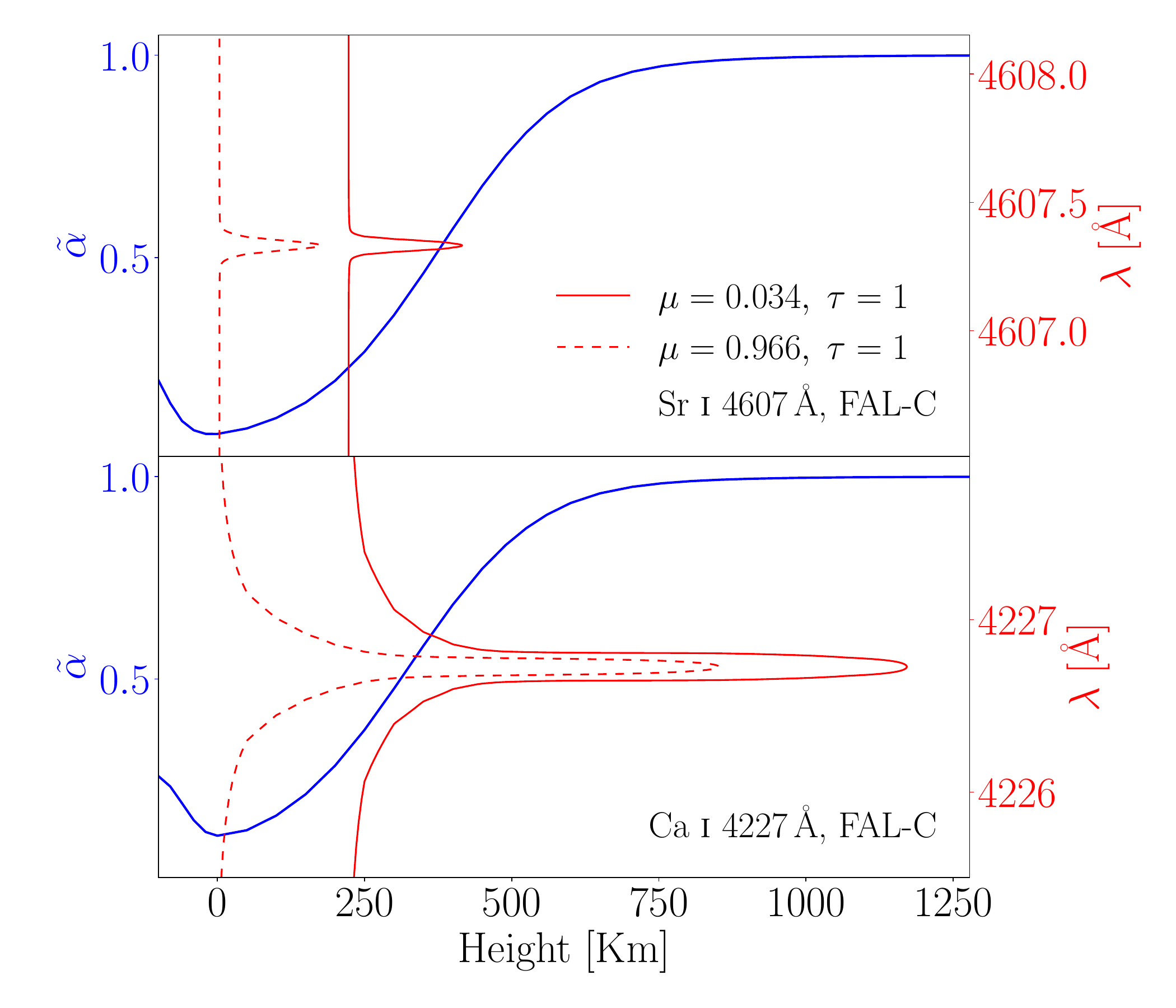}
	\caption{
       Height variation of $\tilde{\alpha}$ (blue line) for Sr~{\sc i} 4607 (\emph{top panel}) and Ca~{\sc i} 4227 (\emph{bottom panel}) in the FAL-C model. The red and dashed red isolines show the height where the optical depth as a function of wavelength is unity (i.e. $\tau=1$) for $\mu=0.034$ and $\mu=0.966$, respectively. 
 }
\label{fig:alphap}
\end{figure}
Figure~\ref{fig:alphap} shows the variation of the coherence fraction $\tilde{\alpha}$ with height in the 1D semi-empirical model C of \citet[][hereafter FAL-C]{fontenla1993} for the two considered spectral lines.
In the same figure, we also plot the height at which the optical depth $\tau$, in the frequency intervals of the considered spectral lines, is unity.
It can be shown \citep[e.g.][]{mihalas1978stellar} that this height provides an approximate estimate of the atmospheric region from which the emergent radiation originates (formation height). 
We recall that the optical depth at the frequency $\nu$ along direction $\vec{\Omega}$ is defined as 
\begin{equation}
    \tau(s,\vec{\Omega},\nu) = \int_0^s \eta_1(x,\vec{\Omega},\nu) \, \mathrm{d} x \, ,
    \label{eq:tau}
\end{equation}
where $s$ is the spatial coordinate along direction $\vec{\Omega}$ (measured with respect to an arbitrary initial point), and $\eta_1$ is the absorption coefficient for the intensity \citep[see also][]{janettI}.
For calculating the formation height, the direction $\vec{\Omega}$ must point inwards in the atmosphere, and the initial point $s=0$ is taken at the upper boundary.
Since the absorption coefficient is largest at the line center and decreases moving to the wings, it can be immediately seen that the formation height is highest in the line core and decreases moving to the wings.
From Eq.~\eqref{eq:tau} it is also clear that, for a given frequency, the formation height is higher for a line of sight (LOS) close to the edge of the solar disk (limb) than for one at the disk center.
Clearly, the formation height is higher for strong lines which have a large absorption coefficient (because the number density of atoms in the lower level of the transition is particularly large in the solar atmosphere) than for weak ones.

\subsection{Ca~{\sc i} 4227\,{\AA} line}

In the intensity spectrum, the Ca~{\sc i} 4227\,{\AA} line shows a very broad and deep absorption profile \citep[e.g.][]{gandorfer2002} with an \textit{equivalent width} of 1426\,m{\AA} \citep{moore1966}, i.e. one of the strongest spectral lines in the visible part of the solar spectrum.
When observed in quiet regions close to the limb, this line shows a large scattering polarization signal with a sharp peak in the line core and broad lobes in the wings \citep[e.g.][]{gandorfer2002}.
This signal, with its peculiar triplet-peak structure, has been extensively observed and modeled in the past 
\citep[e.g.,][and references therein]{stenflo1980,faurobert1992,bianda2003enigmatic,sampoorna2009,anusha2011,supriya2014,carlin17,ballester2018magneto,janett2021a,jaume_bestard2021}.
In particular, it was clearly established that its broad wing lobes are produced by coherent scattering processes with PRD effects.
We recall that the line-core peak is sensitive to the presence of magnetic fields via the Hanle effect. 
The Hanle critical field (i.e., the magnetic field strength for which the sensitivity to the Hanle effect is maximum) is $B_c \approx 25$\,G.
The wing lobes are sensitive to longitudinal magnetic fields of similar strength via the magneto-optical (MO) effects \citep{ballester2018magneto}.

The lower panel of Fig.~\ref{fig:alphap} shows that in the FAL-C atmospheric model, the core of the Ca~{\sc i} 4227\,{\AA} line forms above 800\,km (low chromosphere).
At these heights, the number density of neutral perturbers is very low, the coherence fraction is thus very close to unity, and $R^{\scriptscriptstyle \mathrm{II}}$ dominates with respect to $R^{\scriptscriptstyle \mathrm{III}}$.
On the other hand, as we move from the core to the wings, the formation height quickly decreases to photospheric levels.
At these heights, the coherence fraction is much lower than one and the weight of $R^{\scriptscriptstyle \mathrm{III}}$ is significant.
On the other hand, for the reasons discussed above, its net impact in the line wings is expected to be relatively low.
The most critical region is that of the near wings, where the Ca~{\sc i} 4227\,{\AA} line shows strong scattering polarization signals, and the net contribution from $R^{\scriptscriptstyle \mathrm{III}}$ can be non-negligible. 
The suitability of the $R^{\scriptscriptstyle \mathrm{III-CRD}}$ approximation in this spectral region can only be assessed numerically, and it will be analyzed in Sect.~\ref{sec:FALC_atmos}.

%%%%%%%%%%%%%%%%%%%%%%%%%%%%%%%%%%%%%%%%%%%%%%%%%%%%%
\subsection{Sr~{\sc i} 4607\,{\AA} line}
The Sr~{\sc i} 4607\,{\AA} line is a rather weak absorption line in the intensity spectrum, with an equivalent width of 36\,m{\AA} only \citep[see][]{moore1966}.
Nonetheless, this line shows a prominent scattering polarization signal at the solar limb, characterized by a sharp profile \citep[e.g.][]{gandorfer2002}.
This signal has been also extensively observed and modeled in the past, especially in order to investigate, via the Hanle effect, the small-scale, unresolved magnetic fields that permeate the quiet solar photosphere \citep[e.g.,][]{stenflo1997sss,trujillo2004,delpinoaleman2018,delpinoaleman2021,dhara2019,zeuner2020,zeuner2022}.
%\{Indeed, there are evidences that the quiet solar photosphere is filled with small-scale magnetic fields that vary on spatial scales below the resolution element of current observations}
The Hanle critical field for this line is $B_c \approx 23$\,G.
The limit of CRD has always been considered suited for modeling both the intensity and scattering polarization profiles of this line.

The upper panel of Fig.~\ref{fig:alphap} shows that in the FAL-C atmospheric model, the core of this line forms in the photosphere, below 500\,km.
The curve for the coherence fraction shows that the weight of $R^{\scriptscriptstyle \mathrm{III}}$ is, as expected, significant at these heights.
In the next section, we will explore the suitability of the $R^{\scriptscriptstyle \mathrm{III-CRD}}$ approximation in the %PRD 
modeling of the scattering polarization signal of this line through full PRD RT calculations %for this line, 
in the presence of both deterministic and non-deterministic \citep[e.g.][]{stenflo1982,stenflo1994,degl2006polarization} magnetic fields.
%and we will analyze the suitability of the $R^{\scriptscriptstyle \mathrm{III-CRD}}$ approximation in the PRD modeling of its scattering polarization signal.
%\{We will consider both deterministic and non-deterministic \citep[e.g.,][]{stenflo1982,stenflo1994,degl2006polarization} magnetic fields. 
%As an example of a non-deterministic magnetic field, we consider
%\{Given the interest of this line for investigating small-scale, unresolved (i.e., non-deterministic) magnetic fields \citep[e.g.,][]{stenflo1982,stenflo1994,degl2006polarization}, we also model it in the presence of 
For the latter, we will consider \textit{unimodal micro-structured isotropic} (MSI) magnetic fields, namely magnetic fields with a given strength and an orientation that changes on scales below the mean free path of photons, uniformly distributed over all directions.
The expressions of RT quantities in the presence of MSI magnetic fields are exposed in Appendix~\ref{app:micB} \citep[see also Sect.~4 of][]{ballester2017transfer}.

%
% \textcolor{orange}{[REMOVE NEXT SENTENCES]
% We also note that in the solar photosphere \{in conditions of quite-Sun}, it is common to have magnetic fields that vary on spatial scales below the resolution element of standard observations.
% These magnetic fields are commonly referred to as non-deterministic magnetic fields\{, whose relevance in the interpretation of observational results has been highlighted in (e.g.)} \cite{trujillo2004,manso2004concerning}.
% %
% For this reason, the photospheric Sr~{\sc i} 4607\,{\AA} line is also modeled in the presence of \textit{unimodal micro-structured isotropic} (MSI) magnetic fields 
% % \{which is fully described in Sect.~4 of \cite{ballester2017transfer}}, 
% that is a magnetic field with a given strength and an orientation that changes on scales below the mean free path of photons, uniformly distributed over all directions.
% The expressions of RT quantities in the presence of MSI magnetic fields are exposed in Appendix~\ref{app:micB}, and Sect. 4 of \cite{ballester2017transfer}.}

%%%%%%%%%%%%%%%%%%%%%%%%%%%%%%%%%%%%%%%%%%%%%%%%%%%%

%%%%%%%%%%%%%%%%%%%%%%%%%%%%%%%%%%%%%%%%%%%%%%%%%%%%%%%%%%%%%%%%%%%%%%%%%%%%%%%%%%%%%%%%%%%%%%%%%%%%%%%%%%%%%%%%%%%%

% %%%%%%%%%%%%%%%%%%%%%%%%%%%%%%%%%%%%%%%%%%%%%%%%%%%%%%%%%%%%%%%%%%%%%%%%%%%%%%%%%%%%%

\section{Numerical results: FAL-C atmospheric model}
\label{sec:FALC_atmos}

In this section and in the next one, we present the numerical results\Computer{} of non-LTE RT calculations of the scattering polarization profiles of the Ca~{\sc i} 4227\,{\AA} and Sr~{\sc i} 4607\,{\AA} lines, performed with the numerical solution strategy described in Sect.~\ref{sec:discretization}.
All calculations are performed using the general (angle-dependent) expression of the $R^{\scriptscriptstyle \mathrm{II}}$ redistribution matrix, while considering both the exact form of the $R^{\scriptscriptstyle \mathrm{III}}$ matrix and its $R^{\scriptscriptstyle \mathrm{III-CRD}}$ approximation.
The lower level population, which we keep fixed in the problem (see the end of Sect.~\ref{sec:transfer}), is pre-computed with the RH code \citep{uitenbroek2001multilevel}, which solves the nonlinear unpolarized non-LTE RT problem.
For the Ca~{\sc i} 4227\,{\AA} line, we run RH using a model for calcium composed of 25 levels, including 5 levels of Ca~{\sc ii} and the ground level of Ca~{\sc iii}.
For the Sr~{\sc i} 4607\,{\AA} line, we considered a model composed of 34 levels, including the ground level of Sr~{\sc ii}.
These preliminary computations also provided the rates for elastic and inelastic collisions, as well as the continuum quantities. 

For a quantitative comparison between the emergent Stokes profiles obtained using $R^{\scriptscriptstyle \mathrm{III}}$ and $R^{\scriptscriptstyle \mathrm{III-CRD}}$,
we plot the error defined by
\begin{equation}
    \text{Error}\left(\vec{a},\, \vec{b}\right) = \frac{\left|  \vec{a} \,  (\max{\vec{a}})^{-1}  -  \vec{b} \,  (\max{\vec{a}})^{-1} \right|}{1+\left| \vec{a} \,  (\max{\vec{a}})^{-1} \right|},
    \label{eq:relErr}
\end{equation}
where $\vec{a}$ and $\vec{b}$ represent the values of a given Stokes parameter of the emergent radiation, for a given direction and all considered wavelengths, obtained considering $R^{\scriptscriptstyle \mathrm{III}}$ and $R^{\scriptscriptstyle \mathrm{III-CRD}}$, respectively.
The maximum is calculated with respect to wavelength, over the considered interval.
% \cred{The error definition \eqref{eq:relErr} prevents us from overestimating the differences between $\vec{a}$ and $\vec{b}$ where their values are close to zero.}
%
The error definition in Eq.~\eqref{eq:relErr} does not correspond to the standard relative error and it was introduced to prevent amplifying the discrepancies where $\vec{a}$ and $\vec{b}$ are close to zero and consequently to the numerical noise.
Where the signals $\vec{a}$ and $\vec{b}$ are relevant, this definition provides a value that is smaller, by a factor of two in the worst case (i.e., where $\vec{a}$ takes its maximum value), than the usual relative error.
This definition is thus justified as it provides the correct order of magnitude of the error where the signal is relevant while damping it when the signal becomes negligible.

In this section, we show calculations performed in the FAL-C atmospheric model ($N_z=70$ height nodes), in the presence of constant (i.e., height-independent) deterministic magnetic fields (i.e., magnetic fields having a well-defined strength and orientation at each spatial point), in the absence of bulk velocity fields. 
% \{Besides, the atmosphere becomes transparent at 1065 Km in the Ca~{\sc i} 4227\,{\AA} line, and 400 Km in the Sr~{\sc i} 4607\,{\AA} line, and the line formation freezes above these altitudes (as explained in Section \ref{sec:prior_analysis}), in the FAL-C model, for the sake of completeness, calculations were performed in the full height range of the atmospheric model.} 
%
%
The LOS towards the observer is taken on the $x-z$ plane of the considered reference system (see Fig.~\ref{fig:ref_sys}) and is specified by $\mu=\cos{\theta}$, with $\theta$ the inclination with respect to the vertical.
Typically, we present the emergent Stokes profiles for two specific directions: $\mu=0.034$, which represents radiation coming from the solar limb (nearly horizontal LOS), and $\mu=0.996$, which represents radiation coming from near the center of the solar disk (nearly vertical LOS). 
\begin{figure*}[!hbt]
	\centering
	\includegraphics[page=1, width=\textwidth]
							{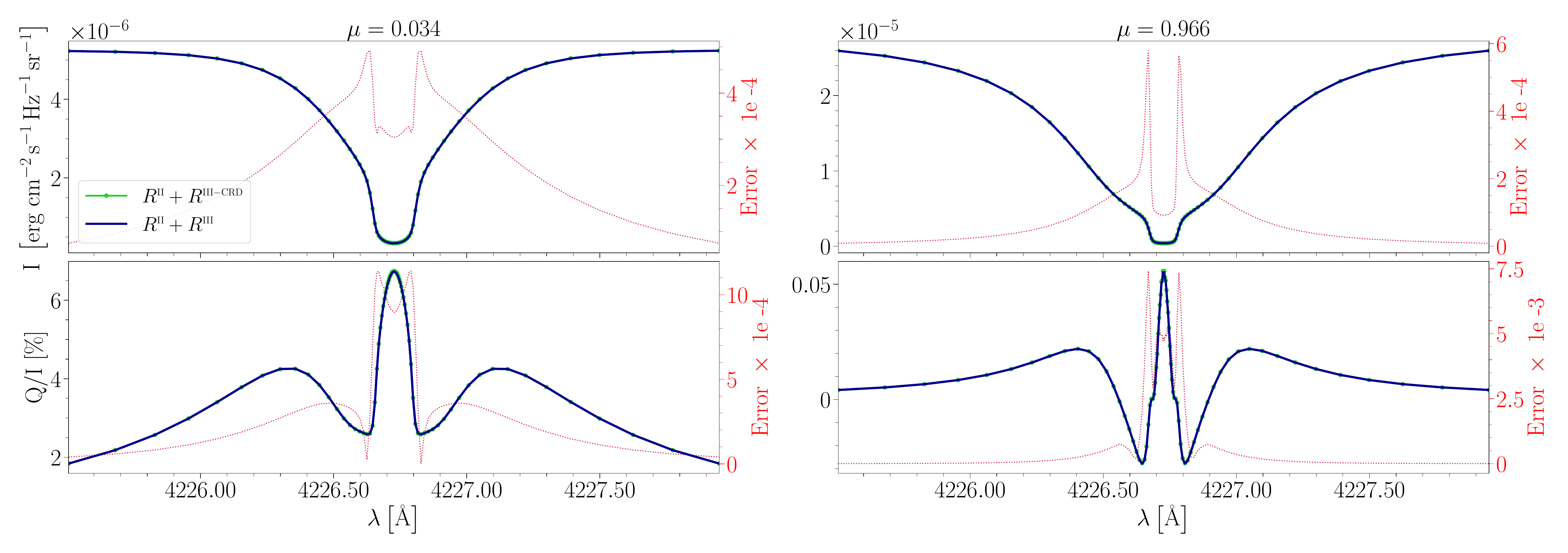}
	\caption{
		Emergent Stokes $I$ (\emph{top panels}) and $Q/I$ (\emph{bottom panels}) profiles of the Ca~{\sc i} 4227\,{\AA} line at $\mu = 0.034$ (\emph{left panels}) and $\mu = 0.966$ (\emph{right panels}) calculated in the FAL-C atmospheric model in the absence of magnetic fields.
		Calculations take into account PRD effects considering the exact expression of $R^{\scriptscriptstyle \mathrm{III}}$ (blue lines) and the $R^{\scriptscriptstyle \mathrm{III-CRD}}$ approximation (green marked lines). 
		The reference direction for positive $Q$ is taken parallel to the limb.
		The red dotted lines report the error between $R^{\scriptscriptstyle \mathrm{III}}$ and $R^{\scriptscriptstyle \mathrm{III-CRD}}$ calculations, given by Eq.~\eqref{eq:relErr}.
	}
	\label{fig:CaI0G}
\end{figure*}

%%%%%%%%%%%%%%%%%%%%%%%%%%%%%%%%%%%%%%%%%%%%%%%%%%%%%%%%%%
\subsection{Ca~{\sc i} 4227\,{\AA} line}
\label{sec:resCaI}

We first consider the modeling of the Stokes profiles of the Ca~{\sc i} 4227\,{\AA} line, which is discretized with $N_\nu=99$ unevenly spaced nodes.
The characteristics of this spectral line
are adequately reproduced by our calculations.
The triple-peak structure in $Q/I$ is evident in the non-magnetized model shown in Figure \ref{fig:CaI0G},
while the magnetized case shown in Figure \ref{fig:CaI20G} displays a clear depolarization of the core as a result of the Hanle effect and a mild depolarization of the lobes due to the magneto-optical effects.

Figure~\ref{fig:CaI0G} shows the Stokes $I$ and $Q/I$ profiles for a non-magnetic scenario, where we observe no significant difference between $R^{\scriptscriptstyle \mathrm{III}}$ and $R^{\scriptscriptstyle \mathrm{III-CRD}}$ calculations, with an error that is always smaller than $7.5\times10^{-3}$.
In the absence of magnetic fields, the $U/I$ and $V/I$ profiles vanish and are consequently not shown.

\begin{figure}[ht!]
    \centering
    \includegraphics[width=\columnwidth]{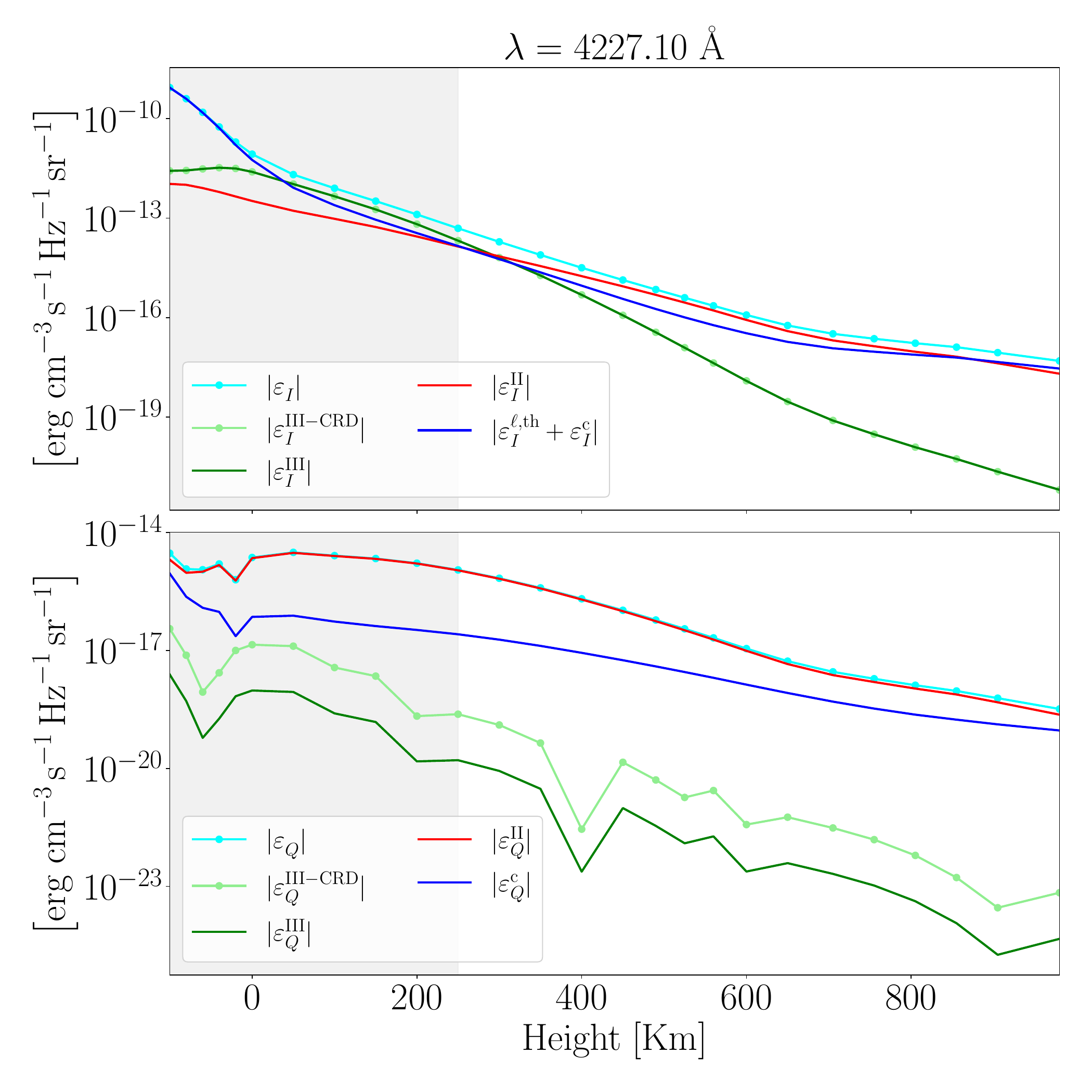}
    \caption{%Contribution of the various components of
    Various contributions (see legend) to the emission coefficients for the Stokes parameters $I$ (\textit{top panel}) and $Q$ (\textit{bottom panel}) as a function of height in the FAL-C model, in the absence of magnetic fields. The emission coefficients are evaluated at the wavelength $\lambda= 4227.1$\,{\AA}, corresponding to the maximum of the $Q/I$ lobe in the red wing of the line, for a LOS with $\mu=0.17$. 
    The shaded area in the panels highlights the atmospheric region where the optical depth at this wavelength and LOS is greater than 1. 
    $\varepsilon_{X}$ \mbox{(where $X=$ $I$ or $Q$)} is the total emissivity, $\varepsilon_{X}^{\scriptscriptstyle\mathrm{II}}$, $\varepsilon_{X}^{\scriptscriptstyle\mathrm{III}}$, and $\varepsilon_{X}^{\scriptscriptstyle\mathrm{III-CRD}}$ are the contributions from $R^{\scriptscriptstyle \mathrm{II}}$, $R^{\scriptscriptstyle \mathrm{III}}$, and $R^{\scriptscriptstyle \mathrm{III-CRD}}$, respectively, while $\varepsilon_{X}^{\ell,\mathrm{th}}$ and $\varepsilon_{X}^{c}$  are the contributions from the line thermal emissivity and continuum, respectively.}
    \label{fig:epslobes}
\end{figure}

\begin{figure*}[hbt!]
        \centering
        \includegraphics[page=1, width=\textwidth]{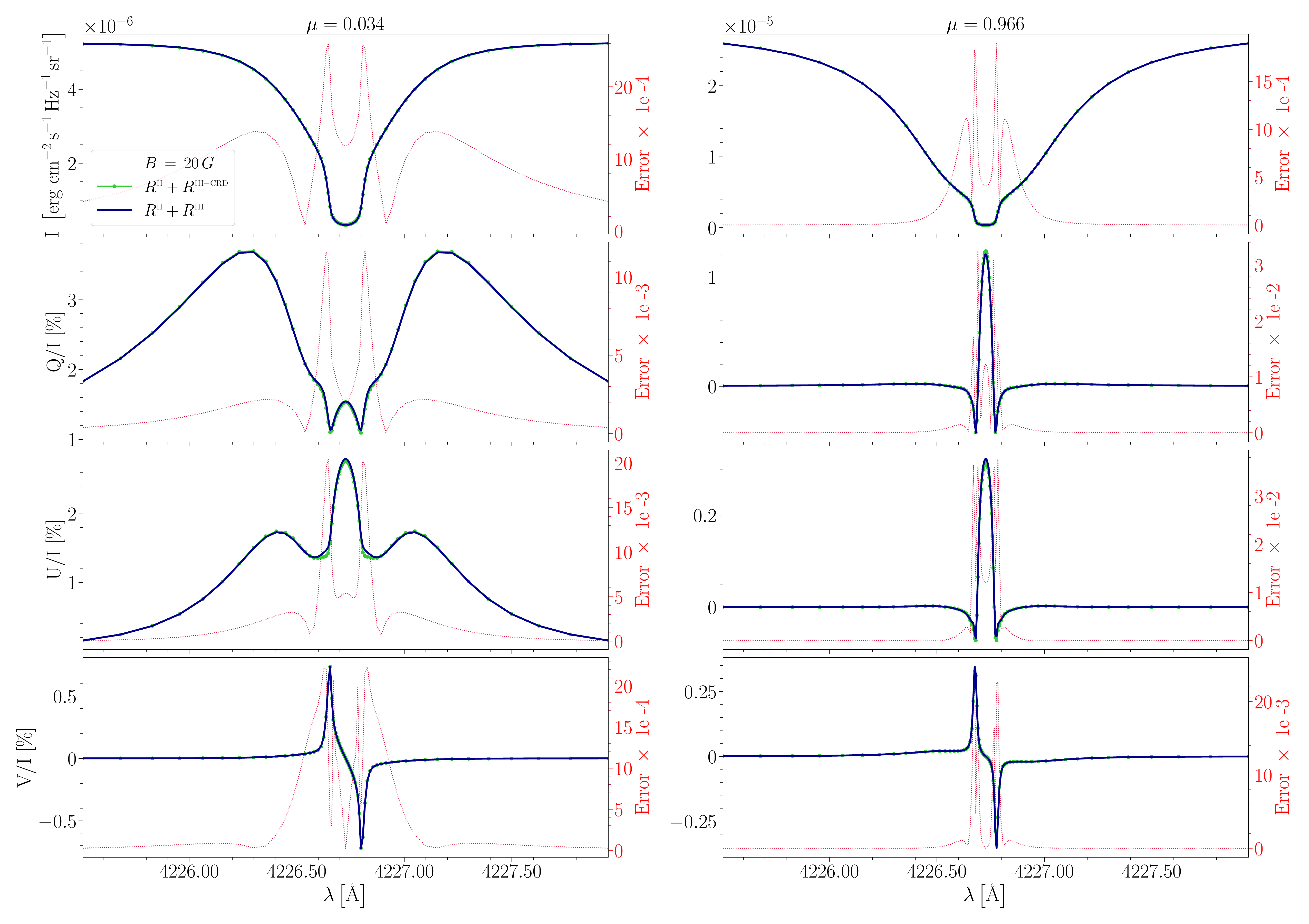}
\caption{
Emergent Stokes $I$ (\textit{first row}), $Q/I$ (\textit{second row}), $U/I$ (\emph{third row}),
and $V/I$ (\textit{fourth row}) profiles for the Ca~{\sc i} 4227\,{\AA}
line at $\mu = 0.034$ (\textit{left column}) and $\mu = 0.966$ (\textit{right column})
calculated in the FAL-C atmospheric model in the presence of a horizontal ($\theta_B=\pi/2, \, \chi_B=0$) magnetic field
with $B = 20$~G. 
Calculations take into account PRD effects considering the exact expression of $R^{\scriptscriptstyle \mathrm{III}}$ (blue lines) and the $R^{\scriptscriptstyle \mathrm{III-CRD}}$ approximation (green marked lines). 
The reference direction for positive $Q$ is taken parallel to the limb.
The red dotted lines report the error between $R^{\scriptscriptstyle \mathrm{III}}$ and $R^{\scriptscriptstyle \mathrm{III-CRD}}$ calculations, given by Eq.~\eqref{eq:relErr}.}
\label{fig:CaI20G}
\end{figure*}

Figure~\ref{fig:epslobes} shows the contributions from $R^{\scriptscriptstyle \mathrm{II}}$, $R^{\scriptscriptstyle \mathrm{III}}$, and $R^{\scriptscriptstyle \mathrm{III-CRD}}$ to $| \varepsilon_{I} |$  and $| \varepsilon_{Q} |$ for the non-magnetized case as a function of height. 
For the sake of completeness, the figure also shows the contributions from continuum and line thermal emissivities to Stokes $I$ and from continuum to Stokes $Q$ (see blue lines).
%\{Considering that in Fig.~\ref{fig:RIII_KKpQZero} we show that in near wings $R^{\scriptscriptstyle \mathrm{III}}$ and $R^{\scriptscriptstyle \mathrm{III-CRD}}$ are nearly equivalent so that their contribution to line formation is in practice identical, then we can omit to report both cases.}
%
In this figure, we consider $\mu=0.17$ and $\lambda=4227.1$~\AA, which corresponds to the wavelength at the maximum of the red $Q/I$ lobe in Figure~\ref{fig:CaI0G} (left bottom panel).
The limit of the gray region represents the height where the optical depth at the considered wing wavelength and LOS is unity.
%In the top panel of Figure~\ref{fig:epslobes}, we can see that at this height $R^{\scriptscriptstyle \mathrm{II}}$ and $R^{\scriptscriptstyle\mathrm{III}}$ (green and red lines, respectively) provide nearly equivalent contributions to the intensity $\varepsilon_I$.
The top panel of Fig.~\ref{fig:epslobes} shows that the contributions to $\varepsilon_I$ from $R^{\scriptscriptstyle \mathrm{III}}$ and $R^{\scriptscriptstyle \mathrm{III-CRD}}$ practically coincide at all heights, and that at the height where the optical depth is unity, they are very similar to that from $R^{\scriptscriptstyle \mathrm{II}}$.
By contrast, in the bottom panel, we see that the contribution to $\varepsilon_Q$ of $R^{\scriptscriptstyle\mathrm{II}}$ dominates over that of $R^{\scriptscriptstyle\mathrm{III}}$ at all heights, and it can be thus considered the only relevant contributor to the formation of the $Q/I$ wing lobe.
In this case, the contribution from $R^{\scriptscriptstyle\mathrm{III-CRD}}$ is different from that of $R^{\scriptscriptstyle\mathrm{III}}$, but it remains negligible with respect to that of $R^{\scriptscriptstyle\mathrm{II}}$.
This explains why the computations of Fig.~\ref{fig:CaI0G} do not show any appreciable differences between $R^{\scriptscriptstyle \mathrm{III}}$ and $R^{\scriptscriptstyle \mathrm{III-CRD}}$ in the line wings.

Finally, Figure~\ref{fig:CaI20G} displays all the Stokes profiles in the presence of a height-independent, horizontal magnetic field with $B=20$\,G.
An overall good agreement between $R^{\scriptscriptstyle \mathrm{III}}$ and $R^{\scriptscriptstyle \mathrm{III-CRD}}$ settings is observed in all profiles.
We note that the error is generally larger in the core and near wings of the $Q/I$, $U/I$, and $V/I$  profiles. 
This conclusion remains valid for magnetic fields ranging from $\rm 10~G$ to $\rm 200~G$.
Indeed, these results show no significant differences and are therefore not reported.

 %%%%%%%%%%%%%%%%%%%%%%%%%%%%%%%%%%%%%%%%%
 %%%%%%%%%%%%%%%%%%%%%%%%%%%%%%%%%%%%%%%%%
%%%%%%%%%%%%%%%%%%%%%%%%%%%%%%%%%%%%%%%%%
 %%%%%%%%%%%%%%%%%%%%%%%%%%%%%%%%%%%%%%%%%

\begin{figure*}[!ht]
	\centering
			\includegraphics[page=1, width=\textwidth]{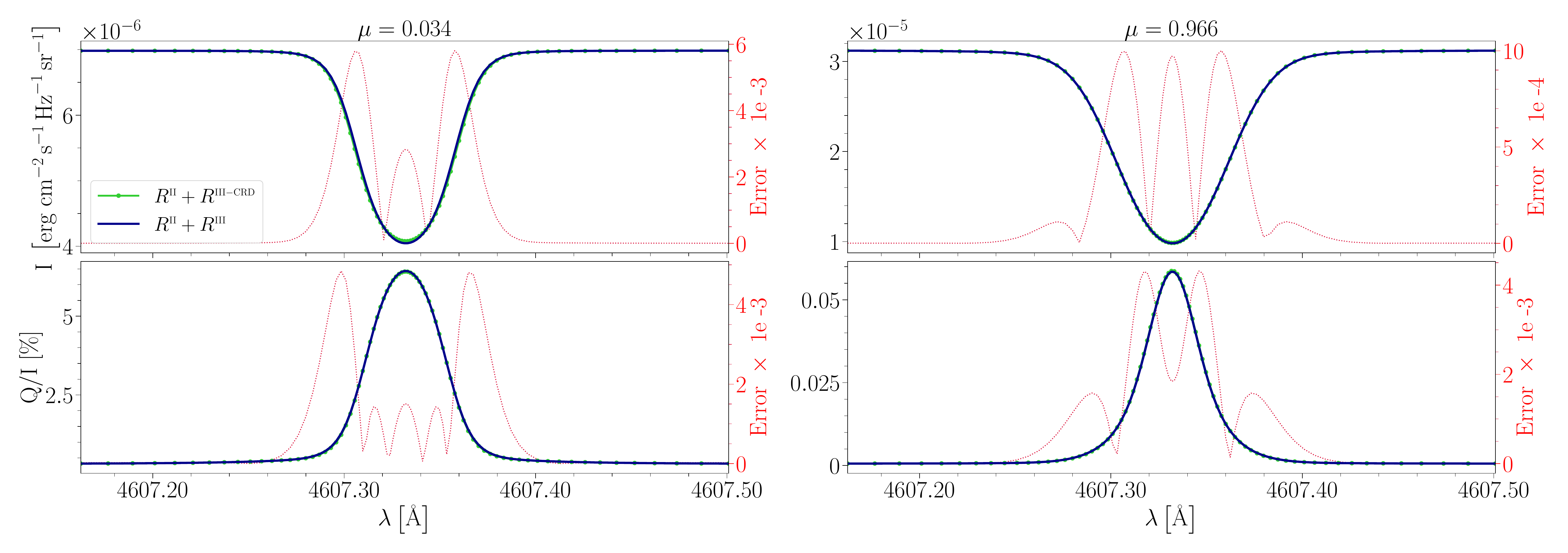}
   			\caption{Same as Fig.~\ref{fig:CaI0G}, but for the Sr~{\sc i}~4607~\AA~line.}
			\label{fig:SrI0G}
\end{figure*}

\begin{figure*}[!ht]
	\centering
	\includegraphics[page=1, width=\textwidth]{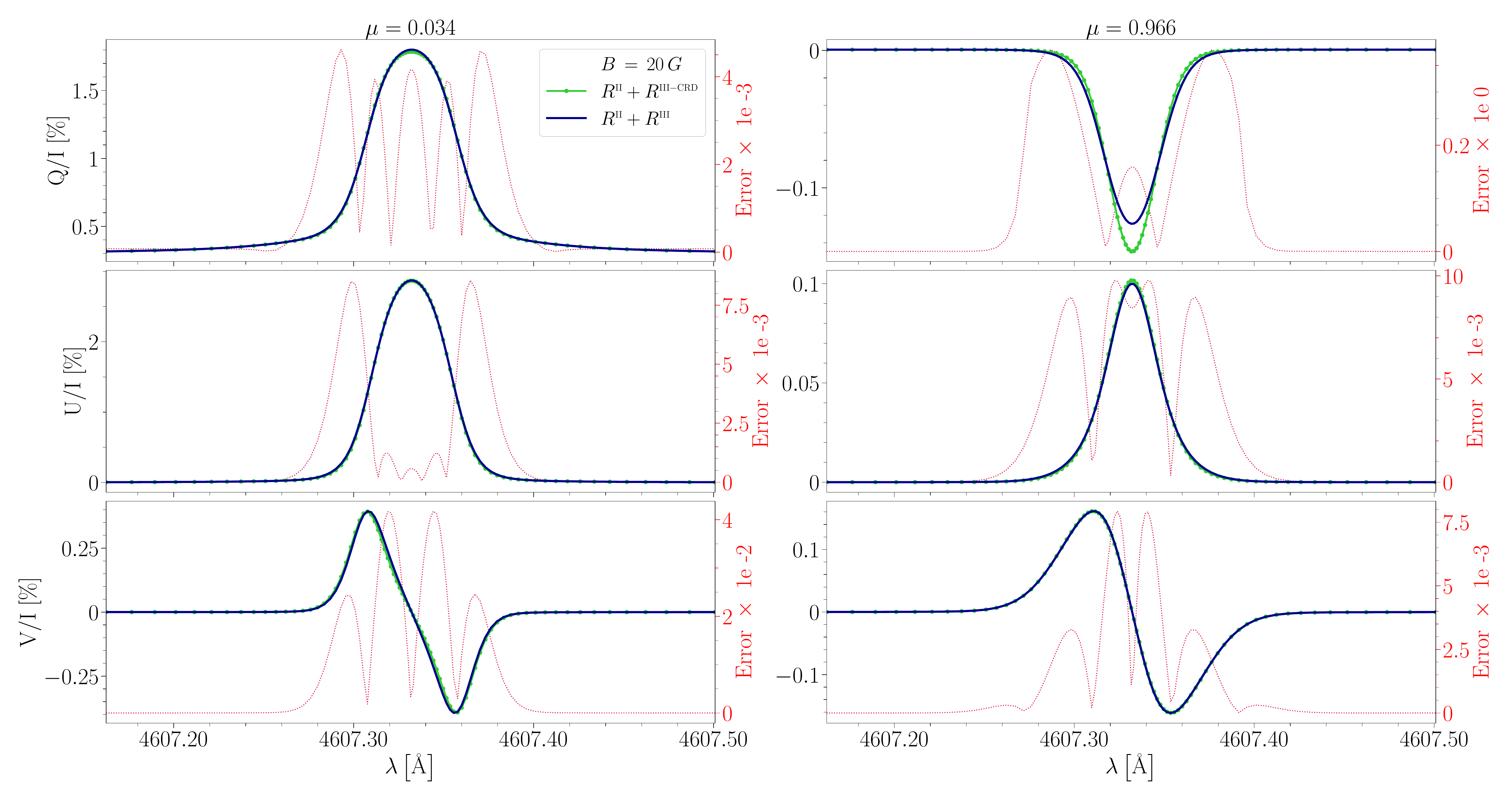}
      			\caption{Same as Fig.~\ref{fig:CaI20G}, but for the Sr~{\sc i}~4607~\AA~line.}
			\label{fig:SrI20G}
\end{figure*}

\begin{figure*}[!ht]
\centering
\includegraphics[page=1, width=\textwidth]{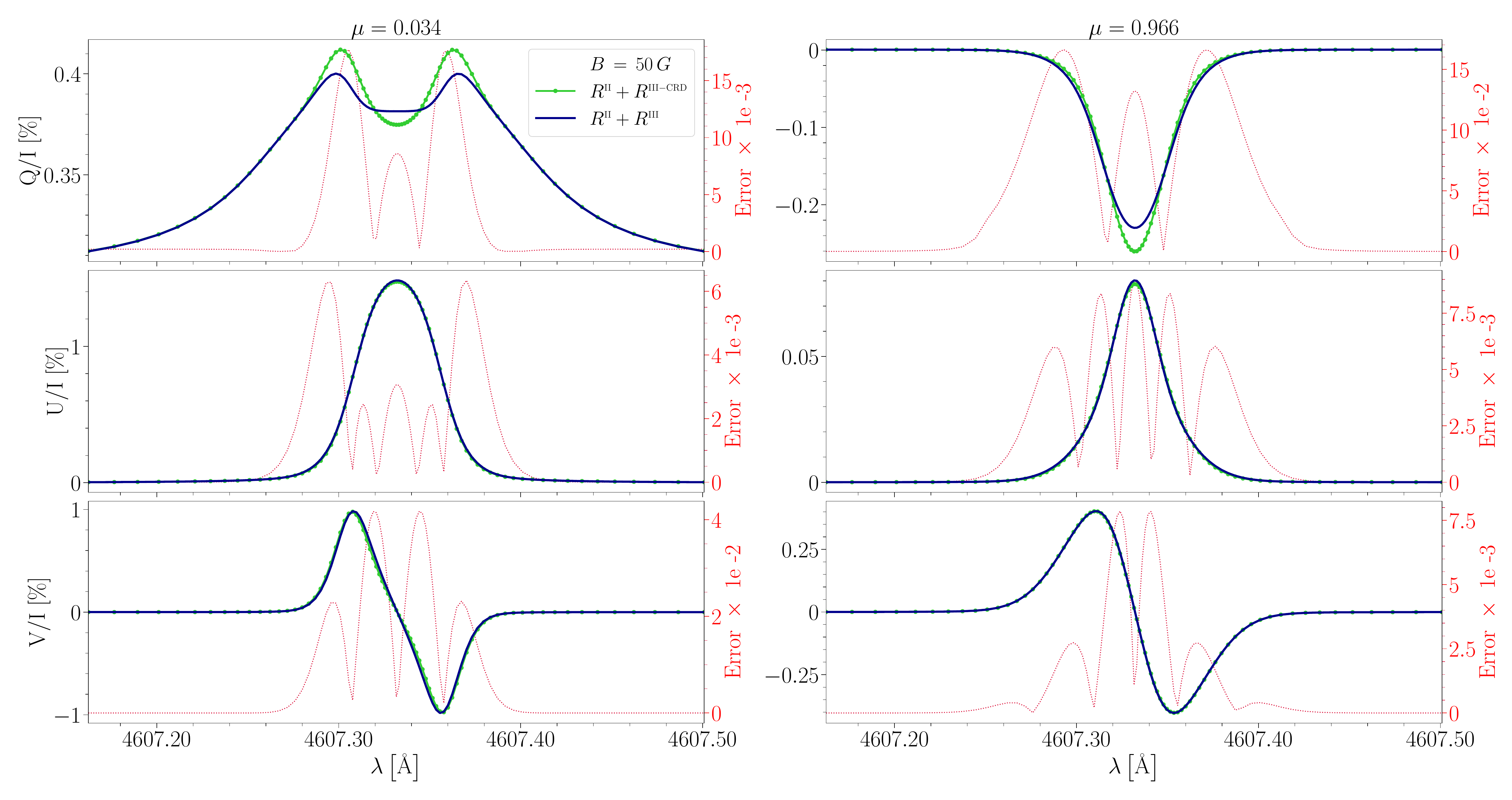}
\caption{Same as Fig.~\ref{fig:SrI20G}, but for $B = 50$~G.}

\label{fig:SrI50G}
\end{figure*}

\begin{figure*}[!ht]
	\centering
	\includegraphics[page=1, width=\textwidth]{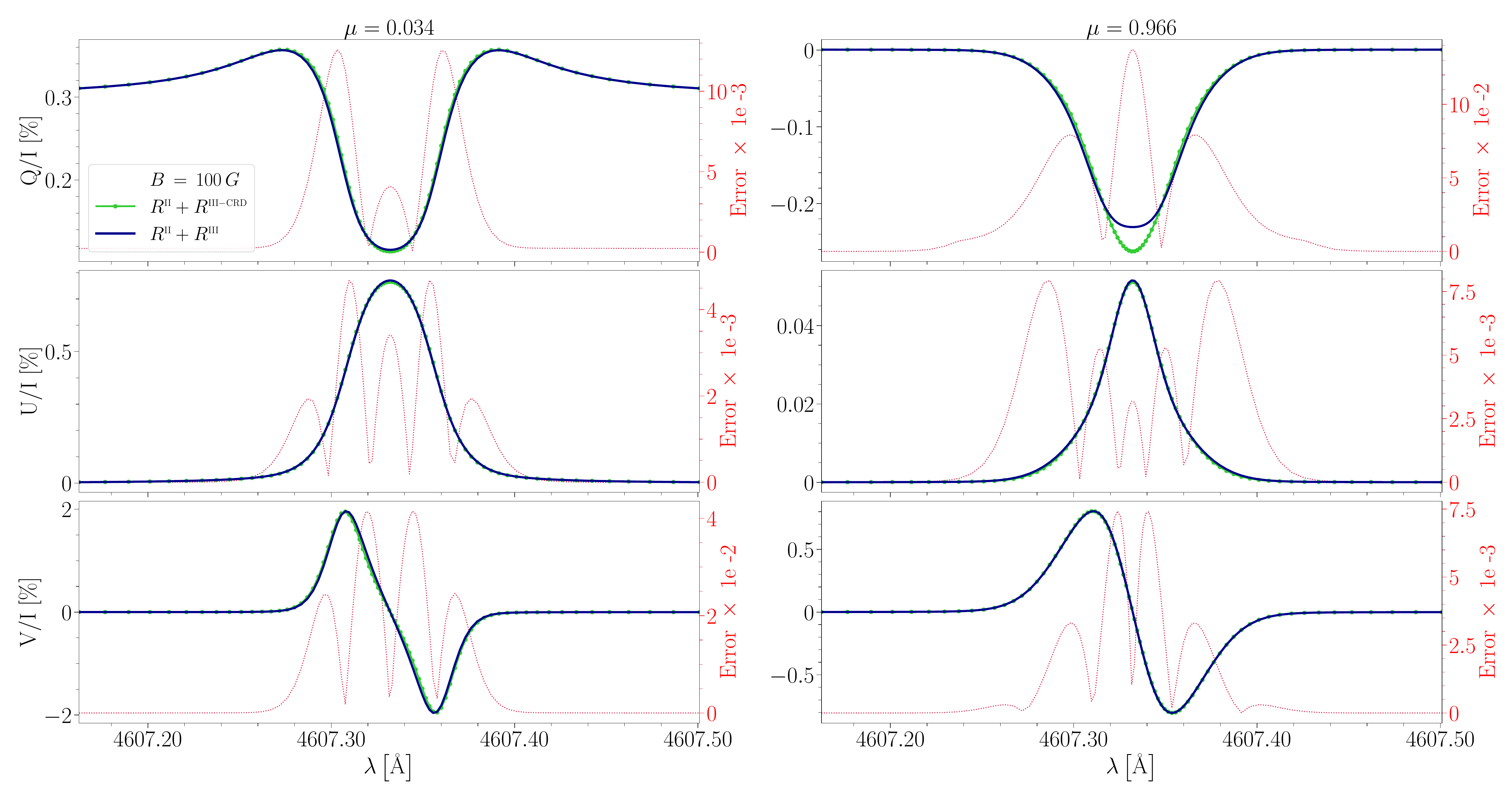}
   \caption{Same as Fig.~\ref{fig:SrI20G}, but for $B = 100$~G.}
			\label{fig:SrI100G}
\end{figure*}

%%%%%%%%%%%%%%%%%%%%
\subsection{Sr~{\sc i} 4607\,\AA~line}
\label{sec:resSrI}
 
We now consider the modeling of the Sr~{\sc i} 4607\,{\AA} line, which is discretized with $N_\nu=130$ unevenly spaced nodes.
Our non-LTE RT calculations adequately reproduce this \emph{weak} line, showing a small absorption profile in the intensity spectrum without wings, and a prominent and sharp $Q/I$ scattering polarization peak (see Fig.~\ref{fig:SrI0G}).

\begin{figure}[!ht]
    \centering
    \includegraphics[page=1, width=\columnwidth]{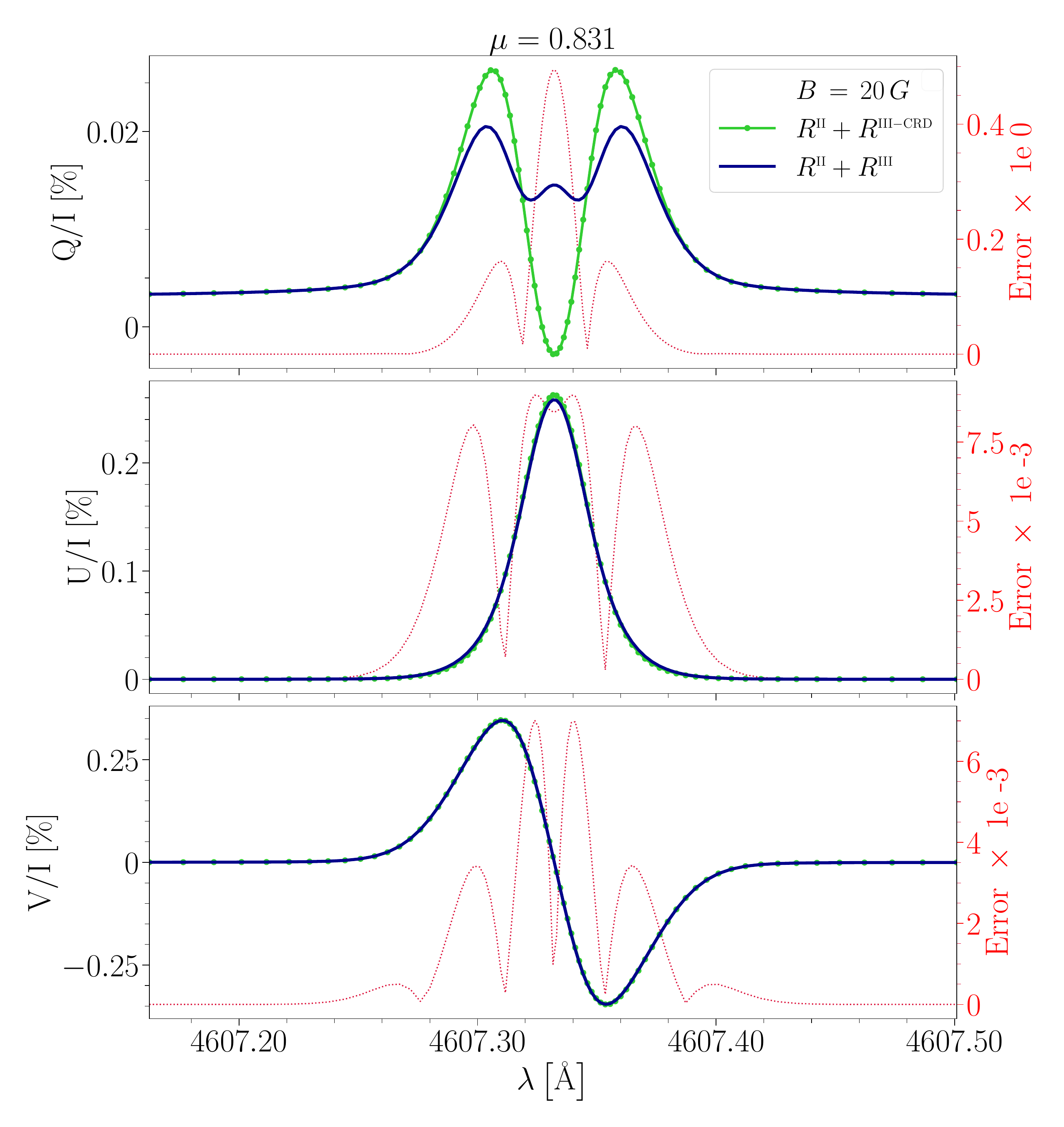}
    \caption{Same as Fig.~\ref{fig:SrI20G}, but for $\mu=0.83$.
    In this $Q/I$ profile, we observed the maximal error between $R^{\scriptscriptstyle \mathrm{III}}$ and $R^{\scriptscriptstyle \mathrm{III-CRD}}$ calculations.}
    \label{fig:SrI_diff}
\end{figure}

In the $I$ and  $Q/I$ profiles shown in Figure~\ref{fig:SrI0G}
for a non-magnetic scenario, we only note minor differences between $R^{\scriptscriptstyle \mathrm{III}}$ and $R^{\scriptscriptstyle \mathrm{III-CRD}}$ calculations.
By contrast, we note some relevant discrepancies between
$R^{\scriptscriptstyle \mathrm{III}}$ and $R^{\scriptscriptstyle \mathrm{III-CRD}}$ cases
when a deterministic magnetic field is considered. 
Figures~\ref{fig:SrI20G}, \ref{fig:SrI50G},
and~\ref{fig:SrI100G} display the
 $Q/I$,  $U/I$, and  $V/I$  profiles in the presence of a height-independent, horizontal magnetic field
with $B=20$~G, $B=50$~G, and $B=100$~G, respectively.
We omit to report the intensity $I$ profiles,
because they are essentially identical
to those exposed in Figure~\ref{fig:SrI0G}
for the non-magnetized case.
We first note that, in all cases, there are no observable discrepancies between
$R^{\scriptscriptstyle \mathrm{III}}$ and $R^{\scriptscriptstyle \mathrm{III-CRD}}$ calculations in the $U/I$  and  $V/I$  profiles. 
By contrast, some relevant differences appear in the
 $Q/I$  profiles, such as the one
shown in the top-left panel of Figure~\ref{fig:SrI50G}.
Moreover, $R^{\scriptscriptstyle \mathrm{III-CRD}}$ calculations generally present larger $Q/I$ line-core signals at $\mu=0.966$ with respect to $R^{\scriptscriptstyle \mathrm{III}}$ ones.
For completeness, Figure~\ref{fig:SrI_diff} displays the $Q/I$ profile where we observed the maximal error, corresponding to the $B=20$~G case at $\mu=0.83$.

It is worth observing that for $\mu=0.966$ and a magnetic field with $B=20$\,G (i.e., a forward scattering Hanle effect geometry), the Ca~{\sc i} 4227\,{\AA} line shows a positive $Q/I$ peak (see Fig.\ref{fig:CaI20G}), while the Sr~{\sc i} 4607\,{\AA} line a negative one (see Fig.~\ref{fig:SrI20G}).
Bearing in mind that both spectral lines originate from a $J_\ell = 0 \rightarrow J_u = 1$ transition and have similar Hanle critical fields, we may suggest that this inversion is due to their different formation heights and properties. An in-depth analysis of this result goes beyond the scope of this paper and will be the object of a future investigation. 

\begin{figure*}[ht!]
    \centering
    \includegraphics[page=1, width=\textwidth]{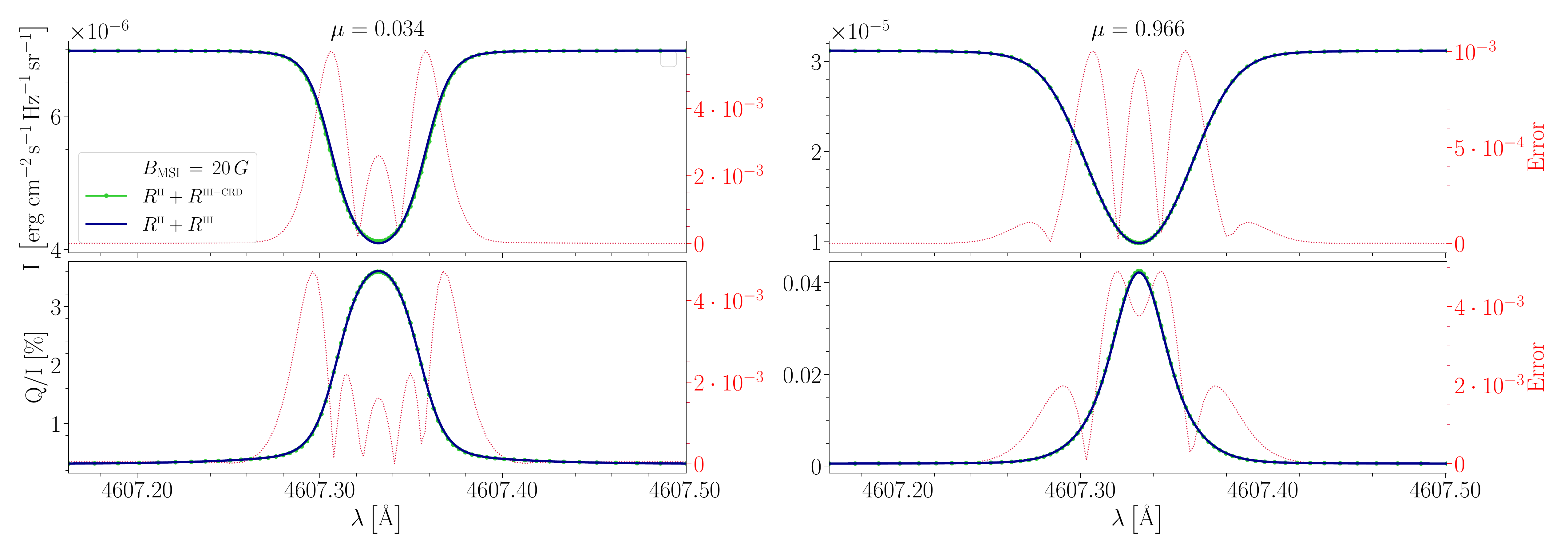}
    \caption{
    Same as Fig.~\ref{fig:SrI0G}, but in the presence of a uniform \textit{micro-structured isotropic} magnetic field with $B=20\,G$
    }
    \label{fig:SrI20MSI}
\end{figure*}

As anticipated, the Sr~{\sc i} 4607\,{\AA} line has been extensively exploited to investigate the small-scale unresolved magnetic fields that fill the quiet solar photosphere.
For this reason, we have also analyzed the case of unimodal MSI magnetic fields.
We recall that in the presence of such magnetic fields, the signatures of Zeeman and magneto-optical effects vanish due to cancellation effects, and the only impact of the magnetic field is the depolarization of $Q/I$ due to the Hanle effect, which depends on the field strength.
Figure \ref{fig:SrI20MSI} shows the $I$ and $Q/I$ profiles for a height-independent MSI magnetic field with $B=20$ G.
No significant differences between $R^{\scriptscriptstyle \mathrm{III}}$ and $R^{\scriptscriptstyle \mathrm{III-CRD}}$ calculations are visible, and the error is always smaller than $5\times10^{-3}$.
This result suggests that the $R^{\scriptscriptstyle \mathrm{III-CRD}}$ approximation provides accurate results when the problem is characterized by cylindrical symmetry, while it can introduce appreciable discrepancies when the direction of a non-vertical, deterministic magnetic field breaks this symmetry.

%%%%%%%%%%%%%%%%%%%%%%%%%%%%%%%%%%%%%%%%%%%%%%%%%%%%%%

%%%%%%%%%%%%%%%%%%%%%%%%%%%%%%%%%%%%%%%%%%%%%%%%%%%%%%%%%%%%%%%%%%%%%%%%%%%%%%
\section{Numerical results: 1D atmospheric model from 3D MHD simulation}
\label{sec:bifrost_atmos}

In Sect.~\ref{sec:FALC_atmos}, we limited our calculations to the semi-empirical FAL-C atmospheric model, possibly including spatially uniform magnetic fields, in the absence of bulk velocities.
In this section, we compare the impact of $R^{\scriptscriptstyle \mathrm{III}}$ and $R^{\scriptscriptstyle \mathrm{III-CRD}}$ in PRD calculations in a 1D atmospheric model extracted from a 3D MHD simulation, which includes height-dependent magnetic and bulk velocity fields.
As we have seen in previous sections, the magnetic field impacts the polarization profiles through the Hanle, magneto-optical, and Zeeman effects. 
On the other hand, bulk velocities 
% \textcolor{red}{\st{generally}}
introduce Doppler shifts as well as amplitude enhancements and asymmetries in the polarization profiles \cite[e.g.,][]{sampoorna2015,sampoorna2016movingmedia, carlin2012scattering,carlin2013,jaume_bestard2021}.
The joint impact of height-dependent magnetic fields and bulk velocities, together with the inherent thermodynamic structure of the atmospheric model, results in complex emergent Stokes profiles, in which it is generally difficult to distinguish the contribution of each factor.

\begin{figure}[!ht]
    \centering
    \includegraphics[width=\columnwidth]{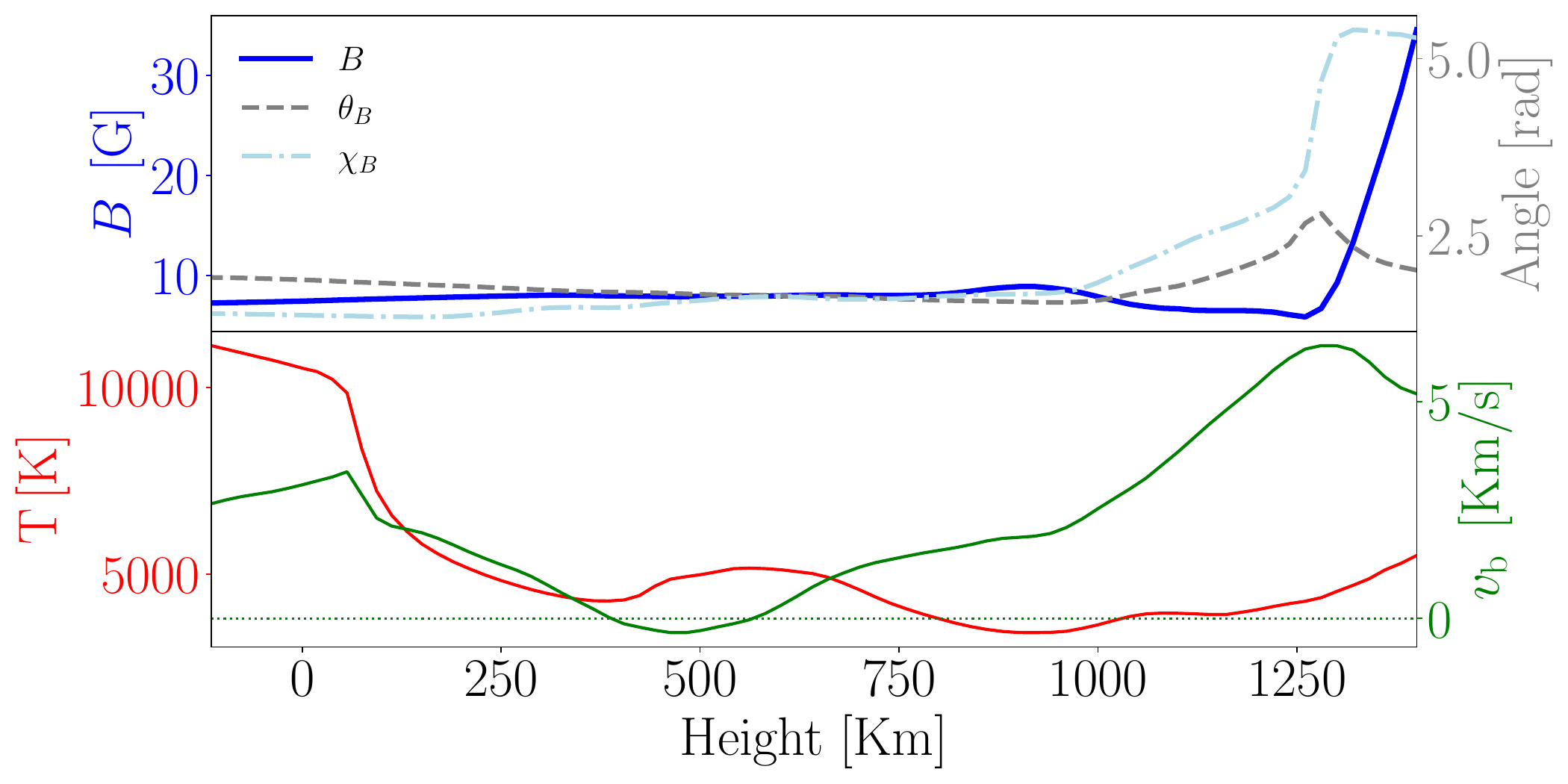}
    \caption{
     Physical quantities of the considered 1D atmospheric model extracted from the 3D MHD Bifrost simulation \textit{en024048\_hion} (snapshot 385, column 120 $\times$ 120). \textit{Upper panel}:
     % \textst{Variation of the magnetic field}
     strength (blue solid line), inclination (gray dashed line), and azimuth (light-blue dashed-dotted line) of the magnetic field as a function of height. 
    \textit{Lower panel:} vertical component of bulk velocity (green solid line)    % \textst{(\emph{lower panel})}
    and temperature (red solid line) as a function of height.
    % \textst{in the Bifrost model}.
    We adopt the convention that the velocity is positive if pointing outwards in the atmosphere 
    % (i.e. $\theta_{v_{\rm b}} =0$
    % $\chi_{v_{\rm b}} = 0$)
    and negative if pointing inwards.
    % (i.e. $\theta_{v_{\rm b}} =\pi$, and $\chi_{v_{\rm b}} = 0$).
    For clarity, the horizontal green dotted line indicates zero velocity.
	% \{The vertical dashed lines are the Line Formation (L.F.) thershold beyond which the atmosphere is transparent for the Ca~{\sc i} 4227\,{\AA}, and Sr~{\sc i} 4607\,{\AA} lines respectively at central lines frequencies, for a LOS of $\mu=0.034$ as decribed in Sect. \ref{sec:prior_analysis} and in Fig~\ref{fig:alphapBiFrost}.}
    }
    \label{fig:MagVelBf}
\end{figure}

\begin{figure}[!ht]
    \centering
    \includegraphics[width=\columnwidth]{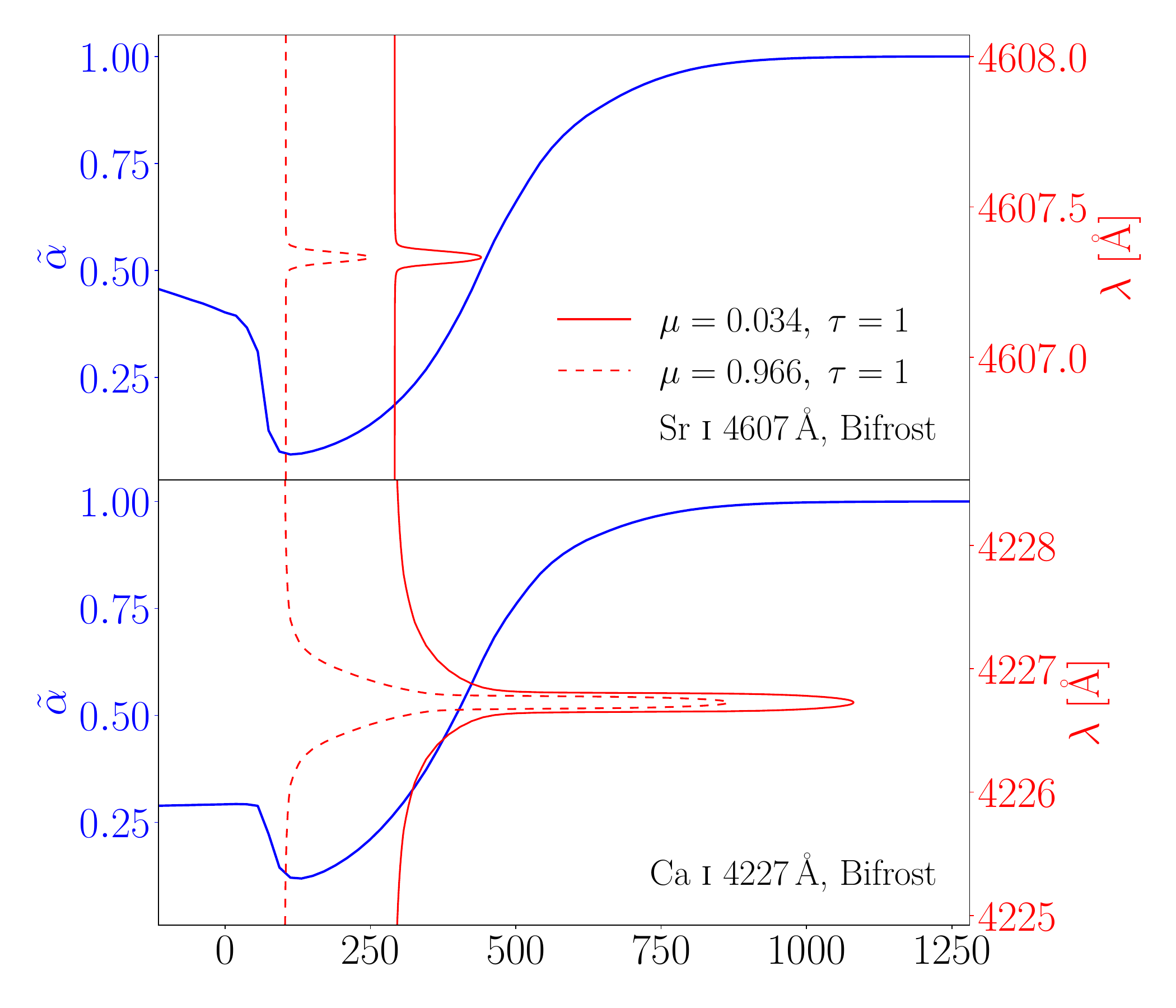}
    \caption{Same as Fig. \ref{fig:alphap} but for the Bifrost atmospheric model.}
    \label{fig:alphapBiFrost}
\end{figure}

In this work, we adopted a 1D atmospheric model extracted from the 3D magnetohydrodynamic (MHD) simulation of \mbox{\citet{carlsson2016}}, performed with the Bifrost code \citep{gudiksen2011stellar}.
In practice, we took a vertical column of such 3D simulation, clipped in the height interval
between -100 and 1400\,km, which includes the region where the considered spectral lines form,
% where the considered spectral lines form, i.e.,
discretized with $N_z=79$ nodes.
% while the spectral interval with $N_\nu=141$ discrete frequencies.
% in both cases.
% \{As can been seen in figures \ref{fig:alphapBiFrost}, and \ref{fig:MagVelBf} the line formation in the \CaILA{} line freezes at 1085\,Km of altitudes, therefore considering heights nodes abobe this thershold does not change the resulting profiles}.
The corresponding vertical resolution ranges between $19~\rm km$ and $100~\rm km$.
Figure~\ref{fig:MagVelBf} shows the variation of the magnetic field, temperature, and vertical component of the bulk velocity as a function of height in the considered model (hereafter, Bifrost model), which shows relatively quiet conditions.
Since the 1D module of the RH code (used to calculate the lower-level population) can only handle vertical bulk velocities, in this study we only considered the vertical component of the model's velocity, although our code can in principle take into account velocities of arbitrary direction.
As additional information, Fig.~\ref{fig:alphapBiFrost} shows the variation of the coherence fraction $\tilde{\alpha}$ with height in the Bifrost model for the two considered spectral lines, as well as the height at which the optical depth $\tau$, in the frequency intervals of the two spectral lines, is unity (see Sect.~\ref{sec:prior_analysis}).
% \{Since, as can been seen in Fig.~\ref{fig:alphapBiFrost} and explained in Section \ref{sec:prior_analysis}, in the Bifrost case the athmosphere become transparent at 1085 Km in the \CaILA{} line and at 450 Km in the \SrILA{} line, for this motivation in the effective calculations, we used, in both cases, an athmospheric model cutted at 1400 Km with $N_z = 79$ spatial nodes.}

\begin{figure*}[!ht]
	\centering
	\includegraphics[page=1, width=\textwidth]{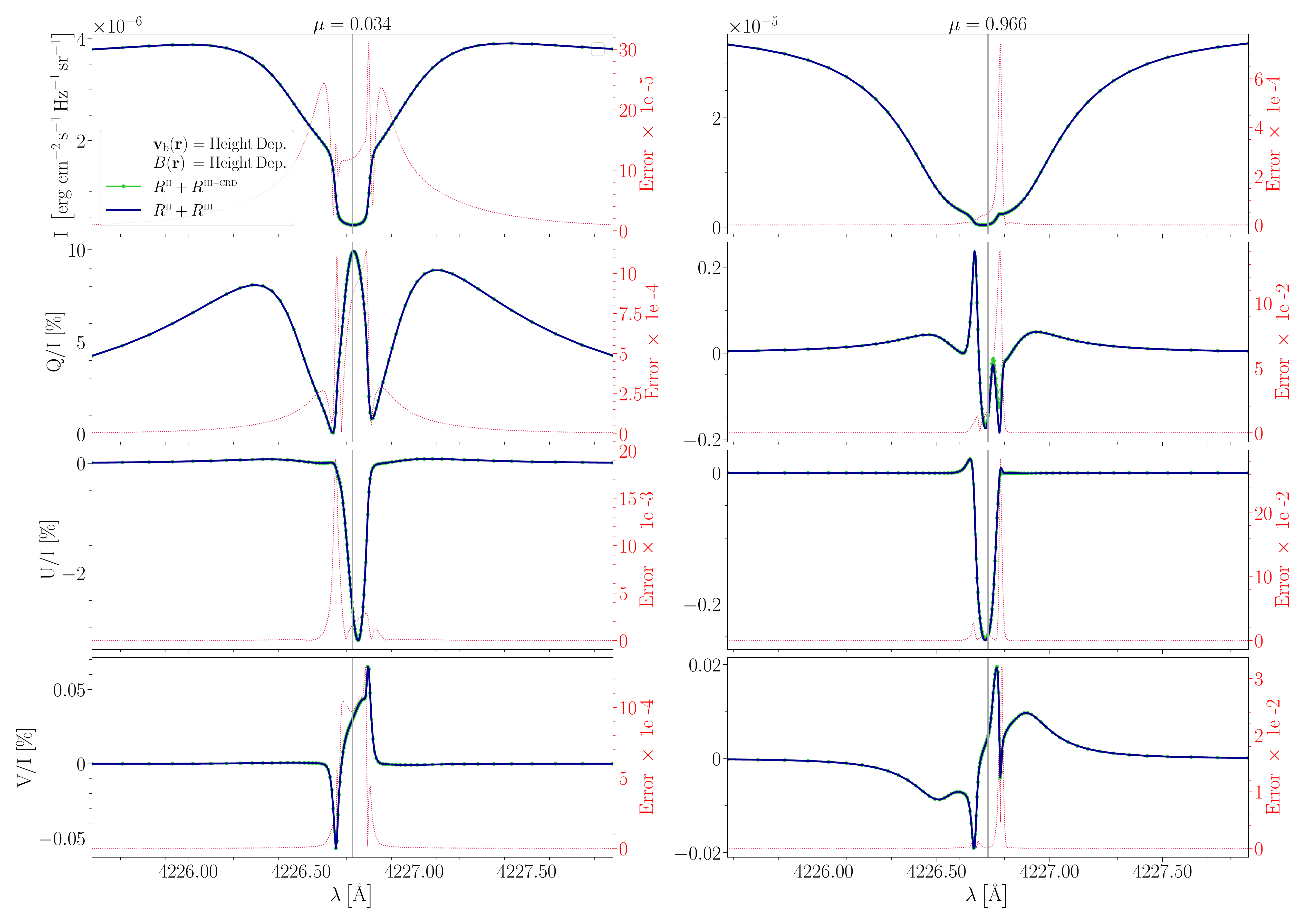}
    \caption{
        Emergent Stokes profiles for the Ca~{\sc i}~4227~\AA~line calculated in the Bifrost model (see Section \ref{sec:bifrost_atmos}), which includes height-dependent magnetic and (vertical) bulk velocity fields. The vertical gray lines show the central line wavelength.
    }
\label{fig:CaIbifrost_result}
\end{figure*}

\begin{figure*}
    \centering
    \includegraphics[page=1,width=\textwidth]{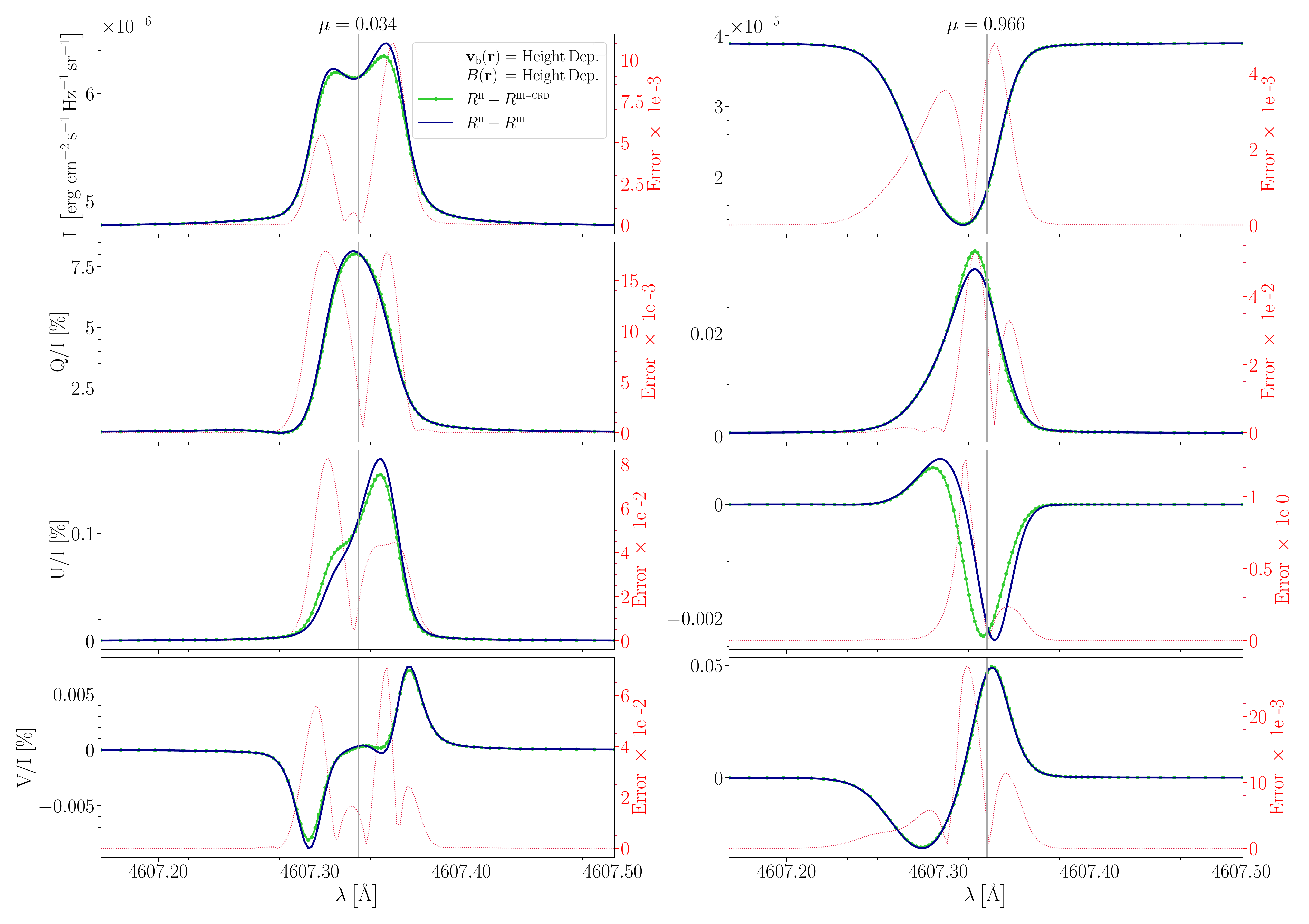}
    \caption{
    Same as Fig.~\ref{fig:CaIbifrost_result} but for the Sr~{\sc i} 4607\,{\AA} line.
    }
    \label{fig:SrIbifrost_result}
\end{figure*}

%%%%%%%%%%%%%%%%%%%%%%
\subsection{Ca~{\sc i} 4227\,{\AA} line}
\label{sec:resCaI_bifrost}

We first consider the modeling of the Ca~{\sc i} 4227\,{\AA} line, which is now discretized with $N_\nu=141$ unevenly spaced nodes.
Figure~\ref{fig:CaIbifrost_result} displays all the Stokes profiles, showing an overall good agreement between $R^{\scriptscriptstyle \mathrm{III}}$ and $R^{\scriptscriptstyle \mathrm{III-CRD}}$ calculations for a LOS with $\mu=0.034$. %in the $I$ and $V/I$ profiles.
By contrast, for a LOS with $\mu=0.996$, we observe some appreciable discrepancies in the $Q/I$ and $U/I$ profiles, with an error up to $3\times10^{-1}$.
We recall that these linear scattering polarization signals, obtained for a LOS close to the disk center, are due to the forward-scattering Hanle effect \mbox{\citep[e.g.][]{bueno2001Atomic}}.
Interestingly, such discrepancies disappear in the \textit{absence} of bulk velocities (profiles not reported here). 
This suggests that the differences between $R^{\scriptscriptstyle \mathrm{III}}$ and $R^{\scriptscriptstyle \mathrm{III-CRD}}$ calculations are amplified in the presence of bulk velocities, and they are larger for a LOS close to the vertical because in this case, the Doppler shifts are more pronounced.
We finally note that the error affects the amplitudes of the main peaks of the profiles but not their shape.

%%%%%%%%%%%%%%%%%%%%%%
\subsection{Sr~{\sc i} 4607\,\AA~line}
\label{sec:resSrI_bifrost}

We now consider the modeling of the Sr~{\sc i} 4607\,\AA~line, which is discretized with $N_\nu=141$ unevenly spaced nodes.
First, we note that in this Bifrost model, this line shows an emission profile in intensity for a LOS of $\mu=0.034$.
We verified that this is due to the thermodynamic structure of this atmospheric model, and it can be explained from the behavior of the source function at the formation height of the line for this limb LOS (see Fig.~\ref{fig:alphapBiFrost}).
Figure~\ref{fig:SrIbifrost_result} shows that appreciable discrepancies between $R^{\scriptscriptstyle \mathrm{III}}$ and $R^{\scriptscriptstyle \mathrm{III-CRD}}$ calculations are  found in all Stokes profiles for $\mu=0.034$, and in $Q/I$ and $U/I$ for $\mu=0.966$.
The maximal error is found in the  $U/I$  profile at $\mu=0.966$, where however the polarization signal is very weak and thus of limited practical interest.
The discrepancies between $R^{\scriptscriptstyle \mathrm{III}}$ and $R^{\scriptscriptstyle \mathrm{III-CRD}}$ computations become slightly larger in the Bifrost model,
where also a bulk velocity field is included. 
On the other hand, observing that the most significant errors appear in very weak polarization signals, we can conclude that for practical applications the $R^{\scriptscriptstyle \mathrm{III-CRD}}$ approximation can be safely used to obtain reliable results.

%%%%%%%%%%%%%%%%%%%%%%%%%%
\section{Conclusions}
\label{sec:conclusions}
%%%%%%%%%%%%%
In this work, we assessed the suitability and range of validity of a
widely used approximation for $R^{\scriptscriptstyle \mathrm{III}}$,
that is its expression under the assumption of CRD in the observer's frame, labeled with $R^{\scriptscriptstyle \mathrm{III-CRD}}$.
To this aim, we solved the full non-LTE RT problem for polarized radiation in 1D models of the solar atmosphere, considering both
the exact expression of $R^{\scriptscriptstyle \mathrm{III}}$ and
its $R^{\scriptscriptstyle \mathrm{III-CRD}}$ approximation.
With respect to the previous work of \mbox{\citet{sampoorna2017}}, we considered semi-empirical models, as well as 1D models extracted from 3D MHD simulations, which provide a reliable approximation of the solar atmosphere.
The analysis was focused on the chromospheric Ca~{\sc i} line at 4227\,{\AA} and the photospheric Sr~{\sc i} line at 4607\,{\AA}, accounting for the impact of magnetic fields (both deterministic and micro-structured) and bulk velocities.

We first compared the analytical forms of $R^{\scriptscriptstyle \mathrm{III}}$ and $R^{\scriptscriptstyle \mathrm{III-CRD}}$, showing that they correspond when the scattering angle $\Theta$ is equal to $\pi/2$, while their difference increases as $\Theta$ approaches 0 (forward scattering) or $\pi$ (backward scattering).
The numerical results for the Ca~{\sc i} 4227\,{\AA} line showed that the $R^{\scriptscriptstyle \mathrm{III-CRD}}$ approximation is suited to model the scattering polarization signals of strong chromospheric lines. 
This is not surprising, considering that in these lines the contribution of $R^{\scriptscriptstyle \mathrm{II}}$ dominates with respect to that of $R^{\scriptscriptstyle \mathrm{III}}$, both in the core (which forms high in the atmosphere) and in the far wings (where scattering is basically coherent). 
On the other hand, we verified that the $R^{\scriptscriptstyle \mathrm{III-CRD}}$ approximation is also adequate in the near wings, where the contribution of $R^{\scriptscriptstyle \mathrm{III}}$ cannot be neglected a-priori. 
The numerical results for the Sr~{\sc i} 4607\,{\AA} line showed that the $R^{\scriptscriptstyle \mathrm{III-CRD}}$ approximation provides reliable results also in photospheric lines, for which the contribution of $R^{\scriptscriptstyle \mathrm{III}}$ is significant.
Observing that the scattering polarization signal of the Sr~{\sc i} 4607\,{\AA} line is extensively used to investigate the small-scale magnetic fields that fill the quiet solar photosphere, we verified that the $R^{\scriptscriptstyle \mathrm{III-CRD}}$ approximation is also suitable in the presence of micro-turbulent magnetic fields.
On the other hand, some appreciable discrepancies (i.e., relative errors larger than 0.01 according to the error definition \eqref{eq:relErr}) are found when deterministic magnetic fields or bulk velocities are included, and the polarization signals are below 0.4\,\%.
However, the qualitative agreement between the two settings remains in general satisfactory; differences in the overall shapes and signs are exceptions and appear when the signals are very weak.
These discrepancies may suggest that the limit of CRD (i.e., to fully neglect PRD effects), which is generally adopted to model the intensity and polarization of these weak lines, can possibly introduce some appreciable inaccuracies with respect to a PRD treatment.
A quantitative comparison between PRD and CRD calculations for this line is ongoing,
and it will be presented in a separate publication.

In conclusion, we can state that in optically-thick media, the use of the lightweight $R^{\scriptscriptstyle \mathrm{III-CRD}}$ approximation guarantees reliable results in the modeling of scattering polarization in strong and weak resonance lines, thus confirming and generalizing the results of \citet{sampoorna2017}. 
This assesses the results of previous studies performed under this approximation and facilitates the development of 3D non-LTE RT simulations modeling the intensity and polarization of solar spectral lines, considering PRD effects \citep[see][]{benedusi3Drt2023}.

%%%%%%%%%%%%%%%%%%%%%%%%%%%%%%%%%%%%%%%%%%%%
\begin{acknowledgements}
    We thank the anonymous referee for carefully reading the manuscript and for many useful comments, suggestions, and corrections.
	The financial support by the Swiss National Science Foundation (SNSF) through grant $\rm CRSII5\_180238$ is gratefully acknowledged.
\end{acknowledgements}

\bibliographystyle{aa}
\bibliography{bibfile}

\begin{thebibliography}{72}
\expandafter\ifx\csname natexlab\endcsname\relax\def\natexlab#1{#1}\fi

\bibitem[{Alsina~Ballester {et~al.}(2017)Alsina~Ballester, Belluzzi, \& Trujillo~Bueno}]{ballester2017transfer}
Alsina~Ballester, E., Belluzzi, L., \& Trujillo~Bueno, J. 2017, \apj, 836, 6

\bibitem[{{Alsina Ballester} {et~al.}(2018){Alsina Ballester}, {Belluzzi}, \& {Trujillo Bueno}}]{ballester2018magneto}
{Alsina Ballester}, E., {Belluzzi}, L., \& {Trujillo Bueno}, J. 2018, \apj, 854, 150

\bibitem[{{Alsina Ballester} {et~al.}(2021){Alsina Ballester}, {Belluzzi}, \& {Trujillo Bueno}}]{alsina2021}
{Alsina Ballester}, E., {Belluzzi}, L., \& {Trujillo Bueno}, J. 2021, \prl, 127, 081101

\bibitem[{{Alsina Ballester} {et~al.}(2022){Alsina Ballester}, {Belluzzi}, \& {Trujillo Bueno}}]{alsina2022}
{Alsina Ballester}, E., {Belluzzi}, L., \& {Trujillo Bueno}, J. 2022, \aap, 664, A76

\bibitem[{{Anusha} \& {Nagendra}(2012)}]{anusha2012}
{Anusha}, L.~S. \& {Nagendra}, K.~N. 2012, \apj, 746, 84

\bibitem[{{Anusha} {et~al.}(2011){Anusha}, {Nagendra}, {Bianda}, {Stenflo}, {Holzreuter}, {Sampoorna}, {Frisch}, {Ramelli}, \& {Smitha}}]{anusha2011}
{Anusha}, L.~S., {Nagendra}, K.~N., {Bianda}, M., {et~al.} 2011, \apj, 737, 95

\bibitem[{Belluzzi \& Trujillo~Bueno(2012)}]{Belluzzi2012}
Belluzzi, L. \& Trujillo~Bueno, J. 2012, \apj, 750, L11

\bibitem[{{Benedusi} {et~al.}(2021){Benedusi}, {Janett}, {Belluzzi}, \& {Krause}}]{benedusi2021}
{Benedusi}, P., {Janett}, G., {Belluzzi}, L., \& {Krause}, R. 2021, \aap, 655, A88

\bibitem[{{Benedusi} {et~al.}(2022){Benedusi}, {Janett}, {Riva}, {Krause}, \& {Belluzzi}}]{benedusi2022}
{Benedusi}, P., {Janett}, G., {Riva}, S., {Krause}, R., \& {Belluzzi}, L. 2022, \aap, 664, A197

\bibitem[{{Benedusi} {et~al.}(2023){Benedusi}, {Riva}, {Zulian}, {{\v{S}}t{\v{e}}p{\'a}n}, {Belluzzi}, \& {Krause}}]{benedusi3Drt2023}
{Benedusi}, P., {Riva}, S., {Zulian}, P., {et~al.} 2023, J. Comput. Phys., 479, 112013

\bibitem[{Bianda {et~al.}(2003)Bianda, Stenflo, Gandorfer, \& Gisler}]{bianda2003enigmatic}
Bianda, M., Stenflo, J., Gandorfer, A., \& Gisler, D. 2003, in Current Theoretical Models and Future High Resolution Solar Observations: Preparing for ATST, Vol. 286, 61

\bibitem[{{Bommier}(1997{\natexlab{a}})}]{bommier1997masterI}
{Bommier}, V. 1997{\natexlab{a}}, \aap, 328, 706

\bibitem[{{Bommier}(1997{\natexlab{b}})}]{bommier1997masterII}
{Bommier}, V. 1997{\natexlab{b}}, \aap, 328, 726

\bibitem[{{Carlin} {et~al.}(2013){Carlin}, {Asensio Ramos}, \& {Trujillo Bueno}}]{carlin2013}
{Carlin}, E.~S., {Asensio Ramos}, A., \& {Trujillo Bueno}, J. 2013, \apj, 764, 40

\bibitem[{{Carlin} \& {Bianda}(2017)}]{carlin17}
{Carlin}, E.~S. \& {Bianda}, M. 2017, \apj, 843, 64

\bibitem[{{Carlin} {et~al.}(2012){Carlin}, {Manso Sainz}, {Asensio Ramos}, \& {Trujillo Bueno}}]{carlin2012scattering}
{Carlin}, E.~S., {Manso Sainz}, R., {Asensio Ramos}, A., \& {Trujillo Bueno}, J. 2012, \apj, 751, 5

\bibitem[{{Carlsson} {et~al.}(2016){Carlsson}, {Hansteen}, {Gudiksen}, {Leenaarts}, \& {De Pontieu}}]{carlsson2016}
{Carlsson}, M., {Hansteen}, V.~H., {Gudiksen}, B.~V., {Leenaarts}, J., \& {De Pontieu}, B. 2016, \aap, 585, A4

\bibitem[{Davis \& Rabinowitz(2007)}]{davis2007methods}
Davis, P.~J. \& Rabinowitz, P. 2007, Methods of numerical integration (Courier Corporation)

\bibitem[{{del Pino Alem{\'a}n} \& {Trujillo Bueno}(2021)}]{delpinoaleman2021}
{del Pino Alem{\'a}n}, T. \& {Trujillo Bueno}, J. 2021, \apj, 909, 180

\bibitem[{{del Pino Alem{\'a}n} {et~al.}(2020){del Pino Alem{\'a}n}, {Trujillo Bueno}, {Casini}, \& {Manso Sainz}}]{delpinoaleman2020}
{del Pino Alem{\'a}n}, T., {Trujillo Bueno}, J., {Casini}, R., \& {Manso Sainz}, R. 2020, \apj, 891, 91

\bibitem[{{del Pino Alem{\'a}n} {et~al.}(2018){del Pino Alem{\'a}n}, {Trujillo Bueno}, {{\v{S}}t{\v{e}}p{\'a}n}, \& {Shchukina}}]{delpinoaleman2018}
{del Pino Alem{\'a}n}, T., {Trujillo Bueno}, J., {{\v{S}}t{\v{e}}p{\'a}n}, J., \& {Shchukina}, N. 2018, \apj, 863, 164

\bibitem[{{Dhara} {et~al.}(2019){Dhara}, {Capozzi}, {Gisler}, {Bianda}, {Ramelli}, {Berdyugina}, {Alsina}, \& {Belluzzi}}]{dhara2019}
{Dhara}, S.~K., {Capozzi}, E., {Gisler}, D., {et~al.} 2019, \aap, 630, A67

\bibitem[{{Domke} \& {Hubeny}(1988)}]{domke88redMat}
{Domke}, H. \& {Hubeny}, I. 1988, \apj, 334, 527

\bibitem[{Faddeeva \& Terent'ev(1961)}]{faddeeva1961tables}
Faddeeva, V.~N. \& Terent'ev, N.~M. 1961, Tables of Values of the Function W(z) (Pergamon Press)

\bibitem[{{Faurobert}(1987)}]{faurobert1987}
{Faurobert}, M. 1987, \aap, 178, 269

\bibitem[{{Faurobert}(1988)}]{faurobert1988}
{Faurobert}, M. 1988, \aap, 194, 268

\bibitem[{{Faurobert-Scholl}(1992)}]{faurobert1992}
{Faurobert-Scholl}, M. 1992, \aap, 258, 521

\bibitem[{{Fontenla} {et~al.}(1993){Fontenla}, {Avrett}, \& {Loeser}}]{fontenla1993}
{Fontenla}, J.~M., {Avrett}, E.~H., \& {Loeser}, R. 1993, \apj, 406, 319

\bibitem[{{Gandorfer}(2000)}]{gandorfer2000}
{Gandorfer}, A. 2000, {The Second Solar Spectrum: A high spectral resolution polarimetric survey of scattering polarization at the solar limb in graphical representation. Volume I: 4625\,{\AA} to 6995\,{\AA}} (Zurich: vdf ETH)

\bibitem[{{Gandorfer}(2002)}]{gandorfer2002}
{Gandorfer}, A. 2002, {The Second Solar Spectrum: A high spectral resolution polarimetric survey of scattering polarization at the solar limb in graphical representation. Volume II: 3910\,{\AA} to 4630\,{\AA}} (Zurich: vdf ETH)

\bibitem[{Gudiksen {et~al.}(2011)Gudiksen, Carlsson, Hansteen, Hayek, Leenaarts, \& Mart{\'\i}nez-Sykora}]{gudiksen2011stellar}
Gudiksen, B.~V., Carlsson, M., Hansteen, V.~H., {et~al.} 2011, \aap, 531, A154

\bibitem[{{Holzreuter} {et~al.}(2005){Holzreuter}, {Fluri}, \& {Stenflo}}]{Holzreuter+al2005}
{Holzreuter}, R., {Fluri}, D.~M., \& {Stenflo}, J.~O. 2005, \aap, 434, 713

\bibitem[{Hummer(1962)}]{hummer1962non}
Hummer, D.~G. 1962, MNRAS, 125, 21

\bibitem[{{Janett} {et~al.}(2021{\natexlab{a}}){Janett}, {Alsina Ballester}, {Guerreiro}, {Riva}, {Belluzzi}, {del Pino Alem{\'a}n}, \& {Trujillo Bueno}}]{janett2021a}
{Janett}, G., {Alsina Ballester}, E., {Guerreiro}, N., {et~al.} 2021{\natexlab{a}}, \aap, 655, A13

\bibitem[{{Janett} {et~al.}(2021{\natexlab{b}}){Janett}, {Benedusi}, {Belluzzi}, \& {Krause}}]{janett2021b}
{Janett}, G., {Benedusi}, P., {Belluzzi}, L., \& {Krause}, R. 2021{\natexlab{b}}, \aap, 655, A87

\bibitem[{{Janett} {et~al.}(2017){Janett}, {Carlin}, {Steiner}, \& {Belluzzi}}]{janettI}
{Janett}, G., {Carlin}, E.~S., {Steiner}, O., \& {Belluzzi}, L. 2017, \apj, 840, 107

\bibitem[{Janett \& Paganini(2018)}]{janettIII}
Janett, G. \& Paganini, A. 2018, \apj, 857, 91

\bibitem[{Janett {et~al.}(2018)Janett, Steiner, \& Belluzzi}]{janettIV}
Janett, G., Steiner, O., \& Belluzzi, L. 2018, \apj, 865, 16

\bibitem[{{Jaume Bestard} {et~al.}(2021){Jaume Bestard}, {Trujillo Bueno}, {{\v{S}}t{\v{e}}p{\'a}n}, \& {del Pino Alem{\'a}n}}]{jaume_bestard2021}
{Jaume Bestard}, J., {Trujillo Bueno}, J., {{\v{S}}t{\v{e}}p{\'a}n}, J., \& {del Pino Alem{\'a}n}, T. 2021, \apj, 909, 183

\bibitem[{{Kano} {et~al.}(2017){Kano}, {Trujillo Bueno}, {Winebarger}, {Auch{\`e}re}, {Narukage}, {Ishikawa}, {Kobayashi}, {Bando}, {Katsukawa}, {Kubo}, {Ishikawa}, {Giono}, {Hara}, {Suematsu}, {Shimizu}, {Sakao}, {Tsuneta}, {Ichimoto}, {Goto}, {Belluzzi}, {{\v{S}}t{\v{e}}p{\'a}n}, {Asensio Ramos}, {Manso Sainz}, {Champey}, {Cirtain}, {De Pontieu}, {Casini}, \& {Carlsson}}]{kano2017}
{Kano}, R., {Trujillo Bueno}, J., {Winebarger}, A., {et~al.} 2017, \apjl, 839, L10

\bibitem[{Kronrod(1965)}]{kronrod1965nodes}
Kronrod, A.~S. 1965, Nodes and weights of quadrature formulas: sixteen-place tables. (New York: Consultants Bureau)

\bibitem[{Landi~Degl'Innocenti \& Landolfi(2004)}]{degl2006polarization}
Landi~Degl'Innocenti, E. \& Landolfi, M. 2004, ASSL, Vol. 307, Polarization in spectral lines (Springer Science \& Business Media)

\bibitem[{{Leenaarts} {et~al.}(2012){Leenaarts}, {Pereira}, \& {Uitenbroek}}]{leenaarts2012}
{Leenaarts}, J., {Pereira}, T., \& {Uitenbroek}, H. 2012, \aap, 543, A109

\bibitem[{Mihalas(1978)}]{mihalas1978stellar}
Mihalas, D. 1978, Stellar atmospheres (W.H. Freeman)

\bibitem[{{Moore} {et~al.}(1966){Moore}, {Minnaert}, \& {Houtgast}}]{moore1966}
{Moore}, C.~E., {Minnaert}, M.~G.~J., \& {Houtgast}, J. 1966, {The solar spectrum 2935 A to 8770 A, Monograph. 61} (National Bureau of Standards, U.S.)

\bibitem[{{Nagendra} {et~al.}(2002){Nagendra}, {Frisch}, \& {Faurobert}}]{nagendra2002Hanle}
{Nagendra}, K.~N., {Frisch}, H., \& {Faurobert}, M. 2002, \aap, 395, 305

\bibitem[{{Nagendra} \& {Sampoorna}(2011)}]{nagendra2011Spectal}
{Nagendra}, K.~N. \& {Sampoorna}, M. 2011, \aap, 535, A88

\bibitem[{{Nagendra} {et~al.}(2020){Nagendra}, {Sowmya}, {Sampoorna}, {Stenflo}, \& {Anusha}}]{nagendra2020}
{Nagendra}, K.~N., {Sowmya}, K., {Sampoorna}, M., {Stenflo}, J.~O., \& {Anusha}, L.~S. 2020, \apj, 898, 49

\bibitem[{Oeftiger {et~al.}(2016)Oeftiger, Aviral, De~Maria, Deniau, Hegglin, Li, McIntosh, \& Moneta}]{oeftiger2016review}
Oeftiger, A., Aviral, A., De~Maria, R., {et~al.} 2016, in Proc. of International Particle Accelerator Conference (IPAC'16), Busan, Korea, May 8-13, 2016, International Particle Accelerator Conference No.~7 (Geneva, Switzerland: JACoW), 3090--3093

\bibitem[{Piessens {et~al.}(2012)Piessens, de~Doncker-Kapenga, {\"U}berhuber, \& Kahaner}]{piessens2012quadpack}
Piessens, R., de~Doncker-Kapenga, E., {\"U}berhuber, C.~W., \& Kahaner, D.~K. 2012, Quadpack: a subroutine package for automatic integration, Vol.~1 (Springer Science \& Business Media)

\bibitem[{{Rachmeler} {et~al.}(2022){Rachmeler}, {Trujillo Bueno}, {McKenzie}, {Ishikawa}, {Auch{\`e}re}, {Kobayashi}, {Kano}, {Okamoto}, {Bethge}, {Song}, {Alsina Ballester}, {Belluzzi}, {del Pino Alem{\'a}n}, {Asensio Ramos}, {Yoshida}, {Shimizu}, {Winebarger}, {Kobelski}, {Vigil}, {De Pontieu}, {Narukage}, {Kubo}, {Sakao}, {Hara}, {Suematsu}, {{\v{S}}t{\v{e}}p{\'a}n}, {Carlsson}, \& {Leenaarts}}]{rachmeler2022}
{Rachmeler}, L.~A., {Trujillo Bueno}, J., {McKenzie}, D.~E., {et~al.} 2022, \apj, 936, 67

\bibitem[{{Rees} \& {Saliba}(1982)}]{rees1982}
{Rees}, D.~E. \& {Saliba}, G.~J. 1982, \aap, 115, 1

\bibitem[{{Sampoorna} \& {Nagendra}(2015)}]{sampoorna2015}
{Sampoorna}, M. \& {Nagendra}, K.~N. 2015, \apj, 812, 28

\bibitem[{Sampoorna \& Nagendra(2016)}]{sampoorna2016movingmedia}
Sampoorna, M. \& Nagendra, K.~N. 2016, \apj, 833, 32

\bibitem[{{Sampoorna} {et~al.}(2011){Sampoorna}, {Nagendra}, \& {Frisch}}]{sampoorna2011spectral}
{Sampoorna}, M., {Nagendra}, K.~N., \& {Frisch}, H. 2011, \aap, 527, A89

\bibitem[{{Sampoorna} {et~al.}(2017){Sampoorna}, {Nagendra}, \& {Stenflo}}]{sampoorna2017}
{Sampoorna}, M., {Nagendra}, K.~N., \& {Stenflo}, J.~O. 2017, \apj, 844, 97

\bibitem[{{Sampoorna} {et~al.}(2009){Sampoorna}, {Stenflo}, {Nagendra}, {Bianda}, {Ramelli}, \& {Anusha}}]{sampoorna2009}
{Sampoorna}, M., {Stenflo}, J.~O., {Nagendra}, K.~N., {et~al.} 2009, \apj, 699, 1650

\bibitem[{Sipser(1996)}]{sipser1996introduction}
Sipser, M. 1996, Introduction to the Theory of Computation (Cengage Learning, USA)

\bibitem[{{Stenflo}(1994)}]{stenflo1994}
{Stenflo}, J. 1994, {Solar Magnetic Fields: Polarized Radiation Diagnostics}, Vol. 189 (Springer)

\bibitem[{{Stenflo}(1982)}]{stenflo1982}
{Stenflo}, J.~O. 1982, \solphys, 80, 209

\bibitem[{{Stenflo} {et~al.}(1980){Stenflo}, {Baur}, \& {Elmore}}]{stenflo1980}
{Stenflo}, J.~O., {Baur}, T.~G., \& {Elmore}, D.~F. 1980, \aap, 84, 60

\bibitem[{{Stenflo} \& {Keller}(1997)}]{stenflo1997sss}
{Stenflo}, J.~O. \& {Keller}, C.~U. 1997, \aap, 321, 927

\bibitem[{{Supriya} {et~al.}(2012){Supriya}, {Nagendra}, {Sampoorna}, \& {Ravindra}}]{supriya2012effect}
{Supriya}, H.~D., {Nagendra}, K.~N., {Sampoorna}, M., \& {Ravindra}, B. 2012, \mnras, 425, 527

\bibitem[{{Supriya} {et~al.}(2013){Supriya}, {Sampoorna}, {Nagendra}, {Ravindra}, \& {Anusha}}]{supriya2013efficient}
{Supriya}, H.~D., {Sampoorna}, M., {Nagendra}, K.~N., {Ravindra}, B., \& {Anusha}, L.~S. 2013, \jqsrt, 119, 67

\bibitem[{{Supriya} {et~al.}(2014){Supriya}, {Smitha}, {Nagendra}, {Stenflo}, {Bianda}, {Ramelli}, {Ravindra}, \& {Anusha}}]{supriya2014}
{Supriya}, H.~D., {Smitha}, H.~N., {Nagendra}, K.~N., {et~al.} 2014, \apj, 793, 42

\bibitem[{{Trujillo Bueno}(2001)}]{bueno2001Atomic}
{Trujillo Bueno}, J. 2001, in ASP Conf. Ser., Vol. 236, Advanced Solar Polarimetry -- Theory, Observation, and Instrumentation, ed. M.~{Sigwarth}, 161

\bibitem[{{Trujillo Bueno}(2014)}]{trujillo2014}
{Trujillo Bueno}, J. 2014, in ASP Conf. Ser., Vol. 489, Solar Polarization 7, ed. K.~N. {Nagendra}, J.~O. {Stenflo}, Z.~Q. {Qu}, \& M.~{Sampoorna}, 137

\bibitem[{{Trujillo Bueno} {et~al.}(2017){Trujillo Bueno}, {Landi Degl'Innocenti}, \& {Belluzzi}}]{trujillo2017}
{Trujillo Bueno}, J., {Landi Degl'Innocenti}, E., \& {Belluzzi}, L. 2017, \ssr, 210, 183

\bibitem[{{Trujillo Bueno} {et~al.}(2004){Trujillo Bueno}, {Shchukina}, \& {Asensio Ramos}}]{trujillo2004}
{Trujillo Bueno}, J., {Shchukina}, N., \& {Asensio Ramos}, A. 2004, \nat, 430, 326

\bibitem[{{Uitenbroek}(2001)}]{uitenbroek2001multilevel}
{Uitenbroek}, H. 2001, \apj, 557, 389

\bibitem[{{Zeuner} {et~al.}(2022){Zeuner}, {Belluzzi}, {Guerreiro}, {Ramelli}, \& {Bianda}}]{zeuner2022}
{Zeuner}, F., {Belluzzi}, L., {Guerreiro}, N., {Ramelli}, R., \& {Bianda}, M. 2022, \aap, 662, A46

\bibitem[{{Zeuner} {et~al.}(2020){Zeuner}, {Manso Sainz}, {Feller}, {van Noort}, {Solanki}, {Iglesias}, {Reardon}, \& {Mart{\'\i}nez Pillet}}]{zeuner2020}
{Zeuner}, F., {Manso Sainz}, R., {Feller}, A., {et~al.} 2020, \apjl, 893, L44

\end{thebibliography}

%%%%%%%%%%%%%%%%%%%%%%%%%%%%%%%%%%%%%%%%%%%%%%%%%%%%%%%%%%%%%%%%%%%%%%%%%%%%%%%%%%%
%%%%%%%%%%%%%%%%%%%%%%%%%%%%%%%%%%%%%%%%%%%%%%%%%%%%%%%%%%%%%%%%%%%%%%%%%%%%%%%%%%
%%%%%%%%%%%%%%%%%%%%%%%%%%%%%%%%%%%%%%%%%%%%%%%%%%%%%%%%%%%%%%%%%%%%%%%%%%%%%%%%%%
%% Appendix
\onecolumn 
\appendix

\section{Atomic model and data}
\label{sec:atomic_model}

\newcommand{\CGSLarmor}{\footnote{In the cgs system, the Larmor frequency is numerically approximated by $\nu_L = 1.3996 \times 10^6 \, B$, 
with $B$ expressed in G and $\nu_L$ in s$^{-1}$.}}

We consider an atomic system composed of two levels (two-level atom).
Each level is characterized by the energy $E$, the quantum number for the total angular momentum $J$ (positive integer or semi-integer values only), 
and the Land\'e factor $g$. 
Hereafter, the physical quantities referring to the upper and lower levels will be labeled with subscripts $u$ and $\ell$, respectively. 
Each level is composed of $2J+1$ magnetic sublevels characterized by the magnetic quantum number $M$ ($M = -J, -J+1, ..., J$).
The magnetic sublevels are degenerate in the absence of magnetic fields, while they split in energy when a magnetic field $\vec{B}$ is present (Zeeman effect).
Their energies are given by $E(M) = E_J + h  \, \nu_L  \, g  \, M$, where $E_J$ is the energy of the considered $J$-level, $h$ the Planck constant, and $ \nu_L =  e B \left( 4 \pi m_e c \right)^{-1}$ (with $e$ the elementary charge, $m_e$ the electron mass, and $c$ the speed of light) the Larmor frequency\CGSLarmor{}. %and $\nu_L$ in s$^{-1}$).
The line-center frequency is $\nu_0 = (E_u - E_\ell) / h$.
The frequency of the Zeeman component corresponding to the transition between the upper magnetic sublevel $M_u$ and the lower magnetic sublevel $M_\ell$ is
\begin{equation*}
	\nu_{M_u M_\ell} = \frac{E(M_u) - E(M_\ell )}{h} = \nu_0 + \nu_L 
	(g_u M_u - g_\ell M_\ell) \, .
	% \label{eq:nu_ul}
\end{equation*}
The energies, angular momenta, Land\'e factors, and Einstein coefficients 
for spontaneous emission $A_{u\ell}$ for the two-level atomic models considered in this work to 
synthesize the intensity and polarization of the Ca~{\sc i} 4227\,{\AA} and 
Sr~{\sc i} 4607\,{\AA} lines are summarized in Table~\ref{tab:atomic-data}. 
The Land\'e factors of levels with $J_\ell=0$ are formally taken to be zero.
\begin{table}[ht!]
	\begin{tabular}{ccccccccc}
		\hline
		Element & Air wavelength [\AA] & $E_u$ [cm$^{-1}$] & $J_u$ & $g_u$
		& $E_\ell$ [cm$^{-1}$] & $J_\ell$ & $g_\ell$ & 
		$A_{u \ell}$ [s$^{-1}$] \\
		\hline
		Ca~{\sc i} & 4226.73 & $23652.304$ & 1 & $1.0$ & $0.00$ & 0 & $0.$ & 
		$2.18 \times 10^8$ \\
		Sr~{\sc i} & 4607.33 & $21698.452$ & 1 & $1.0$ & $0.00$ & 0 & $0.$ & 
		$2.01 \times 10^8$ \\
		\hline\\
	\end{tabular}
 \caption{Spectral lines and corresponding atomic data.}
 \label{tab:atomic-data}
\end{table}

%%%%%%%%%%%%%%%%%%%%%%%%%%%%%%%%%%%%%%%%%%%%%%%%%%%%%%%%%%%

\section{Redistribution matrix}
\label{sec:R_analytic}

All the results presented in this work are obtained considering the redistribution matrix for a two-level atom with an unpolarized and infinitely-sharp lower level, in the presence of magnetic fields, as derived by \citet{bommier1997masterII}.
This redistribution matrix is given by the sum of two terms, commonly referred to as $R^{\scriptscriptstyle \mathrm{II}}$ and 
$R^{\scriptscriptstyle \mathrm{III}}$ \citep[e.g.,][]{hummer1962non}.
In this appendix, we provide their analytic expressions, written with a slightly different notation than the one used in the original 
paper or in subsequent works in which the same redistribution matrices were reported and applied \citep[e.g.,][]{ballester2017transfer}.
We first present their expressions in the atomic reference frame, taking the quantization axis along the magnetic field 
(Sect.~\ref{app:R_atomic}).
We briefly comment on the branching ratios (Sect.~\ref{sec:branch_ratio}) and we derive their expressions in a new reference system with the quantization axis along the vertical (Sect.~\ref{app:R_vertical}). 
Subsequently, we transform them in the observer's reference frame, including bulk velocities (Sect.~\ref{sec:R_observer}).
Finally, we provide their expressions under a series of simplifying approximations, including the one for 
$R^{\scriptscriptstyle \mathrm{III}}$ analyzed in this work (Sect.~\ref{sec:R_approx}).

%%%%%%%%%%%%%%%%%%%%%%%%%%%%%%%%%%%%%%%%%%%%%%%%%%%%%%%%%%%

\subsection{Expression in the atomic reference frame}
\label{app:R_atomic}

In the atomic reference frame, taking the quantization axis parallel to the magnetic field (magnetic reference system), the $R^{\scriptscriptstyle \mathrm{II}}$ and $R^{\scriptscriptstyle \mathrm{III}}$ redistribution matrices are respectively given by Eqs.~(51) and (49) in \citet{bommier1997masterII}.
We refer to these redistribution matrices with the symbol 
$\hat{R}^{\scriptscriptstyle \mathrm{X}}$ (with 
${\scriptstyle \mathrm{X=II,III}}$) to distinguish them from those expressed 
taking the quantization axis along a different direction.
After some variable and index renaming,\footnote{Variable and index 
renaming for Eqs.~(49) and (51) in \citet{bommier1997masterII}:
\begin{align*}
	& \vec{\Omega}_1 \rightarrow \vec{\Omega}' \, , \quad 
	\nu_1 \rightarrow \xi' \, , \quad \nu \rightarrow \xi \, , \quad 
	K \rightarrow K'' \, , \quad K' \rightarrow K \, , \quad 
	K'' \rightarrow K' \, , \\
	& J' \rightarrow J_u \, , \quad J \rightarrow J_\ell \, , \quad 
	M \rightarrow M_u \, , \quad N \rightarrow M_\ell \, , \quad
	M' \rightarrow M_u' \, , \quad N'' \rightarrow M_\ell' \, .
	\label{eq:RX_atom_mag}
\end{align*}}
we rewrite them in the following equivalent form
\begin{equation}
	\hat{R}^{\scriptscriptstyle \mathrm{X}}_{ij}(\vec{r},\vec{\Omega},
	\vec{\Omega}',\xi,\xi') = 
	\sum_{K,K'=0}^2 \sum_{Q=-K_{\rm min}}^{K_{\rm min}} 
	\mathcal{R}^{{\scriptscriptstyle \mathrm{X}},KK'}_Q(\vec{r},\xi,\xi') \, 
	(-1)^Q \, 
	\hat{\mathcal{T}}^{K'}_{Q,i}(\vec{r},\vec{\Omega}) \, 
	\hat{\mathcal{T}}^{K}_{-Q,j}(\vec{r},\vec{\Omega}') \, ,
	\label{eq:RX_atom_magnetic}
\end{equation}
where $\vec{r}$ is the spatial point, $\vec{\Omega}$ the propagation direction, and $\xi$ the radiation frequency in the atomic reference frame. The convention that primed and unprimed quantities refer to the incident and scattered radiation, respectively, is used.
The indices $i$ and $j$ can take values 1, 2, 3, and 4, while $K_{\mathrm{min}} = \min(K,K')$.
The quantity $\hat{\mathcal{T}}^K_{Q,i}$ (with $K=0,...,2$ and $Q=-K,...,K$) is the geometrical tensor defined in Sect.~5.11 of LL04, evaluated in the magnetic reference system. 
Given that the direction of the magnetic field may vary with the spatial point, this tensor depends in general on $\vec{r}$.

The function $\mathcal{R}^{{\scriptscriptstyle \mathrm{III}},KK'}_{Q}$ is 
given by \citep[see Eq.~(49) of][]{bommier1997masterII}
\begin{equation}
	\mathcal{R}^{{\scriptscriptstyle \mathrm{III}},KK'}_{Q}(\vec{r},\xi,\xi') = 
	\sum_{K''=|Q|}^{2J_u} 
	\left(\beta^{K''}_Q(\vec{r}) - \alpha_Q(\vec{r}) \right) \, 
	\Phi^{K''K'}_Q \left( \vec{r},\xi \right) \, 
	\Phi^{K''K}_Q \left( \vec{r},\xi' \right) \, ,
\label{eq:RIII_freq_atom}
\end{equation}
where
\begin{equation}
	\beta^K_Q(\vec{r}) = 
	\frac{\Gamma_R}{\Gamma_R + \Gamma_I(\vec{r}) + D^{(K)}(\vec{r}) + 
	2 \pi \mathrm{i} \nu_L(\vec{r}) g_u Q} \, , 
	\label{eq:beta_KQ}
\end{equation}
and
\begin{equation}
	\alpha_Q(\vec{r}) = \frac{\Gamma_R}
	{\Gamma_R + \Gamma_I(\vec{r}) + \Gamma_E(\vec{r}) + 
	2 \pi \mathrm{i} \nu_L(\vec{r}) g_u Q} \, .
	\label{eq:alpha_Q}
\end{equation}
The quantities $\Gamma_R$, $\Gamma_I$, and $\Gamma_E$ are the line broadening 
constants due to radiative decays, inelastic collisions, and elastic 
collisions, respectively:
\begin{equation*}
	\Gamma_R = A_{u\ell} \, , \quad
	\Gamma_I = C_{u\ell} \, , \quad
	\Gamma_E = Q_{\mathrm{el}} \, ,
\end{equation*}
where $C_{u\ell}$ is the rate of inelastic collisions inducing transitions from 
the upper to the lower level and $Q_{\mathrm{el}}$ is the rate of elastic 
collisions with neutral perturbers (mainly hydrogen and helium atoms).
The quantity $D^{(K)}$ is the depolarizing rate due to elastic collisions.
In the absence of experimental data or detailed theoretical calculations for the rates $D^{(K)}$, the approximate relation $D^{(2)}=0.5 \, Q_{\mathrm{el}}$ is generally used.
Under the assumption that the interaction between the atom and the perturber is described by a single tensor operator of rank $K'$, the rates with $K \ne 2$ can be obtained from $D^{(2)}$ as discussed in Sect.~7.13 of LL04 (see equation after (7.108) and Eq.~(7.109), which relates $D^{(1)}$ and $D^{(2)}$ for the case of $K'=2$).
The generalized profile $\Phi^{KK'}_Q$ is given by Eq.~(10.40) of LL04 and can be equivalently written as
\begin{equation}
	\Phi^{KK'}_Q \left( \vec{r},\xi \right) =
	\sum_{M_u,M_u'=-J_u}^{J_u} \sum_{M_\ell=-J_\ell}^{J_\ell} \sum_{q,q'=-1}^1 
	\mathcal{B}_{K K' Q M_u M_u' M_\ell q q'} \,
	\frac{1}{2} \, \left[ 
	\Phi_{M_u M_\ell}(\vec{r},\xi) + 
	\overline{\Phi}_{M_u' M_\ell}(\vec{r},\xi)
	\right] \, ,
	\label{eq:gen_prof_atom}
\end{equation}
where $K=0,...,2J_u$, $K'=0,1,2$, and $Q=-K_{\mathrm{min}},...,K_{\mathrm{min}}$ with $K_{\mathrm{min}}=\mathrm{min}(K,K')$. 
The notation $\overline{f}(\cdot)$ refers to the complex conjugate function.
The quantity $\mathcal{B}_{KK'Q M_u M_u' M_\ell qq'}$ is given by
\begin{align*}
	\mathcal{B}_{K K' Q M_u M_u' M_\ell q q'} & =
	(-1)^{1+J_u-M_u+q'}
	\sqrt{3 (2J_u+1)(2K+1)(2K'+1)} 
	\nonumber \\
	& \quad \times 
	\begin{pmatrix}
		J_u & J_\ell & 1 \\
		-M_u & M_\ell & -q
	\end{pmatrix}
	\begin{pmatrix}
		J_u & J_\ell & 1 \\
		-M_u' & M_\ell & -q'
	\end{pmatrix}
	\begin{pmatrix}
		J_u & J_u & K \\
		M_u' & -M_u & -Q
	\end{pmatrix}
	\begin{pmatrix}
		1 & 1 & K' \\
		q & -q' & -Q
	\end{pmatrix} \, ,
	% \label{eq:BKKpQ}
\end{align*}
where the quantities in parentheses are the Wigner's 3-$j$ symbols (e.g., 
Sect.~2.2 of LL04).
It must be observed that the 3-$j$ symbols are non-zero only if the sum of the arguments of the lower raw is zero. 
Because of this, the values of $q$ and $q'$ are uniquely determined once 
$M_u$, $M_u'$, and $M_\ell$ are assigned. Thus, the sums over these indices 
in Eq.~(\ref{eq:gen_prof_atom}) are in practice redundant.
The complex profile $\Phi_{M_u M_\ell}$ is defined as
\begin{equation}
	\Phi_{M_u M_\ell}(\vec{r},\xi) = \frac{1}{\pi} \, 
	\frac{1}{\Gamma(\vec{r}) - \mathrm{i}(\nu_{M_u M_\ell}(\vec{r}) -\xi)} \, ,
	\label{eq:Phi_atom}
\end{equation}
with $\Gamma = (\Gamma_R + \Gamma_I + \Gamma_E)/4\pi$.

\medbreak\noindent
The function $\mathcal{R}^{{\scriptscriptstyle \mathrm{II}},KK'}_{Q}$ is given 
by (see Eq.~(51) of \cite{bommier1997masterII})
\begin{align}
	\mathcal{R}^{{\scriptscriptstyle \mathrm{II}},KK'}_{Q}(\vec{r},\xi,\xi') & 
	= \alpha_Q(\vec{r}) \, \sum_{M_u, M_u'=-J_u}^{J_u} 
	\sum_{M_\ell, M_\ell'=-J_\ell}^{J_\ell} \sum_{p, p', p'', p'''=-1}^1
	\mathcal{C}_{K K' Q M_u M_u' M_\ell M_\ell' p p' p'' p'''} \nonumber \\
	& \quad \times 
	\delta(\xi - \xi' - \nu_{M_\ell M_\ell'}(\vec{r})) \, \frac{1}{2} \, \left[
	\Phi_{M_u' M_\ell}(\vec{r},\xi') + 
	\overline{\Phi}_{M_u M_\ell}(\vec{r},\xi') \right] \, ,
\label{eq:RII_freq_atom}
\end{align}
where $\alpha_Q$ is given by Eq.~(\ref{eq:alpha_Q}), $\delta(\cdot)$ is the 
Dirac delta, and $\Phi_{M_u M_\ell}$ is the complex profile defined
in Eq.~(\ref{eq:Phi_atom}).
The quantity $\nu_{M_\ell M_\ell'}$ is given by
\begin{equation*}
	\nu_{M_\ell M_\ell'}(\vec{r})=\nu_L(\vec{r})(g_\ell M_\ell - 
	g_\ell M_\ell') \, ,
	% \label{eq:nu_llp}
\end{equation*}
and $\mathcal{C}_{K K' Q M_u M'_u M_\ell M'_\ell p p' p'' p'''}$ is defined 
as \citep[see Eq.~(12) of][]{bommier1997masterII}
\begin{align*}
	{\mathcal C }_{K K' Q M_u M'_u M_\ell M'_\ell \ p p' p'' p'''} & = 
	3 \, (2J_u + 1) \, \sqrt{2K + 1} \, \sqrt{2K' + 1}  \, 
	(-1)^{2 J_u - M_\ell - M'_\ell} 
	\nonumber \\
 	& \quad \times 
	\left( \begin{array}{c c c} 
  		J_u & J_\ell & 1 \\
  		M_u & -M_\ell & -p 
	\end{array} \right)
	\left( \begin{array}{c c c} 
		J_u & J_\ell & 1 \\
		M'_u & -M_\ell & -p' 
	\end{array} \right)
	\left( \begin{array}{c c c}
		J_u & J_\ell & 1\\
		M_u & -M'_\ell & -p'' 
	\end{array} \right) 
	\left( \begin{array}{c c c}
		J_u & J_\ell & 1\\
		M'_u & -M'_\ell & -p'''
	\end{array} \right) 
	\notag \\
 	& \quad \times 
	\left( \begin{array}{c c c} 
		1 & 1 & K \\
		-p & p' & Q 
	\end{array} \right)
	\left( \begin{array}{c c c}
		1 & 1 & K' \\
		-p'' & p''' & Q
	\end{array} \right) \, .
	% \label{eq:C}
\end{align*}
As for $q$ and $q'$ in Eq.~\eqref{eq:gen_prof_atom}, the sums over $p$, $p'$, $p''$, and 
$p'''$ in Eq.~(\ref{eq:RII_freq_atom}) are in practice redundant.

\subsubsection{Branching ratios}
\label{sec:branch_ratio}
The quantities $(\beta^K_Q - \alpha_Q)$ and $\alpha_Q$ appearing in 
Eqs.~(\ref{eq:RIII_freq_atom}) and (\ref{eq:RII_freq_atom}), 
respectively, are the branching ratios between 
$R^{{\scriptscriptstyle \mathrm{II}}}$ and  $R^{{\scriptscriptstyle \mathrm{III}}}$.
These terms also contain the branching ratio for the scattering contribution 
to the total emissivity, $1 - \epsilon$, with
\begin{equation}
\label{eq:epsilon_pdp}
	\epsilon(\vec{r}) = \frac{\Gamma_I(\vec{r})}{\Gamma_R + \Gamma_I(\vec{r})}
\end{equation}
the photon destruction probability.
Factorizing this term (i.e., writing $\alpha_Q = (1 - \epsilon) \, \tilde{\alpha}_Q$ 
and $\beta^K_Q = (1 - \epsilon) \, \tilde{\beta}^K_Q$), the net branching ratios
for $R^{{\scriptscriptstyle \mathrm{II}}}$ and $R^{{\scriptscriptstyle \mathrm{III}}}$ 
are therefore
\begin{align}
\label{eq:branching_ratio_RII_RIII}
	\tilde{\alpha}_Q(\vec{r}) & = \frac{\Gamma_R + \Gamma_I(\vec{r})}
	{\Gamma_R + \Gamma_I(\vec{r}) + \Gamma_E(\vec{r}) + 2 \pi \mathrm{i}
	\nu_L(\vec{r}) g_u Q} \, ,	\\\nonumber
	\tilde{\beta}^K_Q(\vec{r}) - \tilde{\alpha}_Q(\vec{r}) & = 
	\frac{\Gamma_R + \Gamma_I(\vec{r})}
	{\Gamma_R + \Gamma_I(\vec{r}) + D^{(K)}(\vec{r}) + 2 \pi \mathrm{i}
	\nu_L(\vec{r}) g_u Q} -
	\frac{\Gamma_R + \Gamma_I(\vec{r})} 
	{\Gamma_R + \Gamma_I(\vec{r}) + \Gamma_E(\vec{r}) + 2 \pi \mathrm{i}
	\nu_L(\vec{r}) g_u Q} \, .
\end{align}
It can be observed that the branching ratios for the multipolar component $K=Q=0$ coincide with those for the unpolarized case.
If there are no elastic collisions ($\Gamma_E = D^{(K)} =0$), the branching ratio for $R^{{\scriptscriptstyle \mathrm{III}}}$ vanishes and that for $R^{{\scriptscriptstyle \mathrm{II}}}$ goes to unity (limit of coherent scattering in the atomic frame).
If elastic collisions are very efficient ($\Gamma_E \gg \Gamma_R, \Gamma_I$), the branching ratio for $R^{{\scriptscriptstyle \mathrm{II}}}$ becomes negligible with respect to that for $R^{{\scriptscriptstyle \mathrm{III}}}$. 
However, in this case also $D^{(K)}$ takes very large values and atomic polarization also becomes negligible.

\subsubsection{Rotation of the quantization axis}
\label{app:R_vertical}
The expressions of $R^{\scriptscriptstyle \mathrm{II}}$ and $R^{\scriptscriptstyle \mathrm{III}}$ provided above can be transformed into a new reference system with the quantization axis directed along any arbitrary direction by rotating the tensors $\hat{\mathcal{T}}^K_{Q,i}$ in Eq.~(\ref{eq:RX_atom_magnetic}).
In this work, the problem is formulated considering the Cartesian reference system of Fig.~\ref{fig:ref_sys}, with the $z$-axis (quantization axis) directed along the vertical (vertical reference system). %and the $x$-axis directed so that the LOS towards the observer lies in the $x-z$ plane \textbf{see Fig.~\ref{fig:ref_sys})}.
The relation between the geometrical tensors in the magnetic reference system ($\hat{\mathcal{T}}^K_{Q,i}$) and in the vertical 
reference system ($\mathcal{T}^K_{Q,i}$) is
\begin{equation}
	\hat{\mathcal{T}}^K_{Q,i}(\vec{r},\vec{\Omega}) = \sum_{Q'=-K}^K 
	\mathcal{T}^K_{Q',i}(\vec{\Omega}) \, 
	\overline{\mathcal{D}}^K_{QQ'}(R_B(\vec{r})) \, ,
 \label{eq:TauKQ}
\end{equation}
where $\mathcal{D}^K_{QQ'}$ is the \textit{rotation matrix} (e.g., Sect.~2.6 of LL04), and $R_B$ is the rotation that brings the magnetic reference system onto the vertical one.
A bar over a quantity indicates the complex conjugate. 
It must be noticed that the tensor $\mathcal{T}^K_{Q,i}$ defined in the vertical reference system
only depends on the propagation direction of the incident (or scattered) radiation, and does not 
depend on the spatial point $\vec{r}$.
The rotation $R_B$ is specified by 
the Euler angles $R_B(\vec{r})=(0,-\theta_B(\vec{r}),-\chi_B(\vec{r}))$, where $\theta_B$ and $\chi_B$ 
are the inclination and azimuth, respectively, of the magnetic field in the vertical reference system.
In the vertical reference system, the redistribution matrices are thus given by
\begin{equation}
	R^{\scriptscriptstyle \mathrm{X}}_{ij}(\vec{r},\vec{\Omega},\vec{\Omega}',
	\xi,\xi') = 
	\sum_{K,K'=0}^2 \sum_{Q=-K_{\rm min}}^{K_{\rm min}} 
	\mathcal{R}^{{\scriptscriptstyle \mathrm{X}},KK'}_Q(\vec{r},\xi,\xi') \, 
	\mathcal{P}^{KK'}_{Q,ij}(\vec{r},\vec{\Omega},\vec{\Omega}') \, ,
	\label{eq:RX_atom_vert}
\end{equation}
with
\begin{equation}
	\mathcal{P}^{KK'}_{Q,ij}(\vec{r},\vec{\Omega},\vec{\Omega}') = 
	\sum_{Q'=-K}^K \sum_{Q''=-K'}^{K'} (-1)^{Q'} \,
	\mathcal{T}^{K'}_{Q'',i}(\vec{\Omega}) \,
	\mathcal{T}^{K}_{-Q',j}(\vec{\Omega}') \,
	\overline{\mathcal{D}}^{K'}_{QQ''}(\vec{r}) \, 
	\mathcal{D}^{K}_{QQ'}(\vec{r}) \, .
	\label{eq:PKKpQ}
\end{equation}
For notational simplicity, we have only included the spatial point dependency of the rotation matrices, leaving implicit that the rotation $R_B$ is always considered. Furthermore, we have used the relation
\begin{equation*}
	\overline{\mathcal{D}}^K_{QQ'}(R) = (-1)^{Q-Q'}\mathcal{D}^K_{-Q -Q'}(R) \, .
\end{equation*}

\subsection{Expression in the observer's reference frame}
\label{sec:R_observer}
We now present the expressions of $R^{\scriptscriptstyle \mathrm{II}}$ and $R^{\scriptscriptstyle \mathrm{III}}$ in the observer's frame, where it is assumed that the atom is moving with velocity $\vec{v}$.
Considering the Doppler effect, the frequencies measured in the atomic frame, 
$\xi'$ and $\xi$, and those measured in the observer's frame, $\nu'$ and 
$\nu$, are related by:
\begin{align}
	\xi' & = \nu' - \frac{\nu_0}{c} \vec{v} \cdot \vec{\Omega}' \, , 
	\label{eq:xip_nup} \\
	\xi & = \nu - \frac{\nu_0}{c} \vec{v} \cdot \vec{\Omega} \, ,
	\label{eq:xi_nu} 
\end{align}
where $c$ is the speed of light.
The velocity $\vec{v}$ is generally given by the sum of two terms, namely,
\begin{equation}
	\vec{v}(\vec{r}) = \vec{v}_{\mathrm{th}}(\vec{r}) +
	\vec{v}_{\mathrm{b}}(\vec{r}) \, ,
\end{equation}
where $\vec{v}_{\mathrm{th}}$ is the thermal component and 
$\vec{v}_{\mathrm{b}}$ is the bulk component.
The thermal component is generally well described by a Maxwellian distribution
\begin{equation}
	\mathcal{P}(\vec{v}_{\mathrm{th}}(\vec{r})) = 
	\left( \frac{m}{2 \pi k_B T(\vec{r})} \right)^{3/2} \, 
	\mathrm{exp} \left( -\frac{m v_{\mathrm{th}}(\vec{r})^2}{2 k_B T(\vec{r})} 
	\right) \, ,
\end{equation}
where $k_B$ is the Boltzmann constant, $T$ the temperature, and $m$ the mass 
of the considered atom or ion.
Let $\check{\mathcal{R}}^{{\scriptscriptstyle \mathrm{X}},KK'}_Q$ be the 
frequency-dependent part of the redistribution matrix (see 
Eq.~(\ref{eq:RII_freq_atom}) for $R^{\scriptscriptstyle \mathrm{II}}$ and 
Eq.~(\ref{eq:RIII_freq_atom}) for $R^{\scriptscriptstyle \mathrm{III}}$), 
expressed in terms of the frequencies $\nu$ and $\nu'$ through 
Eqs.~(\ref{eq:xi_nu}) and (\ref{eq:xip_nup}), for an atom moving with 
velocity $\vec{v}$.
The expression of $\mathcal{R}^{{\scriptscriptstyle \mathrm{X}},KK'}_Q$ in the observer's frame is obtained by locally averaging 
$\check{\mathcal{R}}^{{\scriptscriptstyle \mathrm{X}},KK'}_Q$ over the 
distribution of thermal velocities:
\begin{equation}
	\mathcal{R}^{{\scriptscriptstyle \mathrm{X}},KK'}_Q(\vec{r},\vec{\Omega},
	\vec{\Omega}',\nu,\nu') = 
	\int \mathrm{d}^3 \vec{v}_{\mathrm{th}}(\vec{r}) \, 
	\mathcal{P}(\vec{v}_{\mathrm{th}}(\vec{r})) \, 
	\check{\mathcal{R}}^{{\scriptscriptstyle \mathrm{X}},KK'}_Q(\vec{r},
	\vec{\Omega},\vec{\Omega}',\nu,\nu') \, .
\end{equation}
This average is performed following the same approach as in the unpolarized 
case \cite[e.g.,][]{hummer1962non,mihalas1978stellar}. %(e.g., Hummer 1962, Chapt.~13 of Mihalas 1978).
We provide the final expressions, which are better formulated by defining the Doppler width 
\begin{equation}
	\Delta \nu_D(\vec{r})=\frac{\nu_0}{c} \, 
    \sqrt{\frac{2 k_B T(\vec{r})}{m} 
    %+ {v_{mt}}^2 
    }\, ,
\label{eq:dnd}
\end{equation}
%
% where $v_{mt}$ is the micro-turbulent velocity.
the damping constant $a(\vec{r}) = \Gamma(\vec{r}) / \Delta \nu_D(\vec{r})$, and the reduced frequency
\begin{equation}
\label{eq:reduced_freq}
	u(\vec{r},\nu) = \frac{(\nu_0 - \nu)}{\Delta \nu_D(\vec{r})} \, . 
\end{equation}
Moreover, we introduce the reduced magnetic and bulk velocity shifts, respectively defined as
\begin{equation}
\label{eq:reduced_shifts}
	u_{M_u M_\ell}(\vec{r}) = \frac{\nu_L(\vec{r}) 
	(g_u M_u - g_\ell M_\ell)}{\Delta \nu_D(\vec{r})} \, , \, {\rm and} \;\;
	u_b(\vec{r},\vec{\Omega}) = \frac{\nu_0}{c} \, 
	\frac{\vec{v}_b(\vec{r}) \cdot \vec{\Omega}}{\Delta \nu_D(\vec{r})} \, . 
\end{equation}
In order to make the notation lighter, in the following equations, we will also use the variable
\begin{equation}
	\tilde{u}_{M_u M_\ell}(\vec{r},\vec{\Omega},\nu) = u(\vec{r},\nu) + 
	u_{M_u M_\ell}(\vec{r}) + u_b(\vec{r},\vec{\Omega}) \, . 
\label{eq:tilde_u}
\end{equation}

\subsubsection{$R^{\scriptscriptstyle \mathrm{II}}$ redistribution matrix}
\label{sec:RII_observer}

The $R^{\scriptscriptstyle \mathrm{II}}$ redistribution matrix in the 
observer's frame, taking the quantization axis along the vertical, is given by
\begin{align}
	R^{\scriptscriptstyle \mathrm{II}}_{ij}(\vec{r},\vec{\Omega},
	\vec{\Omega}',\nu,\nu') = 
	\sum_{K,K'=0}^2 \sum_{Q=-K_{\rm min}}^{K_{\rm min}} 
	\mathcal{R}^{{\scriptscriptstyle \mathrm{II}},KK'}_Q(\vec{r},\vec{\Omega}, 
	\vec{\Omega}',\nu,\nu') \, 
	\mathcal{P}^{KK'}_{Q,ij}(\vec{r},\vec{\Omega},\vec{\Omega}') \, ,
	\label{eq:RII_obs_vertical}
\end{align}
where $\mathcal{P}^{KK'}_{Q,ij}$ is given by Eq.~(\ref{eq:PKKpQ}) and 
$\mathcal{R}^{{\scriptscriptstyle \mathrm{II}},KK'}_Q$ takes different 
analytic expressions depending on relative orientation of $\vec{\Omega}$ 
and $\vec{\Omega}'$.\footnote{For notational simplicity, the ranges of the 
various sums are not explicitly indicated.}

\begin{itemize}
\item{If $\vec{\Omega}' \ne \vec{\Omega}, -\vec{\Omega}$:
\begin{align}
	\label{eq:RII_KKpQ}
	\mathcal{R}^{{\scriptscriptstyle \mathrm{II}},KK'}_{Q}(\vec{r},
	\vec{\Omega},\vec{\Omega}',\nu,\nu') & =  
	 \frac{1}{\Delta \nu_D^2(\vec{r})}  \, \alpha_Q(\vec{r}) \, 
	\sum_{M_u, M'_u} \sum_{M_\ell, M'_\ell} \sum_{p, p', p'', p'''}
	{\mathcal C}_{K K' Q M_u M'_u M_\ell M'_\ell \, p p' p'' p'''} 
	\nonumber \\
	& \quad \times 
	\frac{1}{2 \pi \sin{\Theta}} \, 
	\mathrm{exp} \left[ - \left( 
	\frac{\tilde{u}_{M_u M_\ell'}(\vec{r},\vec{\Omega},\nu) - 
	\tilde{u}_{M_u M_\ell}(\vec{r},\vec{\Omega}',\nu')}
	{2 \sin(\Theta/2)} \right)^2 \right] 
	\nonumber \\
	& \quad \times 
	\Bigg[ 
	W\left(\frac{a(\vec{r})}{\cos(\Theta/2)}, 
	\frac{\tilde{u}_{M_u' M_\ell'}(\vec{r},\vec{\Omega},\nu) + 
	\tilde{u}_{M_u' M_\ell}(\vec{r},\vec{\Omega}',\nu')} 
	{2 \cos(\Theta/2)} \right) 
	\nonumber \\ 
	& \qquad +
	\overline{W}\left(\frac{a(\vec{r})}{\cos(\Theta/2)}, 
	\frac{\tilde{u}_{M_u M_\ell'}(\vec{r},\vec{\Omega},\nu) + 
	\tilde{u}_{M_u M_\ell}(\vec{r},\vec{\Omega}',\nu')} 
	{2 \cos(\Theta/2)} \right) \Bigg] 
	\, ,
\end{align}
where $\Theta$ is the angle between the directions $\vec{\Omega}$ and 
$\vec{\Omega}'$ (scattering angle)
\begin{equation}
\label{Eq:scattering_angle}
	\cos{\Theta} = \vec{\Omega} \cdot \vec{\Omega}' = 
	\cos{\theta} \, \cos{\theta'} + \sin{\theta} \, \sin{\theta'} \, 
	\cos{(\chi - \chi')} \, .
\end{equation}
In the previous equation, $\theta$ and $\chi$ are respectively the 
inclination and azimuth of the direction $\vec{\Omega}$, and $\theta'$ and 
$\chi'$ are the same angles for the direction $\vec{\Omega}'$.
The Faddeeva function $W(y,x)$ is defined as (e.g., Sect.~5.4 of LL04):
\begin{equation}
\label{eq:faddeeva}
	W(y,x) = H(y,x) + \mathrm{i}L(y,x) = \mathrm{e}^{-z^2} \, 
	\mathrm{erfc}(-\mathrm{i}z) \, ,
\end{equation}
where $H$ and $L$ are the Voigt and associated dispersion profiles, 
respectively, erfc is the complementary error function, and $z=x+\mathrm{i}y$.
}

\item{If $\vec{\Omega}' = \vec{\Omega}$ (forward scattering):
\begin{align}
	\mathcal{R}^{{\scriptscriptstyle \mathrm{II}},KK'}_{Q}(\vec{r},
	\vec{\Omega},\vec{\Omega},\nu,\nu') & =   
	 \frac{1}{\Delta \nu_D^2(\vec{r})}  \, \alpha_Q(\vec{r}) \, 
	\sum_{M_u,M'_u} \sum_{M_\ell, M'_\ell} \, \sum_{p, p', p'', p'''}
	{\mathcal C}_{K K' Q M_u M'_u M_\ell M'_\ell \, p p' p'' p'''} 
	\nonumber \\
	& \quad \times 
	\frac{1}{2 \pi^{1/2}} \, \left[ 
	W(a(\vec{r}), \tilde{u}_{M'_u M_\ell}(\vec{r},\vec{\Omega},\nu')) +
	\overline{W}(a(\vec{r}), \tilde{u}_{M_u M_\ell}(\vec{r},\vec{\Omega},\nu')) 
	\right]
	\nonumber \\
	& \quad \times
	\delta \left( \tilde{u}_{M_u M_\ell}(\vec{r},\vec{\Omega},\nu') - 
	\tilde{u}_{M_u M'_\ell}(\vec{r},\vec{\Omega},\nu) \right) \, .
\label{eq:RII_KKpQ_T0}
\end{align}
}

\item{If $\vec{\Omega}' = -\vec{\Omega}$ (backward scattering):
\begin{align}
	\mathcal{R}^{{\scriptscriptstyle \mathrm{II}},KK'}_{Q}(\vec{r},
	\vec{\Omega},-\vec{\Omega},\nu,\nu') & =  
	 \frac{1}{\Delta \nu_D^2(\vec{r})}  \, \alpha_Q(\vec{r}) \, 
	\sum_{M_u, M'_u} \sum_{M_\ell, M'_\ell} \, \sum_{p, p', p'', p'''}
	{\mathcal C}_{K K' Q M_u M'_u M_\ell M'_\ell \, p p' p'' p'''} 
	\nonumber \\
	& \quad \times 
	\frac{1}{4 \pi^{3/2}} \, \exp{ \left[ - \left( 
	\frac{\tilde{u}_{M_u M'_\ell}(\vec{r},\vec{\Omega},\nu) - 
	\tilde{u}_{M_u M_\ell}(\vec{r},-\vec{\Omega},\nu')}{2} \right)^2 \right] }
	\nonumber \\
	& \quad \times 
	\Bigg[
	\varphi \left( a(\vec{r}), \frac{
	\tilde{u}_{M'_u M'_\ell}(\vec{r},\vec{\Omega},\nu) + 
	\tilde{u}_{M'_u M_\ell}(\vec{r},-\vec{\Omega},\nu')}{2} \right)
	\nonumber \\
    & \qquad	+ 
	\overline{\varphi} \left( a(\vec{r}), \frac{
	\tilde{u}_{M_u M'_\ell}(\vec{r},\vec{\Omega},\nu) + 
	\tilde{u}_{M_u M_\ell}(\vec{r},-\vec{\Omega},\nu')}{2} \right)
	\Bigg] \, ,
\label{Eq:RII_KKpQ_Tpi}
\end{align}
where $\varphi(y,x)$ is defined as
\begin{equation}
	\varphi(y,x) = \frac{1}{y - \mathrm{i}x} \, .
	\label{eq:phi_obs}
\end{equation}
}
\end{itemize}

\subsubsection{$R^{\scriptscriptstyle \mathrm{III}}$ redistribution matrix}
\label{sec:RIII_observer}

The $R^{\scriptscriptstyle \mathrm{III}}$ redistribution matrix in the 
observer's frame, taking the quantization axis along the vertical, is given by
\begin{align}
	R^{\scriptscriptstyle \mathrm{III}}_{ij}(\vec{r},\vec{\Omega},
	\vec{\Omega}',\nu,\nu') = 
	\sum_{K,K'=0}^2 \sum_{Q=-K_{\rm min}}^{K_{\rm min}} 
	\mathcal{R}^{{\scriptscriptstyle \mathrm{III}},KK'}_Q(\vec{r},
	\vec{\Omega},\vec{\Omega}',\nu,\nu') \, 
	\mathcal{P}^{KK'}_{Q,ij}(\vec{r},\vec{\Omega},\vec{\Omega}') \, ,
	\label{eq:RIII_obs_vertical}
\end{align}
where $\mathcal{P}^{KK'}_{Q,ij}$ is given by Eq.~(\ref{eq:PKKpQ}) and 
$\mathcal{R}^{{\scriptscriptstyle \mathrm{III}},KK'}_Q$ by the following
expression
\begin{align}
	& \mathcal{R}^{{\scriptscriptstyle \mathrm{III}},KK'}_Q(\vec{r},
	\vec{\Omega},\vec{\Omega}',\nu,\nu') = \frac{1}{\Delta \nu_D^2(\vec{r})} \, 
	\sum_{K''=|Q|}^{2J_u} 
	\left( \beta^{K''}_Q(\vec{r}) - \alpha_Q(\vec{r}) \right)
	\nonumber \\
	& \qquad \qquad \times 
	\sum_{M_u, M_u'} \sum_{M_\ell} \sum_{q, q'} 
	\mathcal{B}_{K'' K' Q M_u M_u' M_\ell q q'} \, 
	\sum_{M_u'', M_u'''} \sum_{M_\ell'} \sum_{q'', q'''} 
	\mathcal{B}_{K'' K Q M_u'' M_u''' M_\ell' q'' q'''}	
	\nonumber \\
	& \qquad \qquad \times
	\frac{1}{4} \bigg[
		\mathcal{I}_{(M_u M_\ell),(M_u'' M_\ell')}
		(\vec{r},\vec{\Omega},\vec{\Omega}',\nu,\nu') + 
		\mathcal{I}_{(M_u M_\ell),\overline{(M_u''' M_\ell')}}
		(\vec{r},\vec{\Omega},\vec{\Omega}',\nu,\nu') \nonumber \\
		& \qquad \qquad \qquad +
		\mathcal{I}_{\overline{(M_u' M_\ell)},(M_u'' M_\ell')}
		(\vec{r},\vec{\Omega},\vec{\Omega}',\nu,\nu') +
		\mathcal{I}_{\overline{(M_u' M_\ell)},\overline{(M_u''' M_\ell')}}
		(\vec{r},\vec{\Omega},\vec{\Omega}',\nu,\nu') 
		\bigg] \, .
	\label{eq:RIII_KKpQ_obs}
\end{align}
The quantity $\mathcal{I}_{(M_u M_\ell),(M_u' M_\ell')}$ is given by the integral of the product of three functions: an exponential function, which only depends on the integration variable $y$; a function depending on $y$, the frequency, and direction of the scattered radiation, and the first pair of magnetic quantum numbers in the subscript; and a function depending on $y$, the frequency and direction of the incident radiation, and the second pair of magnetic quantum numbers in the subscript.
A bar over a pair of magnetic quantum numbers means that the complex conjugate of the corresponding function has to be considered.
The explicit expression of $\mathcal{I}_{(M_u M_\ell),(M_u' M_\ell')}$ is
provided below.
\begin{itemize}
\item{If $\vec{\Omega}' \ne \vec{\Omega},-\vec{\Omega}$:
\begin{align}
	\mathcal{I}_{(M_u M_\ell),(M_u' M_\ell')}
	(\vec{r},\vec{\Omega},\vec{\Omega}',\nu,\nu')
	 = \frac{1}{\pi^2 \, \sin{\Theta}} \,  
	& \int \mathrm{d} y \, \mathrm{exp} \left( -y^2 \right)
	\nonumber \\
	& \quad \times 
	W \left( \frac{a(\vec{r})}{\sin{\Theta}} ,
	\frac{
	\tilde{u}_{M_u M_\ell}(\vec{r},\vec{\Omega},\nu) + 
	y \cos{\Theta}}{\sin{\Theta}} \right) \,
 % \times 
 \;
	\varphi\left( a(\vec{r}), \tilde{u}_{M_u' M_\ell'}(\vec{r},
	\vec{\Omega}'\nu') + y \right)
	\, .
	\label{eq:int_final} 
\end{align}

}

\item{If $\vec{\Omega}' = \vec{\Omega}$ (forward scattering):
\begin{align}
	\mathcal{I}_{(M_u M_\ell),(M_u' M_\ell')}
	(\vec{r},\vec{\Omega},\vec{\Omega},\nu,\nu')
	 = \frac{1}{\pi^{5/2}} \,  
	& \int  \mathrm{d} y \, \mathrm{exp} \left( -y^2 \right)
	\nonumber \\
	& \quad \times 
	\varphi \left( a(\vec{r}), \tilde{u}_{M_u M_\ell}(\vec{r},\vec{\Omega},
	\nu) + y\right) \, 
% \times 
\;
	\varphi \left( a(\vec{r}), \tilde{u}_{M_u' M_\ell'}(\vec{r},\vec{\Omega},
	\nu') + y \right) 
	\, .
	\label{eq:int_final_T0}
\end{align}
}

\item{If $\vec{\Omega}' = -\vec{\Omega}$ (backward scattering):
\begin{align}
	\mathcal{I}_{(M_u M_\ell),(M_u' M_\ell')}
	(\vec{r},\vec{\Omega},-\vec{\Omega},\nu,\nu')
	 = \frac{1}{\pi^{5/2}} \,  
	& \int \mathrm{d} y \, \mathrm{exp} \left( -y^2 \right)
	\nonumber \\
	& \quad \times 
	\varphi \left( a(\vec{r}), \tilde{u}_{M_u M_\ell}(\vec{r},\vec{\Omega},
	\nu) - y \right) \,
% \times 
\;
	\varphi \left( a(\vec{r}), \tilde{u}_{M_u' M_\ell'}(\vec{r},-\vec{\Omega},
	\nu') + y \right) \, .
	\label{eq:int_final_Tpi}
\end{align}
}
\end{itemize}
The function $\varphi(y,x)$ is defined in Eq.~(\ref{eq:phi_obs}).

\subsection{Approximate expressions}
\label{sec:R_approx}
The expressions of the redistribution matrices in the observer's frame 
derived in the previous section show the complex coupling between frequencies 
and angles introduced by the Doppler effect. 
This coupling makes the evaluation of the scattering integral 
(\ref{eq:scat_int}) extremely demanding from a computational point 
of view. 
To reduce the computational cost of the problem, approximate expressions in which the frequency and angular dependencies are partially or totally decoupled are often used. %have been proposed in the past (REF).

In the absence of bulk velocities, or when working in a reference frame in 
which the bulk velocity is zero (comoving frame), the arguments of the 
exponential and of the Faddeeva functions do not depend on $\vec{\Omega}$ 
and $\vec{\Omega}'$, and the angular dependence of the  
$\mathcal{R}^{{\scriptscriptstyle \mathrm{II}},KK'}_Q$ function 
(\ref{eq:RII_KKpQ}) is reduced to the scattering angle $\Theta$. 
In this case, a commonly used approximation is the so-called 
\emph{angle-averaged} (AA) one, which consists in averaging this function 
with respect to $\Theta$ \citep[e.g.,][]{rees1982}:
\begin{equation}
	\mathcal{R}^{{\scriptscriptstyle \mathrm{II-AA}},KK'}_Q(\vec{r},\nu,\nu') 
	= \frac{1}{2} \int_0^\pi \mathrm{d} \Theta \, \sin{\Theta} \,
	\mathcal{R}^{{\scriptscriptstyle \mathrm{II}},KK'}_Q(\vec{r},\Theta,
	\nu,\nu') \, .
	\label{eq:RII-AA}
\end{equation}
The function $\mathcal{R}^{{\scriptscriptstyle \mathrm{II}},KK'}_Q$
in the r.h.s. of Eq.~(\ref{eq:RII-AA}) is given by Eq.~(\ref{eq:RII_KKpQ}), 
assuming no bulk velocities.
The ensuing $R^{{\scriptscriptstyle \mathrm{II-AA}}}_{ij}$ redistribution 
matrix in the observer's frame is characterized by a complete decoupling of 
the frequencies and angles. Thus, the computational cost of the 
problem is significantly lowered. 

The AA approximation can in principle be applied also to 
$\mathcal{R}^{{\scriptscriptstyle \mathrm{III}},KK'}_Q$ 
\citep[e.g.,][]{bommier1997masterII}. 
However, an even stronger assumption is often considered for this function, 
namely that there is no correlation between the frequencies of the incident 
and scattered radiation in the observer's frame \citep[e.g.,][]{mihalas1978stellar}. 
Under this approximation, often referred to as the limit of \emph{complete frequency redistribution} 
(CRD) in the observer's frame, we have:
\begin{align}
	\mathcal{R}^{{\scriptscriptstyle \mathrm{III-CRD}},KK'}_Q(\vec{r},
	\vec{\Omega},\vec{\Omega}',\nu,\nu') & = 
	\frac{1}{\Delta \nu_D^2(\vec{r})} \, \sum_{K''=|Q|}^{2J_u} 
	\left( \beta^{K''}_Q(\vec{r}) - \alpha_Q(\vec{r}) \right) \, 
	% \nonumber \\
	% & \quad \times 
	\Phi^{K''K'}_Q(\vec{r},\vec{\Omega},\nu) \, 
	\Phi^{K''K}_Q(\vec{r},\vec{\Omega}',\nu') \, , 
	\label{eq:RIII_KKpQ_CRD}
\end{align}
where $\Phi^{KK'}_Q$ is the generalized profile (see 
Eq.~(\ref{eq:gen_prof_atom})), defined in the observer's frame.
This is obtained by convolving the profiles $\Phi_{M_u M_\ell}$ of 
Eq.~(\ref{eq:Phi_atom}) with a Gaussian function in order to account for the 
thermal %and microturbulent 
velocity distribution, thus obtaining a Faddeeva 
function:
\begin{align}
	\Phi^{KK'}_Q(\vec{r},\vec{\Omega},\nu) & =
	\sum_{M_u, M_u'} \sum_{M_\ell} \sum_{q, q'} 
	\mathcal{B}_{K K' Q M_u M_u' M_\ell q q'} \,
	% \nonumber \\
	% & \quad \times 
	\frac{1}{2 \sqrt{\pi}} \, \left[
	W(a(\vec{r}),\tilde{u}_{M_u M_\ell}(\vec{r},\vec{\Omega},\nu))+
	\overline{W}(a(\vec{r}),\tilde{u}_{M_u' M_\ell}(\vec{r},\vec{\Omega},\nu))
	\right] \, .
	\label{eq:gen_prof_obs}
\end{align}
In the absence of bulk velocities or when working in the comoving frame, the 
$R^{{\scriptscriptstyle \mathrm{III-CRD}}}$ redistribution matrix is 
characterized by a complete decoupling between angles and frequencies.

%%%%%%%%%%%%%%%%%%%%%%%%%%%%%%%%%%%%%%%%%%%%%%%%%%%%%%%%%%%%%%%%

% \section{Propagation matrix and thermal emissivity}
\section{Line contribution to the propagation matrix and thermal emissivity}
\label{sec:thermal}

Neglecting stimulated emission (which is generally an excellent assumption in the solar atmosphere), the elements of the propagation matrix for a two-level atom with an unpolarized lower level, in the observer's reference frame, are given by (see App.~13 of LL04)
\begin{align}
	 \eta^{\ell}_i(\vec{r},\vec{\Omega},\nu)  = & \, k_L(\vec{r}) 
	\sum_{K=0}^2 \Phi^{0K}_0 \! (\vec{r},\vec{\Omega},\nu) \,
    \sum_{Q=-K}^K \mathcal{T}^K_{Q,i}(\vec{\Omega}) \, 
	\overline{\mathcal{D}}^K_{0Q}(\vec{r}) \, ,	\qquad (i=1,2,3,4) \, , \\
	 \rho^{\ell}_i (\vec{r},\vec{\Omega},\nu)  = & \, k_L(\vec{r}) 
	\sum_{K=0}^2 \Psi^{0K}_0 \! (\vec{r},\vec{\Omega},\nu) \,
    \sum_{Q=-K}^K \mathcal{T}^K_{Q,i}(\vec{\Omega}) \, 
	\overline{\mathcal{D}}^K_{0Q}(\vec{r}) \, , \qquad (i=2,3,4) \, ,
\end{align}
where $\mathcal{T}^K_{Q,i}$ is the geometrical tensor evaluated in the vertical reference system
and $\mathcal{D}^K_{QQ'}$ are the rotation matrices (see App.~\ref{app:R_vertical}).
The quantities $\Phi^{0K}_0$ are particular components of the generalized profile of 
Eq.~(\ref{eq:gen_prof_obs}), while $\Psi^{0K}_0$ are particular components of the generalized 
\emph{dispersion} profile, defined as (see App.~13 of LL04)
\begin{align}
	{\mathrm i} \, \Psi^{KK'}_Q(\vec{r},\vec{\Omega},\nu) & =
	\sum_{M_u, M_u'} \sum_{M_\ell} \sum_{q, q'} 
	\mathcal{B}_{K K' Q M_u M_u' M_\ell q q'} \,
	% \nonumber \\
	% & \quad \times 
	\frac{1}{2 \sqrt{\pi}} \, \left[
	W(a(\vec{r}),\tilde{u}_{M_u M_\ell}(\vec{r},\vec{\Omega},\nu)) -
	\overline{W}(a(\vec{r}),\tilde{u}_{M_u' M_\ell}(\vec{r},\vec{\Omega},\nu)) \right] \, .
	\label{eq:gen_disp_prof_obs}
\end{align}
The explicit expression of the frequency-integrated line absorption coefficient $k_L$ is
\begin{equation}
	k_L(\vec{r}) = \frac{h \nu_0}{4 \pi} B_{\ell u} \, \mathcal{N}_{\ell}(\vec{r}) = 
	\frac{c^2}{8 \pi \nu_0^2} \frac{2J_u+1}{2J_\ell+1} A_{u \ell} \,
	\mathcal{N}_{\ell}(\vec{r})	\, ,
\end{equation}
where $A_{u \ell}$ and $B_{\ell u}$ are the Einstein coefficients for spontaneous emission and absorption, respectively, and $\mathcal{N}_{\ell}$ is the population of the lower level.
Under the assumption of isotropic inelastic collisions, the
line thermal contribution to the emissivity is given by 
\citep[e.g.,][]{ballester2017transfer}:
\begin{equation}
\label{eq:eps_th}
     \varepsilon^{\ell, \rm th}_{i} \left( \vec{r}, \vec{\Omega}, \nu \right ) = 
    \epsilon\left( \vec{r} \right) \,
	W \left ( \nu , T (\vec{r}) \right ) \,
    \eta^{\ell}_i(\vec{r}, \vec{\Omega}, \nu)  \, ,	\qquad (i=1,2,3,4) \, ,
\end{equation}
where $W$ is the Planck function in the Wien limit (consistently with the assumption of neglecting stimulated emission), and $\epsilon$ is the photon destruction probability defined in Eq.~(\ref{eq:epsilon_pdp}).
%

%%%%%%%%%%%%%%%%%%%%%%%%%%%%%%%%%%%%%%%%%%%%%%%%%%%%%%%%%%%%%%%%

\section{Continuum contributions}
\label{sec:continuum}
In the visible range of the solar spectrum, continuum processes only contribute to the emission coefficient (with a thermal and a scattering term) and to the absorption coefficient for intensity. 
Labeling continuum contributions with the apex `$c$', we have:
\begin{align}
    & \varepsilon^c_i \left( \vec{r}, \vec{\Omega}, \nu \right) =
    \varepsilon^{c, {\rm th}}_I \left( \vec{r}, \nu \right)  \delta_{i1}  +
    \varepsilon^{c, {\rm sc}}_i \left( \vec{r}, \vec{\Omega}, \nu \right) 
    \quad (i = 1, 2, 3, 4) \, , \label{eq:epsilon_c} \\
    & \eta_i^c \left( \vec{r}, \vec{\Omega}, \nu \right) = k_c \left( \vec{r}, \nu \right)  \delta_{i1}   
    \qquad \qquad \qquad \qquad (i = 1, 2, 3, 4) \, , \\
    & \rho_i^c \left( \vec{r}, \vec{\Omega}, \nu \right) = 0 
    \qquad \qquad \qquad \qquad \qquad \quad \; \; (i = 2, 3, 4) \, ,
\end{align}
where $\varepsilon^{c, {\rm th}}_I$ is the continuum thermal emissivity and $k_c$ the continuum total opacity.
% and the can be evaluated using existing codes (e.g., RH).
%
Under the assumption that continuum scattering processes are coherent in the observer's frame, the scattering contribution to the continuum emission coefficient is given by
\begin{align}
    \varepsilon^{c, {\rm sc}}_i\left( \vec{r}, \vec{\Omega}, \nu \right) = 
    \sigma \left( \vec{r},  \nu \right) \sum_{K=0}^{2} \sum_{Q=-K}^{K} (-1)^Q \,
    \mathcal{T}^{K}_{Q,i} \left(\vec{\Omega}\right)
    \, J^{K}_{-Q}  \left( \vec{r}, \nu +\frac{\nu_0}{c}\vec{v_b}\cdot(\vec{\Omega}'-\vec{\Omega}) \right) \, ,  
\end{align}
where $\sigma$ is the continuum absorption coefficient for scattering and $J^{K}_{Q}$ is 
the radiation field tensor (see Sect. 5.11 of LL04), given by
\begin{equation}
\label{eq:JkqAppendix}
    J^{K}_{Q} \left( \vec{r}, \nu \right) = 
    \oint  \frac{{\mathrm d} \vec{\Omega}}{ 4 \pi } \sum_{j=1}^4 
    \mathcal{T}^{K}_{Q,j}\left(\vec{\Omega}\right)
    I_j \left(\vec{r},\vec{\Omega},\nu \right) \, .
\end{equation}
We note that we neglected the impact of bulk velocities on the $\varepsilon^{c, {\rm th}}_I$, $k_c$, and $\sigma$ coefficients.
Line and continuum contributions to the emissivity and
propagation matrix simply add together.

%%%%%%%%%%%%%%%%%%%%%%%%%%%%%%%%%%%%%%%%%%%%%%%%%%%%%%%%%%
\section{Micro-structured isotropic magnetic field}\label{app:micB}

In this section, we provide the expressions of the various RT coefficients in the presence of an unimodal micro-structured magnetic field, namely a magnetic field with a given intensity and an orientation that changes over scales below the photons' mean-free path.
In particular, we consider a magnetic field whose orientation is isotropically distributed over all possible directions.
% Hereafter, such a magnetic field will be referred to as ... .
To describe this scenario, the RT coefficients must be averaged over this distribution of magnetic field orientations:
\begin{align}
	\tilde{\eta}^{\ell}_i(\vec{r},\vec{\Omega},\nu) & = 
	\langle \eta^{\ell}_i(\vec{r},\vec{\Omega},\nu) \rangle = 
	\frac{1}{4 \pi} \int_0^{2\pi} \mathrm{d}\chi_B \int_0^{\pi} 
	\mathrm{d} \theta_B \sin{\theta_B} \, 
	\eta^{\ell}_i(\vec{r},\vec{\Omega},\nu) \, , \\
	\tilde{\rho}^{\ell}_i(\vec{r},\vec{\Omega},\nu) & = 
	\langle \rho^{\ell}_i(\vec{r},\vec{\Omega},\nu) \rangle = 
	\frac{1}{4 \pi} \int_0^{2\pi} \mathrm{d}\chi_B \int_0^{\pi} 
	\mathrm{d} \theta_B \sin{\theta_B} \, 
	\rho^{\ell}_i(\vec{r},\vec{\Omega},\nu) \, , \\
	\tilde{\varepsilon}^{\ell}_i(\vec{r},\vec{\Omega},\nu) & = 
	\langle \varepsilon^{\ell}_i(\vec{r},\vec{\Omega},\nu) \rangle = 
	\frac{1}{4 \pi} \int_0^{2\pi} \mathrm{d}\chi_B \int_0^{\pi} 
	\mathrm{d} \theta_B \sin{\theta_B} \, 
	\varepsilon^{\ell}_i(\vec{r},\vec{\Omega},\nu) \, ,
\end{align}
Recalling the expressions of the RT coefficients in the presence of a deterministic magnetic field, and observing that all the dependence on the orientation of the magnetic field is contained in the rotation matrices, it can be easily verified that the calculation of the averages above reduces to the evaluation of the following integrals \citep[see Eqs.~(47a) and (47b) of][]{ballester2017transfer}
\begin{align}
	& \frac{1}{4 \pi} \int_0^{2\pi} \mathrm{d}\chi_B \int_0^{\pi} 
	\mathrm{d} \theta_B \sin{\theta_B} \, \overline{\mathcal{D}}^K_{0Q}(\vec{r}) = 
	\delta_{K0} \, \delta_{Q0} \, , \\
	& \frac{1}{4 \pi} \int_0^{2\pi} \mathrm{d}\chi_B \int_0^{\pi} 
	\mathrm{d} \theta_B \sin{\theta_B} \, 
	\overline{\mathcal{D}}^{K'}_{QQ''}(\vec{r}) \, \mathcal{D}^{K}_{QQ'}(\vec{r}) = 
	\delta_{KK'} \, \delta_{Q'Q''} \frac{1}{2K+1} \, .
\end{align}
Considering the integrals above, it can be immediately seen that the only non-zero element of the propagation matrix is:
\begin{equation}
    \tilde{\eta}_1^{\ell} \left( \vec{r}, \vec{\Omega}, \nu \right) = 
    k_L\left( \vec{r} \right) \,
    \Phi^{00}_0 \left( \vec{r}, \vec{\Omega}, \nu \right).
    \label{eq:eta_MSI_B}
\end{equation}
Similarly, the only non-zero element of the thermal emissivity is:
\begin{equation}
\tilde{\varepsilon}_1^{\ell, {\rm th}} \left( \vec{r}, \vec{\Omega}, \nu \right) = 
    \epsilon \left( \vec{r} \right) \,
    W_T \left( \vec{r}, \nu \right) \, \tilde{\eta}_1^{\ell} \left( \vec{r}, \vec{\Omega}, \nu \right).
    \label{eq:eps_term_MSI_B}
\end{equation}
%janettII
Finally, the scattering term of the emissivity is still given by Eq.~(\ref{eq:scat_int}), considering 
the redistribution matrices
\begin{equation}
	\tilde{R}^{\scriptscriptstyle \mathrm{X}}_{ij}(\vec{r},\vec{\Omega},\vec{\Omega}',\nu,\nu') =
	\sum_{K=0}^2 \; \sum_{Q=-K}^{K} 
	\mathcal{R}^{{\scriptscriptstyle \mathrm{X}},KK}_{Q}(\vec{r},\vec{\Omega},\vec{\Omega}',\nu,\nu') \, 
	\tilde{\mathcal{P}}^{K}_{ij}(\vec{r},\vec{\Omega},\vec{\Omega}') \, ,
\end{equation}
where $\mathcal{R}^{{\scriptscriptstyle \mathrm{II}},KK}_{Q}$ and 
$\mathcal{R}^{{\scriptscriptstyle \mathrm{III}},KK}_{Q}$ 
have the expressions provided in Sects.~\ref{sec:RII_observer} and \ref{sec:RIII_observer}, respectively, and
\begin{equation}
    \tilde{\mathcal{P}}^{K}_{ij}\left( \vec{r} , \vec{\Omega}, \vec{\Omega}' \right ) = 
    \frac{1}{2K+1} \sum_{Q'=-K}^{K} (-1)^{Q'} \mathcal{T}^K_{Q',i}\left( \vec{\Omega} \right ) \, 
    \mathcal{T}^K_{-Q',j}\left( \vec{\Omega}' \right ) \, .
    \label{eq:PKKpQ_MSIB}
\end{equation}

\end{document}